\documentclass[12pt]{article}

\usepackage{subfig} 
\usepackage{url}
\usepackage{wrapfig}
\usepackage{eufrak} 
\usepackage{amsmath}   
\usepackage{a4p}
\usepackage{microtype}\usepackage{lmodern}\usepackage[T1]{fontenc}
\usepackage{mathptmx}
\usepackage{graphicx}
\usepackage[normalem]{ulem} 
\usepackage{color}
\definecolor{myred}{rgb}{0.6,0,0} 
\definecolor{myblue}{rgb}{0,0.2,0.4}
\definecolor{mygreen}{rgb}{0,0.9,0.1}
\definecolor{Orange}{rgb}{1.,0.65,0.}
\usepackage{color}
\definecolor{myred}{rgb}{1.0,0,0} 
\definecolor{mygreen}{rgb}{0,0.9,0.1} 
\definecolor{myblue}{rgb}{0,0.2,0.4}
\definecolor{mygray}{rgb}{.8,.8,.8}
\definecolor{darkorange}{rgb}{1, 0.549, 0}
\definecolor{purple}{rgb}{0.6,0.4,0.6}
\definecolor{mymagenta}{rgb}{0.6,0.4,0.6}
 \definecolor{LightCyan} {rgb}{0.88,1.,1.}
 \definecolor{Orange} {rgb}{1.,0.65,0.}
 \definecolor{PaleGreen} {rgb}{0.6,0.98,0.6}
 \definecolor{Pink} {rgb}{1.,0.75,0.8}
\definecolor{Red}{rgb}{1,0,0}
   \definecolor{Blue}{rgb}{0,0,1}
   \definecolor{Yellow}{rgb}{1,1,0}
   \definecolor{Orange}{rgb}{1,0.4,0}
   \definecolor{Pink}{rgb}{1,0,1}
   \definecolor{Purple}{rgb}{0.5,0,0.5}
   \definecolor{Teal}{rgb}{0,0.5,0.5}
   \definecolor{Navy}{rgb}{0,0,0.5}
   \definecolor{Aqua}{rgb}{0,1,1}
   \definecolor{Lime}{rgb}{0,1,0}
   \definecolor{Green}{rgb}{0,0.5,0}
   \definecolor{Olive}{rgb}{0.5,0.5,0}
   \definecolor{Maroon}{rgb}{0.5,0,0}
   \definecolor{Brown}{rgb}{0.6,0.4,0.2}
   \definecolor{Black}{gray}{0}
   \definecolor{Gray}{gray}{0.5}
   \definecolor{Silver}{gray}{0.75}
   \definecolor{White}{gray}{1}

\usepackage{hyperref}  
\definecolor{darkblue}{rgb}{0,0,.5}
\hypersetup{
linktocpage=true,
    bookmarks=true,         
breaklinks=true,
    frenchlinks=false, %
    unicode=false,          
    pdftoolbar=true,        
    pdfmenubar=true,        
    pdffitwindow=false,     
    pdfstartview={FitH},    
    pdftitle={one-loop tensor reduction},    
    pdfauthor={J.Fleischer, T.Riemann},     
    pdfsubject={tensor Feynman integrals },   
    pdfcreator={JF+TR},   
    pdfproducer={JF+TR}, 
    pdfkeywords={}, 
    pdfnewwindow=true,      
    colorlinks=false,       
    linkcolor=red,          
    citecolor=green,        
    filecolor=magenta,      
    urlcolor=cyan           
}
\usepackage[all]{hypcap} 

\usepackage{rotating}

\numberwithin{equation}{section}
\numberwithin{figure}{section}
\numberwithin{table}{section}

\newcommand{\be}{\begin{equation}}
\newcommand{\ee}{\end{equation}}
\newcommand{\bea}{\begin{eqnarray}}
\newcommand{\eea}{\end{eqnarray}}

\newcommand{\nl}{\nonumber \\}

\def \eps {\varepsilon}

\newcommand{\nn}{\nonumber}

\begin{document}
\sloppy

\texttt{
\begin{flushleft}
DESY 10-145
\\
BI-TP 2010/31
\\
HEPTOOLS 10-025
\\
SFB/CPP-10-86
\end{flushleft}
}
\vspace{1cm}

\bigskip

\begin{center}
{\LARGE \bf
A complete algebraic reduction of one-loop tensor
\\[3mm]
Feynman integrals}
\\
\vspace{1.0cm}

\renewcommand{\thefootnote}{\fnsymbol{footnote}}
{\Large J. Fleischer~\footnote{E-mail:~fleischer@physik.uni-bielefeld.de}~${}^{a}$  ~~and~~
        T. Riemann~\footnote{E-mail:~Tord.Riemann@desy.de}~${}^{b}$        }
\\[1cm]
\end{center}

{\noindent
{${}^{a}$~Fakult\"at f\"ur Physik, Universit\"at Bielefeld, Universit\"atsstr. 25,  33615
Bielefeld, Germany\\ }
\smallskip
{${}^{b}$~Deutsches Elektronen-Synchrotron, DESY, Platanenallee
  6, 15738 Zeuthen, Germany \\ }
}

\vspace{1cm}

\begin{center}
{\Large \bf
{Abstract}}
\\[5mm]
\end{center}
We set up a new, flexible
approach for the tensor reduction of one-loop Feynman integrals.
The $5$-point tensor integrals up to rank $R=5$  are expressed by $4$-point
tensor integrals of rank $R-1$,
such that the appearance of the inverse 5-point Gram determinant
is avoided.
The $4$-point tensor coefficients are represented in terms of
4-point integrals, defined in $d$ dimensions, $4-2\epsilon \le d \le 4-2\epsilon+2(R-1)$, with higher
powers of the propagators.
They can be further reduced to expressions which stay free of
the inverse 4-point Gram determinants
but contain higher-dimensional
$4$-point integrals with only the first power of scalar propagators, plus
$3$-point tensor coefficients.
A direct evaluation of the higher dimensional $4$-point functions
would avoid the appearance of inverse powers of
the Gram determinants completely.
The simplest approach, however, is to apply here dimensional recurrence relations in order to reduce  them
to the familiar $2$- to $4$-point functions in generic dimension $d = 4-2\eps$, introducing thereby coefficients with inverse 4-point Gram determinants up to power $R$ for tensors of rank $R$.
For small or vanishing Gram determinants -- where this reduction is not
applicable -- we use
analytic expansions in positive powers of the Gram determinants.
Improving the convergence of the expansions substantially
with Pad\'{e} approximants we close up to the evaluation of the $4$-point tensor coefficients
for larger Gram determinants.
Finally, some relations are discussed which may be useful for analytic simplifications
of Feynman diagrams.

\bigskip

PACS index categories: 12.15.Ji, 12.20.Ds, 12.38.Bx

\setcounter{footnote}{0}
\renewcommand{\thefootnote}{\arabic{footnote}}
\thispagestyle{empty}
\clearpage
\tableofcontents
\clearpage
\listoffigures
\listoftables

\newpage
\section{\label{Intro} Introduction}
The evaluation of tensorial Feynman integrals with $n$ external legs is an important
technical ingredient of perturbative quantum field theoretical calculations with
Feynman diagrams.
They are needed in particular for the fast and efficient numerical evaluation of next-to-leading
order contributions at high energy colliders. Special attention is concentrated these days on experiments performed at the LHC;
{for a snapshot on related activities see~\cite{Binoth:2010ra}.
Of course, there is an unlimited variety of other reasons to use tensor reductions of Feynman integrals with quite diverse requirements in detail,
and a unique all-purpose, final approach does not exist.
}
For a recent overview see~\cite{Denner:THHH2009} and references therein.

The first systematic approach to reduce {tensor components} to
{a basis of} scalar 1-point to 4-point integrals in generic dimension $d=4-2 \eps$
 for $n \le 4$  is the Passarino-Veltman reduction~\cite{Passarino:1978jh}, obtained by solving a system of linear equations.
In fact, this is a unique basis.

{We use, with some sophistication, Davydychev's approach~\cite{Davydychev:1991va}, where  $n$-point tensor coefficients  are represented in terms of scalar Feynman integrals.
For tensors of rank $R$ they are  defined  in space-time dimensions up to  $4-2\epsilon+2R$, with an additional modification:  propagators may appear with higher powers.
These integrals are complicated objects, and an important step towards their evaluation is the application of dimensional recurrence relations, derived for $L$-loop functions in~\cite{Tarasov:1996br}.
They have been systematically worked out for $L=1$ in~\cite{Fleischer:1999hq}, and
in a subsequent article~\cite{Fleischer:2003rm} the evaluation of scalar integrals in $d$ dimensions
with powers $1$ of the scalar propagators is advocated.
Alternatively, the straightforward derivation of  representations in the generic dimension  $4-2\epsilon$, or finally just in four dimensions, by means of recurrence relations
introduces coefficients containing inverse Gram determinants, which may become small in some
kinematical domains and thus raise numerical problems.
For $n \le 4$ this problem was not very severe
\cite{Devaraj:1997es}.
Serious problems arise, however, for $n$-point Feynman integrals with $n \geq 5$.
{In that case the choice of the tensor basis is not unique and the freedom may be used to
completely avoid the appearance of}
inverse Gram determinants for $n\geq 5$~\cite{Bern:1992em,Bern:1993kr,Bern:1994zx,Bern:1994cg,Campbell:1996zw,Binoth:1999sp,Denner:2002ii}.
On the contrary, for $n<5$
one has to find explicit methods to stabilize the numerics for vanishing
or small Gram determinants.

{In view of these facts, one may wonder if the approach of Davydychev can be worked out with an optimization of the handling of exceptional (small or vanishing) Gram determinants.}
This is what has been achieved in the present work.

For non-exceptional kinematics  the
$g^{\mu \nu}$ tensor, considering $5$-point functions, is redundant and may be
expressed in terms of $4$-momenta.
A nicely compact algebraic result is obtained due to this ansatz after applying \emph{symmetrized} dimensional recurrences
, but inverse Gram determinants of $5$-point as well as of $4$-point functions are introduced.
In an earlier attempt~\cite{Fleischer:2007ff,Diakonidis:2008ij,Gluza:2009mj}, tensor ranks until $R=3$ were presented \emph{without} inverse Gram determinants of the $5$-point
function using the \emph{algebra of the signed minors}~\cite{Melrose:1965kb},
but a generalization to higher ranks was not evident.
In the present work we first apply a particular recursion, which was obtained in~\cite{Diakonidis:2009fx}.
It reduces the tensor rank from $R$ to $R-1$ and the further reduction
can be arranged  in a systematic manner without introducing inverse 5-point Gram determinants.
The 4-point tensor coefficients, which are in fact, due to
\cite{Davydychev:1991va},
higher-dimensional $4$-point functions with higher powers of the scalar propagators,
were reduced by lengthy algebraic calculations
to  higher-dimensional $4$-point functions with powers $1$ of the scalar propgators  plus tensor coefficients of 3-point functions.
The latter higher-dimensional $4$-point functions are tensor coefficients of the $g^{\mu_1\mu_2}\cdots g^{\mu_{2l-1}\mu_{2l}}$ terms of 4-point tensors.
At this stage inverse 4-point Gram determinants are avoided completely.
The recursions of all remaining 3-point functions may be performed
{simply} \'{a} la~\cite{Fleischer:1999hq}.
{This is
the way the numerics for this article was performed. Nevertheless, the same approach
as above can also be applied to the $3$-point functions, leaving only
higher dimensional $3$-point functions with powers $1$ of the scalar propagators, thereby avoiding inverse Gram determinants of the $3$-point function.}

Finally, one has to calculate the higher dimensional $4$-point functions with
powers $1$ of the scalar propagators.
This may be done by direct evaluation, see e.g.~\cite{Binoth:2002xh}, or by further reduction to simpler integrals, which in general, however, introduces inverse powers of 4-point Gram determinants.
For small Gram determinants we therefore derive a relatively simple \emph{analytic expansion} in positive powers of the Gram determinant.
Such an infinite series was, to our knowledge, first proposed in equation~(36) of
\cite{Fleischer:2003rm}, but was not numerically applied so far.
In fact, this expansion applies for higher-dimensional integrals with powers $1$
of the scalar propagators only and would not be appropriate for a representation of tensor
coefficients in their original form.

{
Another approach to the problem of small Gram determinants was chosen in  sect.  5.4 of~\cite{Denner:2005nn},
where
relations between different  3- and 4-point tensor coefficients are exploited
and the full set is calculated with increasing iterations.
The $n^{th}$ iteration
requires all $3$-point coefficients of rank $n$.
In contrary to that our expansion is
concerned only with the subset of the $g^{\mu_1\mu_2}\cdots g^{\mu_{2l-1}\mu_{2l}}$
coefficients of the 4-point tensors, which are approximated only by the corresponding
subset of the $3$-point tensor coefficients.
Indeed, our series expansion of the higher dimensional 4-point functions
is useful only since these are embedded in expressions which are already free of
inverse Gram determinants .}

{
In~\cite{Fleischer:2003rm} it was shown that the integrals under consideration
can be expressed in terms of multiple hypergeometric functions.
One can apply their series expansion or, alternatively,
one could use their representations in terms of $1$-dimensional integrals also given in~\cite{Fleischer:2003rm}.
These, in general, present the integrals in rather different domains of phase space, including the case of small Gram determinants of the $4$-point function.
Thus our approach offers a variety of options to adjust to the given kinematical
situation. For the time being we only use our series expansion applying
Pad\'{e} approximants.}
This turns out to be very efficient and allows to obtain high precision for the numerical values of the
tensor coefficients of the $4$-point function.
In an example, we demonstrate that the combination of representations for non-exceptional kinematics with this expansion covers the complete phase-space from medium to vanishing Gram determinants.

In view of the importance of stable numerics  for tensor reductions, it would be welcome to have one or more complete opensource programs for this task, including the treatment of small Gram determinants.
To our knowledge, none is presently available.
Following the approach of this article, a C++ program is under development to close this gap~\cite{c++yundin:2010bb}.

The article is organized as follows.
In sect.~\ref{Tshift}, some definitions and basic formulae are recalled.
{Sect.~\ref{LargeGrams} describes a compact analytical tensor reduction of 5-point functions with non-exceptional kinematics.} In sect.~\ref{5to4}
the $5$-point functions up to rank $R=5$ are reduced to
$4$-point tensor coefficients in terms of
4-point integrals in higher dimensions and with higher powers of the scalar propagators.
In sect.~\ref{4togeneric}, we reduce these 4-point integrals in several steps.
In sect.~\ref{DiffQuo} the integrals are reduced to 4-point integrals in higher dimensions
with powers $1$ of the scalar propagators plus 3-point tensor coefficients. The results
are given in eqns.~\eqref{I4id+2},~\eqref{want1},~\eqref{fulld3} and~\eqref{fulld4}.
{Indeed, these eqns. are the central point of our approach since
they allow to proceed further in different directions.
One might e.g. apply the general method of calculating higher
dimensional integrals of~\cite{Fleischer:2003rm}.
With app.~\ref{App}, alternatively a reduction to the Passarino-Veltman basis is straightforward.
In subsect.~\ref{Gram}, we recall how to expand the
higher-dimensional integrals with powers $1$ of the scalar propagators for the case of vanishing
or small Gram determinants, the result being given in~(\ref{final})}.
The symmetrized recursion relations, useful
for the $5$-point functions with non-exceptional Gram determinants
are presented in sect.~\ref{LaGra}.
Additionally, in sect.~\ref{Simplify}
we give some relations which may be useful for an analytic simplification of
Feynman diagrams.
We end with conclusions in sect.~\ref{conclude}.
Appendix~\ref{App} contains a list of dimensional recurrences and  app.~\ref{Bpp} collects divergent parts of higher-dimensional integrals. A numerical example is discussed in app.~\ref{Num}.
Appendix~\ref{app-nota} contains notations and some relevant algebraic relations.

\section{\label{Tshift}Tensor integrals in terms of integrals in shifted dimensions}
A tensorial Feynman integral with $n$ external legs is shown in fig.~\ref{fig-n-point} and is defined as
\bea
\label{definition}
 I_{n,\{\nu_j\}}^{\mu_1\cdots\mu_R} &=&~\frac{{(2 \pi \mu)}^{4-d} }{i {\pi}^2}~\int d^d k~~\frac{\prod_{r=1}^{R} k^{\mu_r}}{\prod_{j=1}^{n}c_j^{\nu_j}},
\eea
with denominators $c_j$, having \emph{indices} $\nu_j$ and \emph{chords}
$q_j$,
\begin{eqnarray}\label{propagators}
c_j &=& (k-q_j)^2-m_j^2 +i \epsilon.
\end{eqnarray}
Here, we use the generic dimension $d=4-2\epsilon$ and $\mu =1$.
Reducing the tensors to $1$- to $4$-point scalar functions $I_n^d$, in general their
expansions in terms of $\eps$ is needed. The first expansion terms
can be expressed in terms of  Euler dilogarithmic (or simpler) functions~\cite{'tHooft:1978xw,Denner:1991qq,vanOldenborgh:1990yc,Hahn:1998yk,Ellis:2007qk}.

\begin{figure}[bt]
\begin{center}
\includegraphics[width=.39\textwidth]{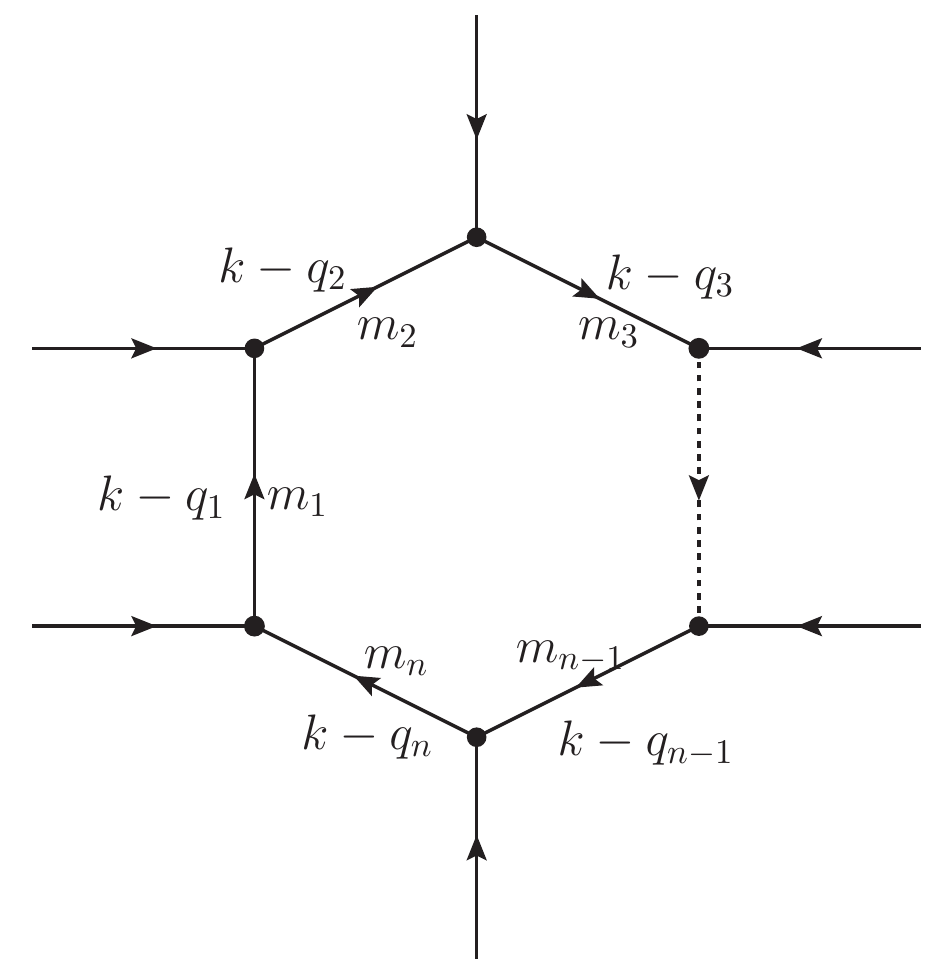}
\end{center}
\caption[Momenta flow of the $n$-point function.]{%
\label{fig-n-point}
Momenta flow of the $n$-point function.}
\end{figure}

The six-point tensor integrals may be expressed in terms of five-point tensor functions~\cite{Fleischer:1999hq,Diakonidis:2008ij,Binoth:2005ff,Denner:2005nn}:
\bea\label{tensor6general}
I_6^{\mu_1 \dots \mu_{R-1} \mu}  =
-  \sum_{s=1}^{6}
I_5^{\mu_1 \dots \mu_{R-1} ,s } \bar{Q}_s^{\mu},
\eea
where the auxiliary vectors $\bar{Q}_s$ are
\bea
 \bar{Q}_s^{\mu}&=&\sum_{i=1}^{6}  q_i^{\mu} \frac{{0s\choose 0i}_6}{{0\choose 0}_6}~~~,~~~ s=1 \dots 6.
\label{Q6}
\eea
A similar formula exists also for five-point tensor integrals~\cite{Diakonidis:2009fx}:
\bea
I_5^{\mu_1  \dots \mu_{R-1} \mu}  =I_5^{\mu_1  \dots \mu_{R-1}} Q_0^{\mu} -  \sum_{s=1}^{5}
I_4^{\mu_1  \dots \mu_{R-1},s } Q_s^{\mu}.
\label{tensor5general}
\eea
The auxiliary vectors here are:
\bea
\label{Qs}
 Q_s^{\mu}&=&\sum_{i=1}^{5}  q_i^{\mu} \frac{{s\choose i}_5}{\left(  \right)_5},~~~ s=0, \dots, 5.
\eea
{For later use, we introduce also
\bea\label{3.11}
Q_s^{t,\mu}&=&\sum_{i=1}^{5} q_i^{\mu} \frac{ {st\choose it}_5}{{t\choose t}_5}~ \nn \\
&=&Q_s^{\mu}~~-\frac{{s\choose t}_5}{{t\choose t}_5}Q_t^{\mu} ,
\\\label{3.12}
Q_s^{tu,\mu}&=&\sum_{i=1}^{5} q_i^{\mu} \frac{ {stu\choose itu}_5}{{tu\choose tu}_5} \nn \\
&=&Q_s^{u,\mu}-\frac{{su\choose tu}_5}{{tu\choose tu}_5}Q_t^{u,\mu},
\eea
and
\be
g^{\mu\, \nu}=2 \sum_{i,j=1}^{5} \frac{{i\choose j}_5}{\left(  \right)_5} \, q_i^{\mu}\, q_j^{\nu} .
\label{gmunu}
\ee
}
In fact,~\eqref{tensor6general} is essentially the same formula as~\eqref{tensor5general}, except
that  ${\left( \right)}_5$ is replaced by ${0\choose 0}_6$ etc. and ${00\choose 0i}_6=0$.
With the definition
\bea
\label{gram}
Y_{ij}=-(q_i-q_j)^2+m_i^2+m_j^2,
\eea
the \emph{modified Cayley determinant} of a topology
with internal lines  $1 \cdots  n $ becomes
\bea\label{gram1}
()_n~\equiv~
\begin{vmatrix}
  0 & 1       & 1       &\ldots & 1      \\
  1 & Y_{11}  & Y_{12}  &\ldots & Y_{1n} \\
  1 & Y_{12}  & Y_{22}  &\ldots & Y_{2n} \\
  \vdots  & \vdots  & \vdots  &\ddots & \vdots \\
  1 & Y_{1n}  & Y_{2n}  &\ldots & Y_{nn}
\end{vmatrix}
 .
\eea
One chord may be chosen arbitrarily to vanish, $q_n=0$, and then this object is the Gram determinant: \footnote{Usually we will use indices $s,t,\cdots = 1,\cdots, n$ for labelling internal lines, and indices $i,j,\cdots = 1,\cdots, n-1$ for labelling the (non-vanishing) chords.}
\bea\label{gram11}
{()_n|_{q_n=0}} &=&{ -~\det ~ G_{n-1},}
\\\label{gram12}
{G_{n-1,ik}}&=& {2q_i q_k,~~ i,k=1,\dots n-1 .}
\eea
The Gram determinant  is independent of the internal masses.
The \emph{signed minors}~\cite{Melrose:1965kb} are denoted as follows:
\bea\label{gram2}
\left(
\begin{array}{ccc}
  j_1 & j_2 & \cdots j_m\\
  k_1 & k_2 & \cdots k_m\\
\end{array}
\right)_n .
\eea
They are determinants, labeled by those rows $j_1,j_2,\cdots j_m$ and columns $k_1,k_2,\cdots k_m$ which have been
excluded from the definition of the Gram determinant $()_n$, with sign
\bea
\label{eq-modc}
\mathrm{sign}
\begin{pmatrix}
  j_1 & j_2 & \cdots & j_m\\
  k_1 & k_2 & \cdots  & k_m\\
\end{pmatrix}_n
=
(-1)^{j_1+j_2+ \cdots +j_m+k_1+k_2+ \cdots +k_m} \cdot S(j_1,  j_2  \cdots  j_m) \cdot
S(k_1 , k_2 , \cdots   k_m).
\eea
Here
$S(j_1 , j_2  \cdots  j_m)$
gives the sign of permutations needed to place the indices in increasing order. \footnote{The definitions are related to similar ones used in the literature, see app.~\ref{app-nota}.}
We have e.g.
\bea\label{gram3}
\Delta_n=  
\begin{vmatrix}
  Y_{11}  & Y_{12}  &\ldots & Y_{1n} \\
  Y_{12}  & Y_{22}  &\ldots & Y_{2n} \\
  \vdots  & \vdots  &\ddots & \vdots \\
  Y_{1n}  & Y_{2n}  &\ldots & Y_{nn}
\end{vmatrix}
= {0\choose 0}_n.
\label{mcd}
\eea
Applying Davydychev's  method~\cite{Davydychev:1991va},  one expresses the tensor integrals $I_n^{\mu_1  \dots \mu_{R}}$ by scalar Feynman integrals
$I_{n,i\cdots}^{(d)}$ in higher dimensions $d$ and with higher indices $\nu_i$.
We reproduce here integrals  with rank $R\leq 5$:
\bea
\label{tensor1}
  I_n^{\mu} & =&  \int \frac{d^d k}{{i\pi}^{d/2}} k^{\mu} \prod_{j=1}^{n} \, {c_j^{-1}}
\nl
&=& - ~ \sum_{i=1}^{n} \, q_i^{\mu} \, I_{n,i}^{[d+]} ,
\\
\label{tensor2}
 I_{n}^{\mu\, \nu}& =& \int \frac{d^d k}{{i\pi}^{d/2}} k^{\mu} \, k^{\nu} \, \prod_{j=1}^{n} \, {c_j^{-1}}
\nl
&=&
   \sum_{i,j=1}^{n} \, q_i^{\mu}\, q_j^{\nu} \, n_{ij} \,  \, I_{n,ij}^{[d+]^2} -\frac{1}{2}
   \, g^{\mu \nu}  \, I_{n}^{[d+]} ,
 \\
\label{tensor3}
I_{n}^{\mu\, \nu\, \lambda}& =& \int \frac{d^d k}{{i\pi}^{d/2}} k^{\mu} \, k^{\nu} \,  k^{\lambda} \, \prod_{j=1}^{n} \, {c_j^{-1}}
\nl
&=&
   -~ \sum_{i,j,k=1}^{n} \, q_i^{\mu}\, q_j^{\nu}\, q_k^{\lambda} \,  n_{ijk} \,  \, I_{n,ijk}^{[d+]^3}
  +\frac{1}{2} \sum_{i=1}^{n} g^{[\mu \nu} q_i^{\lambda]} I_{n,i}^{[d+]^2} ,
\nl \\
\label{tensor4}
I_{n}^{\mu\, \nu\, \lambda\, \rho} &=&
\int \frac{d^d k}{{i\pi}^{d/2}} k^{\mu} \, k^{\nu} \,  k^{\lambda} \, k^{\rho} \, \prod_{j=1}^{n} \, {c_j^{-1}}
\nn \\ \nn \\
&=&
    \sum_{i,j,k,l=1}^{n} \, q_i^{\mu}\, q_j^{\nu}\, q_k^{\lambda}
 \, q_l^{\rho}\,  n_{ijkl} \,  \, I_{n,ijkl}^{[d+]^4}
    -\frac{1}{2} \sum_{i,j=1}^{n} g^{[\mu \nu} q_i^{\lambda} q_j^{\rho]}
\, n_{ij} I_{n,ij}^{[d+]^3}
 +\frac{1}{4} g^{[\mu \nu} g^{\lambda \rho]} I_{n}^{[d+]^2} ,
\nl \\
\label{tensor5}
I_{n}^{\mu\, \nu\, \lambda\, \rho\, \sigma}&=&
\int \frac{d^d k}{i\pi^{d/2}} k^{\mu} \, k^{\nu} \,  k^{\lambda} \, k^{\rho} \,
 k^{\sigma} \,
\prod_{j=1}^{n} \, {c_j^{-1}}
\nn\\
&=&
-~    \sum_{i,j,k,l,m=1}^{n} \, q_i^{\mu}\, q_j^{\nu}\, q_k^{\lambda}
 \, q_l^{\rho}\, q_m^{\sigma}\,  n_{ijklm} \,  \, I_{n,ijklm}^{[d+]^5}
    +\frac{1}{2} \sum_{i,j,k=1}^{n} g^{[\mu \nu} q_i^{\lambda} q_j^{\rho} q_k^{\sigma]}
\, n_{ijk} I_{n,ijk}^{[d+]^4}
\nn\\&&
 -~\frac{1}{4} \sum_{i=1}^{n} g^{[\mu \nu} g^{\lambda \rho}  q_i^{\sigma]} I_{n,i}^{[d+]^3} .
\eea
\newpage
The following symmetrized tensors are used:
\bea
\label{G2V1}
g^{[\mu \nu} q_i^{\lambda]}&=& g^{\mu \nu}  \, q_i^{\lambda} \,+ g^{\mu \lambda}
\, q_i^{\nu} \,+   \, g^{\nu \lambda}  \, q_i^{\mu} ,
\\
\label{G2V2}
g^{[\mu \nu} q_i^{\lambda} q_j^{\rho]}&=&\, g^{\mu \nu}  \, q_i^{\lambda}
\, q_j^{\rho} \,+ g^{\mu \lambda}  \, q_i^{\nu} \, q_j^{\rho} \,+
    \, g^{\nu \lambda}  \, q_i^{\mu} \, q_j^{\rho} \,+
\, g^{\mu \rho}  \, q_i^{\nu}
\, q_j^{\lambda} \,+ g^{\nu \rho}  \, q_i^{\mu} \, q_j^{\lambda} \,+
    \, g^{\lambda \rho}  \, q_i^{\mu} \, q_j^{\nu} ,
\nl \\ ~~
\label{G2G2}
g^{[\mu \nu} g^{\lambda \rho]}~~&=&\, g^{\mu \nu}  \,g^{\lambda \rho}  \,+ g^{\mu \lambda}  \,  g^{\nu \rho} \,+
    \, g^{\mu \rho}  \,  g^{\nu \lambda} ,
\\
\label{G2V3}
g^{[\mu \nu} q_i^{\lambda} q_j^{\rho} q_k^{\sigma]}&=&
g^{\mu \nu} q_i^{\lambda} q_j^{\rho} q_k^{\sigma}+g^{\mu \lambda} q_i^{\nu} q_j^{\rho} q_k^{\sigma}+
g^{\mu \rho} q_i^{\nu} q_j^{\lambda} q_k^{\sigma}+g^{\mu \sigma} q_i^{\nu} q_j^{\lambda} q_k^{\rho}+
g^{\nu \lambda} q_i^{\mu} q_j^{\rho} q_k^{\sigma}
\nn \\
&&+~
g^{\nu \rho} q_i^{\mu} q_j^{\lambda} q_k^{\sigma}+
g^{\nu \sigma} q_i^{\mu} q_j^{\lambda} q_k^{\rho}+g^{\lambda \rho} q_i^{\mu} q_j^{\nu} q_k^{\sigma}+
g^{\lambda \sigma} q_i^{\mu} q_j^{\nu} q_k^{\rho}+g^{\rho \sigma} q_i^{\mu} q_j^{\nu} q_k^{\lambda},~~~~
\\
g^{[\mu \nu} g^{\lambda \rho} q_i^{\sigma]}
&=&
g^{[\mu \nu} g^{\lambda \rho]} q_i^{\sigma}~+g^{[\mu \nu} g^{\lambda \sigma]} q_i^{\rho}+
g^{[\mu \nu} g^{\rho \sigma]} q_i^{\lambda}+g^{[\mu \sigma} g^{\lambda \rho]} q_i^{\nu}~+
g^{[\nu \sigma} g^{\lambda \rho]} q_i^{\mu}.
\label{G4V1}
\eea
The scalar integrals are:
\begin{eqnarray}
  \label{eq:Inij}
   I_{p, \, i\,j \,k\cdots} ^{[d+]^l,stu \cdots} =  \int \frac{d^{[d+]^l}k}{i\pi^{[d+]^l/2}}  \prod_{r=1}^{n} \, \frac{1}{c_r^{1+\delta_{ri} + \delta_{rj}+\delta_{rk}+\cdots
                -\delta_{rs} - \delta_{rt}-\delta_{ru}-\cdots}} ,
\end{eqnarray}
where $[d+]^l=4-2 \eps+2 l $.
The index $p$ is the number of propagators of the $p$-point function.
Note that equal lower and upper indices cancel.
The coefficients $n_{ij}, n_{ijk}$ and $n_{ijkl}$ etc. in~(\ref{tensor2}) to~(\ref{tensor5}) were introduced in~\cite{Diakonidis:2008ij}.
They stand for the product of factorials of the number
of equal indices: e.g. $n_{iiii}=4!, n_{ijii}=3!, n_{iijj}=2! 2!, n_{ijkk}=2!,n_{ijkl}=1!$;
the indices $i,j,k,l$ are assumed here to be different from each other.
The following relations  are of particular relevance
for the successive application of recurrence relations to reduce higher-dimensional integrals:
\bea\label{ndef1}
n_{ij}&=&{\nu}_{ij} ,
\nn\\
n_{ijk}&=&{\nu}_{ij} {\nu}_{ijk} ,
\nn\\
n_{ijkl}&=&{\nu}_{ij} {\nu}_{ijk} {\nu}_{ijkl},
\nn\\
n_{ijklm}&=&{\nu}_{ij} {\nu}_{ijk} {\nu}_{ijkl}{\nu}_{ijklm},
\eea
and:
\bea
\label{nudef1}
{\nu}_{ij}&=&1+{\delta}_{ij} ,
\nn\\
{\nu}_{ijk}&=&1+{\delta}_{ik}+{\delta}_{jk} ,
\nn\\
{\nu}_{ijkl}&=&1+{\delta}_{il}+{\delta}_{jl}+{\delta}_{kl}.
\nn\\
{\nu}_{ijklm}&=&1+{\delta}_{im}+{\delta}_{jm}+{\delta}_{km}+{\delta}_{lm}.
\eea

In a second step, one may choose to express the higher-dimensional scalar integrals in terms of the generic scalar integrals
 The algorithm
is based on recurrence relations with shifts of dimension $d \geq 4-2\eps$  and indices $\nu_s\geq 1$,
\bea
\label{eq:RR1}
  \left( \right)_n \nu_s    \left( \mathbf{s^{+}}  I_{n}^{(d+2)}\right)
&=&
 - {s \choose 0}_n I_n^{(d)} + \sum_{t=1}^{n} {s \choose t}_n \left( \mathbf{t^{-}} I_{n}^{(d)} \right),
\eea
or with a shift of dimension $d$:
\bea
 \left( \right)_n (d-\sum_{s=1}^{n}\nu_s+1)      I_n^{(d+2)}
  &=&
 {{0 \choose 0}_n} I_n^{(d)}
 - \sum_{t=1}^n {0 \choose t}_n   \left( \mathbf{t^{-}} I_{n}^{(d)}  \right).
\label{eq:RR2}\eea
These relations hold for arbitrary index sets $\{\nu_s\}$.
The integrals $\mathbf{s^{+}} I_{n}^{(d)}$ and $\mathbf{t^{-}} I_{n}^{(d)}$ are obtained from $ I_{n}^{(d)}$ by replacing $\nu_s \rightarrow (\nu_s + 1)$ and $\nu_t \rightarrow (\nu_t - 1)$, respectively. For more explicit expressions see app.~\ref{App} .

\section{\label{LargeGrams}
 An efficient reduction of $5$-point tensor integrals
}
The reduction of $5$-point tensor integrals to $4$-point tensor integrals at non-exceptional momenta
may be performed  by iterative application of~\eqref{tensor5general}.
This was exemplified in~\cite{Diakonidis:2009fx}, and an opensource Fortran code \texttt{olotic}~\cite{olotic:2010aa} is available.
In this sect., we derive a very compact, explicit representation of the tensor coefficients for 5-point functions in a minimal basis, chosen to be free of the metric tensor.
This will rely on an exploitation of~\eqref{gmunu} and
\eqref{trick}, a specifically useful relation of the \emph{algebra of the signed minors},
and applying the \emph{symmetrized} dimensional recurrences of sect.~\ref{LaGra}

We investigate the $5$-point tensor integrals step by step.
For the tensor of rank $R=1$ we get from~\eqref{tensor5general}
\bea\label{3.1}
I_5^{\mu}=I_5 \cdot Q_0^{\mu} -\sum_{s=1}^{5} I_4^s \cdot Q_s^{\mu},
\eea
and $I_5$ may be taken from~\eqref{scalar4p}.

Similarly for the tensor of rank $2$,
\bea\label{3.2}
I_5^{\mu \nu}=I_5^{\mu} \cdot Q_0^{\nu} -\sum_{s=1}^{5} I_4^{\mu, s} \cdot Q_s^{\nu} ,
\eea
with four-point integrals from~\eqref{tensor1} and~\eqref{A511},
\bea\label{3.3}
I_4^{\mu,s}&=&-\sum_{i=1}^{5} q_i^{\mu} I_{4,i}^{[d+],s}, \\\label{3.32}
I_{4,i}^{[d+],s}&=&-\frac{ {0s\choose is}_5}{{s\choose s}_5} I_4^s+ \sum_{t=1}^{5}
\frac{ {ts\choose is}_5}{{s\choose s}_5} I_3^{st} ,
\eea
such that we can write the tensor of rank $R=2$ as
\bea\label{3.4}
I_5^{\mu \nu}=I_5^{\mu} \cdot Q_0^{\nu} -\sum_{s=1}^{5} \left\{Q_0^{s,\mu} I_4^s-
\sum_{t=1}^{5} Q_t^{s, \mu} I_3^{st} \right\}
Q_s^{\nu}.
\eea
Compared to~\eqref{tensor2}, this representation and the following ones are free of the metric tensor.
{Further, the compactness relies on the use of the auxiliary vectors $Q_s^{\mu},Q_t^{s, \nu}$ instead of the chords $q_i^{\mu}$.}

{The tensor of rank $R=3$ deserves a bit more effort,}
\bea\label{3.138}
I_5^{\mu \nu \lambda}=I_5^{\mu\nu} \cdot Q_0^{\lambda} -\sum_{s=1}^{5} I_4^{\mu\nu, s} \cdot Q_s^{\lambda}.
\eea
The corresponding $4$-point function
reads now due to~\eqref{tensor3} and with~\eqref{A522}
\bea
\label{TwoTy}
I_4^{\mu \nu ,s}&=&\sum_{i,j=1}^{5} q_i^{\mu} q_j^{\nu} {\nu}_{ij} I_{4,ij}^{[d+]^2,s}-\frac{1}{2} g^{\mu \nu}I_{4}^{[d+],s},
\\\label{TwoTy2}
{\nu}_{ij} I_{4,ij}^{[d+]^2,s}&=&-\frac{ {0s\choose js}_5}{{s\choose s}_5} I_{4,i}^{[d+],s}+
\frac{ {is\choose js}_5}{{s\choose s}_5}I_{4}^{[d+],s}+
\sum_{t=1}^{5}
\frac{ {ts\choose js}_5}{{s\choose s}_5} I_{3,i}^{[d+],st}.
\eea
Observe that for $i,j=s$ 
the integrals $I_{4,i}^{[d+],s}$ and $I_{4,ij}^{[d+]^2,s}$ vanish (due to vanishing signed minors)
such that indeed {a formal} summation over all five values of $i,j$ is possible.
\newpage
Now we use identity~\eqref{r2},
\bea
\frac{ {is\choose js}_5}{{s\choose s}_5}&=&\frac{ {i\choose j}_5}{{\left( \right)}_5}
-\frac{ {s\choose i}_5 {s\choose j}_5 }{{\left( \right)}_5 {s\choose s}_5} \nn \\
&=&
\frac{ {i\choose j}_5}{{\left( \right)}_5}-\frac{{\left( \right)}_5}{{s\choose s}_5} Q_s^i Q_s^j,
\label{trick}
\eea
where the $Q_s^i$ are the vector components of $Q_s^{\mu}$, see~\eqref{Qs}.

Performing summation over $i,j$ in~(\ref{TwoTy}), the first term on the right hand side of
~\eqref{trick} yields
$\frac{1}{2} g^{\mu \nu } I_4^{[d+],s}$ (see~\eqref{gmunu}) and thus cancels against the last term in~(\ref{TwoTy}). Thus
we can write the 4-point tensor of rank $R=2$ as
\bea\label{3.7}
I_4^{\mu \nu ,s}&=&\sum_{i,j=1}^{5} q_i^{\mu} q_j^{\nu} J_{4,ij}^{s},
\\\label{3.111}
J_{4,ij}^{s}&=&-\frac{ {0s\choose js}_5}{{s\choose s}_5} I_{4,i}^{[d+],s}-\frac{ {s\choose i}_5 {s\choose j}_5 }{{\left( \right)}_5 {s\choose s}_5}I_{4}^{[d+],s}+\sum_{t=1}^{5} \frac{ {ts\choose js}_5}{{s\choose s}_5} I_{3,i}^{[d+],st},
\eea
where the metric tensor has again disappeared compared to~\eqref{tensor2} and instead $q_i^{\mu} q_j^{\nu}$ contribute
for $i,j=s$. A compact notation can now be used with~\eqref{3.11} and~\eqref{3.12}:
\bea\label{3.13}
I_4^{\mu \nu ,s}=Q_0^{s,\mu} Q_0^{s,\nu}  I_{4}^{s}-\frac{{\left( \right)}_5}{{s\choose s}_5}
Q_s^{\mu} Q_s^{\nu}I_{4}^{[d+],s}
+\sum_{i=1}^{5} q_i^{\mu} q_j^{\nu}R_{3,ij}^{[d+]^2,s} ,
\eea
with $R_{3,ij}^{[d+]^2,s}$ the scratched version of~\eqref{wanty}.
Inserting~\eqref{3.13} in~\eqref{3.138} yields the  compact expression for $I_5^{\mu\nu\lambda}$, free of the metric tensor.

Next, for the tensor of rank $R=4$ of the $5$-point function,
\bea\label{3.42}
I_5^{\mu \nu \lambda\rho}=
I_5^{\mu\nu\lambda} \cdot Q_0^{\rho}
-\sum_{s=1}^{5} I_4^{\mu\nu\lambda, s} \cdot Q_s^{\rho},
\eea
we need the $4$-point function  of rank $R=3$, according to~\eqref{tensor3}:
\bea\label{3.14}
I_4^{\mu \nu \lambda ,s}&=&-\sum_{i,j,k=1}^{5} q_i^{\mu} q_j^{\nu} q_k^{\lambda} n_{ijk}
I_{4,ijk}^{[d+]^3,s}+\frac{1}{2}\sum_{i=1}^{5} g^{[\mu \nu} q_i^{\lambda ]}I_{4,i}^{[d+]^2,s},
\\
n_{ijk}I_{4,ijk}^{[d+]^3,s}&=&-\frac{{0s\choose is}_5 {0s\choose js}_5 {0s\choose ks}_5}
{{s\choose s}_5^3}I_{4}^s+
\left\{\frac{{is\choose js}_5}{{s\choose s}_5} I_{4,k}^{[d+]^2,s}+
(j \leftrightarrow k) + (i \leftrightarrow k) \right\}+R_{3,ijk}^{[d+]^3,s}.
\label{fullxs}
\eea
Here we have used for $I_{4,ijk}^{[d+]^3,s}$, instead of~\eqref{A533}, the symmetrized form~\eqref{fullx} with
$R_{3,ijk}^{[d+]^3,s}$ being the scratched version of~\eqref{fullxr}.
Applying again~\eqref{trick} we obtain the analogue of~\eqref{3.13},
\bea\label{3.35}
I_4^{\mu \nu \lambda ,s}&=&\sum_{i,j,k=1}^{5} q_i^{\mu} q_j^{\nu} q_k^{\lambda} J_{4,ijk}^{s},
\\\label{3.351}
J_{4,ijk}^{s}&=&\frac{{0s\choose is}_5{0s\choose js}_5{0s\choose ks}_5}{{s\choose s}_5^3}I_{4}^s+\frac{1}{{\left( \right)}_5} \left\{\frac{ {s\choose i}_5 {s\choose j}_5 }{ {s\choose s}_5}I_{4,k}^{[d+]^2,s}+(j \leftrightarrow k) + (i \leftrightarrow k) \right\}-R_{3,ijk}^{[d+]^3,s}.
\eea
For the following it also pays to introduce
\bea
J_4^{\mu, s}&=&\sum_{i=1}^{5} q_i^{\mu} I_{4,i}^{[d+]^2,s} \nn \\
&=&-Q_0^{s,\mu} I_4^{[d+]} +
\sum_{t=1}^{5} Q_t^{s,\mu} I_3^{[d+],st},
\label{J4V}
\eea
so that finally
\bea\label{3.352}
I_4^{\mu \nu \lambda ,s }=Q_0^{s,\mu} Q_0^{s,\nu}Q_0^{s,\lambda} I_4^s+
\frac{{\left( \right)}_5}{{s\choose s}_5} Q_s^{[\mu}Q_s^{\nu}J_4^{\lambda, s]}
-\sum_{i,j,k=1}^{5} q_i^{\mu} q_j^{\nu} q_k^{\lambda} R_{3,ijk}^{[d+]^3,s}.
\eea
with $R_{3,ijk}^{[d+]^3,s}$ the scratched version of~\eqref{fullxr}.
This finishes the determination of $I_5^{\mu \nu \lambda\rho}$.

Finally, for the tensor of rank $R=5$ of the $5$-point function,
\bea\label{3.52}
I_5^{\mu \nu \lambda\rho\sigma}=
I_5^{\mu\nu\lambda\rho} \cdot Q_0^{\sigma}
-\sum_{s=1}^{5} I_4^{\mu\nu\lambda\rho, s} \cdot Q_s^{\sigma},
\eea
we need the scratched tensor of
rank $R=4$ of the $4$-point function. The corresponding symmetrized tensor coefficients are
taken from~\eqref{fully} and~\eqref{I451} by scratching. We begin with
the term  $I_4^{[d+]^2}$ from~\eqref{fully}.
Using again~\eqref{trick} we have
\bea
&&
\frac{{is\choose ls}_5}{{s\choose s}_5}\frac{{js\choose ks}_5}{{s\choose s}_5}+
\frac{{js\choose ls}_5}{{s\choose s}_5}\frac{{is\choose ks}_5}{{s\choose s}_5}+
\frac{{ks\choose ls}_5}{{s\choose s}_5}\frac{{is\choose js}_5}{{s\choose s}_5}=
\frac{{i\choose l}_5}{{\left( \right)}_5}\frac{{j\choose k}_5}{{\left( \right)}_5}+
\frac{{j\choose l}_5}{{\left( \right)}_5}\frac{{i\choose k}_5}{{\left( \right)}_5}+
\frac{{k\choose l}_5}{{\left( \right)}_5}\frac{{i\choose j}_5}{{\left( \right)}_5} \nn \\
-&&\frac{{\left( \right)}_5}{{s\choose s}_5}\left\{
\frac{{i\choose l}_5}{{\left( \right)}_5} Q_s^j Q_s^k+
\frac{{j\choose k}_5}{{\left( \right)}_5} Q_s^i Q_s^l+
\frac{{j\choose l}_5}{{\left( \right)}_5} Q_s^i Q_s^k+
\frac{{i\choose k}_5}{{\left( \right)}_5} Q_s^j Q_s^l+
\frac{{k\choose l}_5}{{\left( \right)}_5} Q_s^i Q_s^j+
\frac{{i\choose j}_5}{{\left( \right)}_5} Q_s^k Q_s^l \right\}\nn \\
+&& 3 \frac{{\left( \right)}_5^2}{{s\choose s}_5^2} Q_s^i Q_s^j Q_s^k Q_s^l.
\label{ijkls}
\eea
The first term on the right hand side of~\eqref{ijkls} yields after summation over $i,j,k,l$
\bea\label{3.357}
\sum_{ijkl=1}^5 q_i^{\mu} q_j^{\nu} q_k^{\lambda} q_l^{\rho}
\left\{
\frac{{i\choose l}_5}{{\left( \right)}_5}\frac{{j\choose k}_5}{{\left( \right)}_5}
+
\frac{{j\choose l}_5}{{\left( \right)}_5}\frac{{i\choose k}_5}{{\left( \right)}_5}
+
\frac{{k\choose l}_5}{{\left( \right)}_5}\frac{{i\choose j}_5}{{\left( \right)}_5} \right\}
=
\frac{1}{4} g^{[ \mu \nu} g^{\lambda \rho ]}.
\eea
The same contribution comes directly from~\eqref{tensor4}. Finally,
the second term of~\eqref{tensor4} contributes with the term $\sim I_4^{[d+]^2,s}$
of~\eqref{I451}
and again the first term on the right hand side of~\eqref{trick}
\bea\label{3.358}
-\frac{1}{2} \sum_{i,j=1}^5 g^{[ \mu \nu} q_i^{\lambda} q_j^{\rho ]} \frac{{i\choose j}_5}{{\left( \right)}_5} \cdot
I_4^{[d+]^2,s}=-\frac{1}{2} g^{[ \mu \nu} g^{\lambda \rho ]} \cdot I_4^{[d+]^2,s},
\eea
i.e. the terms $g^{[ \mu \nu} g^{\lambda \rho ]} \cdot I_4^{[d+]^2,s}$ cancel.
The second
term on the right hand side of~\eqref{ijkls} yields after summation over $i,j$
\bea\label{3.359}
&&
-~\frac{{\left( \right)}_5}{{s\choose s}_5}\sum_{i,j,k,l=1}^5 q_i^{\mu} q_j^{\nu} q_k^{\lambda} q_l^{\rho}
\nn \\
&&~~~
\left\{
\frac{{i\choose l}_5}{{\left( \right)}_5} Q_s^j Q_s^k+
\frac{{j\choose k}_5}{{\left( \right)}_5} Q_s^i Q_s^l+
\frac{{j\choose l}_5}{{\left( \right)}_5} Q_s^i Q_s^k+
\frac{{i\choose k}_5}{{\left( \right)}_5} Q_s^j Q_s^l+
\frac{{k\choose l}_5}{{\left( \right)}_5} Q_s^i Q_s^j+
\frac{{i\choose j}_5}{{\left( \right)}_5} Q_s^k Q_s^l \right\} \cdot I_4^{[d+]^2,s}
\nn \\
&=&
-~\frac{1}{2}\frac{{\left( \right)}_5}{{s\choose s}_5} g^{[\mu \nu} Q_s^{\lambda} Q_s^{\rho ]}
\cdot I_4^{[d+]^2,s}.
\eea
A contribution of this type also comes from the second term of~\eqref{tensor4}
with the term $\sim I_4^{[d+]^2,s}$ of~\eqref{I451}, but now the second part on
the right hand side of~\eqref{trick} is
\bea\label{3.36}
\frac{1}{2}\frac{{\left( \right)}_5}{{s\choose s}_5} g^{[\mu \nu} Q_s^{\lambda} Q_s^{\rho ]}\cdot I_4^{[d+]^2,s},
\eea
which means that also these terms cancel. Finally there remains the last term in~\eqref{ijkls},
which does not cancel but contains no $g^{\mu \nu}$
\bea\label{3.388}
3 \frac{{\left( \right)}_5^2}{{s\choose s}_5^2} Q_s^{\mu}Q_s^{\nu}Q_s^{\lambda}Q_s^{\rho}
\cdot I_4^{[d+]^2,s}.
\eea
The contributions from~\eqref{fully} of the type $I_{4,i}^{[d+]^2,s}$
- after cancelling the first term on the right hand side
of~\eqref{I451} -
can also be written in a compact manner,
\bea\label{3.361}
\frac{{\left( \right)}_5}{{s\choose s}_5} Q_s^{[\mu}Q_s^{\nu} J_4^{\lambda,s} Q_0^{s,\rho ]},
\eea
where the symmetrization in the tensor indices
is understood.
 Introducing like in~\eqref{J4V}
\bea\label{3.362}
J_3^{\mu, st}&=&\sum_{i=1}^{5} q_i^{\mu} I_{3,i}^{[d+]^2,st} \nn \\
&=&-Q_0^{st,\mu} I_3^{[d+],st}+
\sum_{u=1}^5 Q_u^{st,\mu} I_2^{[d+],stu},
\eea
we can finally write
\bea\label{3.363}
I_4^{\mu \nu \lambda \rho, s}&&=Q_0^{s,\mu} Q_0^{s,\nu}Q_0^{s,\lambda}Q_0^{s,\rho}  I_4^s+
3 \frac{{\left( \right)}_5^2}{{s\choose s}_5^2} Q_s^{\mu}Q_s^{\nu}Q_s^{\lambda}Q_s^{\rho}
\cdot I_4^{[d+]^2,s}
+\frac{{\left( \right)}_5}{{s\choose s}_5} Q_{s}^{[\mu}Q_{s}^{\nu}J_4^{\lambda,s } Q_0^{s,\rho ]} \nn \\
&&-\sum_{\substack{t=1 \\ t \neq s}}^{5}\left\{ Q_0^{st,\mu} Q_0^{st,\nu} Q_0^{st,\lambda} I_3^{st}+
 \frac{1}{{st\choose st}_5} \left[
{t\choose t}_5 Q_s^{[\mu}Q_s^{t,\nu}J_3^{\lambda,st ]}+
{s\choose s}_5 Q_t^{[\mu}Q_t^{s,\nu}J_3^{\lambda,st ]} \right] \right\} Q_t^{s,\rho} \nn \\
&&-Q_0^{s,\rho}\sum_{i,j,k=1}^{5} q_i^{\mu} q_j^{\nu} q_k^{\lambda} R_{3,ijk}^{[d+]^3,s}
+\sum_{\substack{t=1 \\ t \neq s}}^{5} Q_t^{s,\rho}\sum_{i,j,k=1}^{5} q_i^{\mu} q_j^{\nu} q_k^{\lambda} R_{2,ijk}^{[d+]^3,st}.
\eea
The $ J_4^{\lambda,s}$  defined in~\eqref{J4V} occurs again, the
 $R_{3,ijk}^{[d+]^3,s}$  is given in~\eqref{fullxr}, and $R_{2,ijk}^{[d+]^3,st}$ in~\eqref{fullx23}.
With the expression for $I_4^{\mu \nu \lambda \rho, s}$, free of the metric tensor, we complete the rewriting of $I_5^{\mu \nu \lambda\rho\sigma}$ with~\eqref{3.52}.

{It is remarkable that all the coefficients of the $g^{\mu \nu}$ terms of the four-point functions in
\eqref{tensor2}-\eqref{tensor4}  completely cancel in a way that the remaining tensor coefficients are much simpler than the original ones.
This is achieved due to the symmetrization of the recurrence relations given in sect.~\ref{LaGra}
and would have been seen less easily with the ``standard'' recursions of app.~\ref{App}.   }
{
The new representations for the tensors may be useful in several respects.
First of all we have here an extremely compact notation, due to the use of auxiliary vectors $Q_s^{\mu}$, which is not evident at the outset.
Further, the representations may be used for a completely independent programming and thus for stringent numerical cross checks.
The latter one is an important aspect because there are not too many opportunities for that in case of the 5-point and 6-point functions.
 Finally, the auxiliary vectors $Q_s^{\mu}$ have some specific properties so that they may be used for simplifying manipulations with physical amplitudes, see sect.~\ref{Simplify}. }
\vspace{0.5cm}
\section{\label{5to4}Reduction of $5$-point tensor coefficients }

The purpose of this sect.  is to express the $5$-point tensor coefficients
in terms of $4$-point tensor coefficients, which will be evaluated in Sec.
\ref{4togeneric} in such a way that also the case of inverse sub-Gram determinants
can be dealt with in an elegant manner. The difference to the former sect.  is
the fact that we avoid all inverse Gram determinants, $1/{\left( \right)}_5$ as well as
$1/{\left( \right)}_4$. {In this case we have to keep the $g^{\mu \nu}$-terms.}

\subsection{\label{degree01}Scalar and vector integrals}
For the  \emph{scalar} 5-point function $I_5$, we use  the recurrence relation~\eqref{eq:RR2}:
\begin{eqnarray}
(d-4) \left(  \right)_5 I_{5}^{[d+]}={0\choose 0}_5 I_{5}
-\sum_{s=1}^{5} {0\choose s}_5 I_{4}^{s} .
\label{scalargn}
\end{eqnarray}
The integral  $I_{5}^{[d+]}$ is finite for $d=4$, and we get in this limit:
\begin{eqnarray}
I_{5} \equiv E =
\frac{1}{{0\choose 0}_5}\sum_{s=1}^{5} {0\choose s}_5 I_{4}^{s},
\label{scalar4p}
\end{eqnarray}
i.e. the scalar $5$-point function is expressed in the limit $d \to 4$
in terms of scalar $4$-point functions, which are obtained by
scratching in the five terms of the sum the $s^{th}$ scalar propagator,
respectively.
This was already derived in~\cite{Melrose:1965kb}, see Eq.~(6.1) there. See also~\cite{Petersson:1965zz}.

The tensor $n$-point integral
of rank $R=1$ in~(\ref{tensor1}) can be  expressed by integrals $I_{n,i}^{[d+]}$,
and  we obtain quite similarly
\begin{eqnarray}\label{i5vc1}
 I_{n}^{\mu}&=&\sum_{i=1}^{n} \, q_i^{\mu} E_i,
\\\label{i5vc2}
E_i&\equiv&
-I_{n,i}^{[d+]}
\nl
&=& (d+1-n) \frac{{0\choose i}_n}{{0\choose 0}_n} I_{n}^{[d+]}-
\frac{1}{{0\choose 0}_n} \sum_{s=1}^{n} {0i\choose 0s}_n I_{n-1}^{s},
\label{first}
\end{eqnarray}
where again for $n=5$ in the limit $d \to 4$ the scalar integral  $I_{5}^{[d+]}$
disappears:
\bea\label{i5vc1a}
E_i &=& \sum_{s=1}^{5} E_i^s,
\\\label{i5vc1b}
E_i^s &=& -\frac{ {0i\choose 0s}_5}{{0\choose 0}_5}  I_{4}^{s}.
\eea

\subsection{\label{central}The recursion formulae}
For the general case, we use as a starting point~\eqref{tensor5general}.
In order to solve this eqn.    recursively, we multiply it with ${0\choose 0}_5$
\footnote{Throughout the present work we assume ${0\choose 0}_5 \ne 0$. In case of vanishing
{and/or small ${0\choose 0}_5$ see the discussion in Sec.~2.2. of~\cite{Fleischer:1999hq}). }}
and use the identity
\bea
{0\choose 0}_5 {s\choose i}_5 = {0s\choose 0i}_5 \left(  \right)_5 + {0\choose i}_5 {s\choose 0}_5 .
\label{product}
\eea
The first term on the right-hand-side can cancel already a Gram determinant $\left(  \right)_5$, and the second
one transforms a vector $Q_s^{\mu}$ into a vector $Q_0^{\mu}$.
As a result, we get from~\eqref{tensor5general}  the general form
\bea
{0\choose 0}_5 I_5^{\mu_1  \dots \mu_{R-1} \mu}  =
T^{\mu_1  \cdots \mu_{R-1}}
Q_0^{\mu} - \sum_{s=1}^{5}I_4^{\mu_1  \cdots \mu_{R-1},s }
{\bar{Q}}_s^{0,\mu} ,
\label{starter}
\eea
with:
\bea
T^{\mu_1  \dots \mu_{R-1}}={0\choose 0}_5 I_5^{\mu_1  \dots \mu_{R-1} }-\sum_{s=1}^{5} {s\choose 0}_5
I_4^{\mu_1  \dots \mu_{R-1},s },
\label{tensorT}
\eea
and
\bea\label{barQ}
{\bar{Q}}_s^{0,\mu}&=&\sum_{i=1}^{5}  q_i^{\mu} {0s\choose 0i}_n,~~~ s=1, \dots, 5.
\eea
The barred vectors are free of the inverse Gram determinant $()_5$.
Evidently, the reduction of $I_4^{\mu_1  \cdots \mu_{R-1},s }$ is also free of $()_5$, and
we have to care only about the product $T^{\mu_1  \cdots \mu_{R-1}} Q_0^{\mu}$.

The following observation will prove to be useful:
 $T^{\mu_1  \dots \mu_{R-1}}$ contains general tensor structures as given in~\eqref{tensor1}--\eqref{tensor5} with chords $q_i$ and the metric tensor.
In fact, when calculating the $5$-point tensor recursively, we  keep
at this stage the $4$-point tensor as given there.
With the  $5$-point tensor of rank $R=1$, given in~(\ref{first}) above,  the recursion is started.

In order to cancel $1/()_5$, in each recursive step a term ${s\choose i}_5$
will be generated and summed over with the corresponding chord $q_i$.
We will apply to such terms the identity
\begin{eqnarray}
{s\choose i}_5 \frac{{0\choose j}_5} {\left(  \right)_5}=
-{0i\choose sj}_5  + {s\choose 0}_5 \frac{{i\choose j}_5}{\left(  \right)_5} .
\label{cancel}
\end{eqnarray}
The ratio ${0\choose j}_5 / ()_5$ comes from $Q_0^{\mu}$, see~\eqref{Qs} .
In the first term of the right-hand-side of~(\ref{cancel})
the Gram determinant $\left(  \right)_5$ has cancelled and the second term
yields a $g^{\mu \nu}$ contribution according to~\eqref{gmunu}.
 The metric tensors in the original $T^{\mu_1  \dots \mu_{R-1}}$
remain unchanged.
From the following examples the scheme will become more evident.

\subsection{\label{degree2}The tensor integral of rank $R=2$}
Equation~(\ref{starter}) reads for the tensor of rank $R=2$:
\bea
{0\choose 0}_5 I_5^{\mu \nu}  =
\left[{0\choose 0}_5 I_5^{\mu}-\sum_{s=1}^{5} {s\choose 0}_5
I_4^{\mu,s } \right] Q_0^{\nu} - \sum_{s=1}^{5}I_4^{\mu,s }
{\bar{Q}}_s^{0,\nu} .
\label{starter2}
\eea
The square bracket,  a special case of~(\ref{tensorT}) for $R=2$, will be rewritten now.
We use~(\ref{tensor1}) and~(\ref{i5vc1b}) for $d=4$ and insert the reduction~(\ref{A511})
with $l=1$:
\bea\label{starter2a}
T^{\mu} &=& \sum_{s=1}^5 T^{\mu,s},
\\\label{starter2b}
T^{\mu,s}&=&
\sum_{i=1}^{5} q_i^{\mu} \left\{{0\choose 0}_5  E_i^s + {s\choose 0}_5 I_{4,i}^{[d+],s} \right\}
\nn \\
&=&\sum_{i=1}^{5} q_i^{\mu} \left\{-{0s\choose 0i}_5 I_4^s+{s\choose 0}_5
\left[-{0s\choose is}_5 I_{4}^{s} +\sum_{t=1,t \ne s}^{5} {ts\choose is}_5 I_{3}^{st} \right]
\frac{1}{{s\choose s}_5} \right\} .
\label{CancEi}
\eea
Using further
\begin{eqnarray}
{s\choose 0}_5{0s\choose is}_5=
 {s\choose i}_5 {0s\choose 0s}_5 -{s\choose s}_5 {0s\choose 0i}_5 ,
\label{Zauberei1}
\end{eqnarray}
we see the cancellation of $E_i^s$. Additionally, it is
\begin{eqnarray}
{s\choose 0}_5 {ts\choose is}_5=
 {s\choose i}_5 {ts\choose 0s}_5 -{s\choose s}_5 {ts\choose 0i}_5.
\label{Zauberei3}
\end{eqnarray}
Here the ${s\choose s}_5$ term cancels and the remaining factor ${ts\choose 0i}_5$
is antisymmetric in $s,t$, yielding a vanishing contribution after summation over
$s,t$.
With~(\ref{A401}), reintroducing $I_4^{[d+],s}$, we obtain
\bea\label{bra1a}
T^{\mu,s}&=&\sum_{i=1}^{5} q_i^{\mu} T_i^{s} ,
\\
T_i^{s}&=& - {s\choose i}_5 I_{4}^{[d+],s} .
\label{bra1}
\eea
Here we observe the first occurrence of a term ${s\choose i}_5$, as mentioned in sect.~\ref{central}.
Using~(\ref{cancel}) and the notation
\begin{eqnarray}
I_{4}^{\mu\, \nu\,}= \sum_{i,j=1}^{5} \, q_i^{\mu}\, q_j^{\nu} E_{ij} +
g^{\mu \nu}  E_{00} ,
\label{final2}
\end{eqnarray}
we finally get, taking into account~(\ref{tensor1}) for $n=4$, the expressions for the tensor coefficients:
\bea\label{E00}
E_{00}&\equiv&
 \sum_{s=1}^5 E_{00}^s
\nn\\&=&
 - \sum_{s=1}^5    \frac{1}{2} \frac{1}{{0\choose 0}_5} {s\choose 0}_5 I_4^{[d+],s},
\\
E_{ij} &\equiv&  \sum_{s=1}^5 E_{ij}^s
\nn\\\label{Exy}
&=&
\sum_{s=1}^5   \frac{1}{{0\choose 0}_5}   \left[{0i\choose sj}_5 I_4^{[d+],s}+
{0s\choose 0j}_5 I_{4,i}^{[d+],s} \right].
\label{Eij}
\eea
The functions  $I_4^{[d+],s}, I_{4,i}^{[d+],s}$ will be further treated in sect.~\ref{4togeneric}.
A comparison shows that the tensor coefficients  $E_{ij}$ given in Eqs. (3.10)--(3.12) and (A.22) of~\cite{Diakonidis:2008ij} are much more involved.

\subsection{\label{appT3}Reduction of integrals with rank $R=3$}
For the tensor integral of rank $R=3$,  Eq.~(\ref{starter}) reads:
\bea
{0\choose 0}_5 I_5^{\mu \nu \lambda}  =
\left[{0\choose 0}_5 I_5^{\mu \nu}-\sum_{s=1}^{5} {s\choose 0}_5
I_4^{\mu \nu,s } \right] Q_0^{\lambda} - \sum_{s=1}^{5}I_4^{\mu \nu,s }
{\bar{Q}}_s^{0,\lambda}.
\label{starter3}
\eea
Investigating the square bracket, i.e. the tensor~(\ref{tensorT}) for $R=3$,
we see that the  corresponding $g^{\mu \nu}$ term vanishes.
Indeed,  from~(\ref{E00}) and~(\ref{tensor2}) we have:
\bea
{0\choose 0}_5  E_{00}^s+\frac{1}{2}{s\choose 0}_5 I_4^{[d+],s} =0.
\label{rank3cancel}
\eea
This is interesting in view of our general scheme, which was described in
Sec.~\ref{central}:
Since there is no vector $q_i$ in this contribution, no ${s\choose i}_5$
is produced, and if we assume that no inverse Gram ${\left( \right)}_5$ should occur
in this case, the contribution must vanish.

Further, from~(\ref{Exy}),~(\ref{tensor2}) and~(\ref{A522}) we obtain
\bea \label{rank3cancel0}
T^{\mu \nu} &=& \sum_{s=1}^5 T^{\mu \nu,s},
\\\label{rank3cancela}
T^{\mu \nu,s} &=&  \sum_{i,j=1}^{5} q_i^{\mu} q_j^{\nu} T^s_{ij},
\eea
and
\bea\label{rank3cancelb}
T^s_{ij}&=&
{0\choose 0}_5  E_{ij}^s -
{s\choose 0}_5 {\nu}_{ij} I_{4,ij}^{[d+]^2,s}  \nn \\
&=&
{s\choose s}_5 \left[{0i\choose sj}_5 I_4^{[d+],s}+
{0s\choose 0j}_5 I_{4,i}^{[d+],s} \right] 
\nn \\
&&-~{s\choose 0}_5 \left[
-{0s\choose js}_5 I_{4,i}^{[d+],s} +{is\choose js}_5 I_{4}^{[d+],s}+
 \sum_{t=1,t \ne s, i}^{5} {ts\choose js}_5 I_{3,i}^{[d+],st}
\right]\frac{1}{{s\choose s}_5}.
\label{toobtain}
\eea
With~(\ref{Zauberei1}) and
\begin{eqnarray}\label{canc3r}
{s\choose 0}_5 {is\choose js}_5=
 {s\choose i}_5 {0s\choose js}_5 +{s\choose s}_5 {0i\choose sj}_5 ,
\label{Zauberei2}
\end{eqnarray}
we see that the complete term $E_{ij}^s$ cancels. As above we use again~(\ref{Zauberei3})
and with the same arguments as before
we see that only ${s\choose i}_5$-type terms remain such that~(\ref{cancel})
can be used again to cancel the Gram determinant.

Before collecting all contributions, we would like to point out
that, after the above manipulations, the expressions are in general not
explicitly symmetric in their indices, although
the original integral \emph{is} symmetric in  $\mu, \nu, \lambda$.
Consequently, our result must also be symmetric in the indices $i,j,k$,
however,  after summation over $s$ and $t$.
For an explicit example see also the discussion after~(\ref{wantx}).
If there is no explicit symmetry before summation over $s$ and $t$ it may be useful to
symmetrize the result.
With this in mind, collecting all contributions, we have
\bea\label{}
T_{ij}^{s}=&&
\left\{-\left[ {s\choose i}_5 {0s\choose js}_5+ {s\choose j}_5 {0s\choose is}_5 \right] I_{4}^{[d+],s}
\right.
\nn \\
&&\left. ~~+\sum_{t=1,t \ne s, i}^{5}\left[ {s\choose i}_5 {ts\choose js}_5+ {s\choose j}_5 {ts\choose is}_5 \right]
\frac{d-2}{2}I_{3}^{[d+],st}
\right\}\frac{1}{{s\choose s}_5}.
\label{squarebr}
\eea
To obtain this result, the vector integral $I_4^{\mu}$, represented by tensor coefficients   $I_{4,i}^{[d+],s}$,  and the vector integral $I_3^{\mu}$, represented by tensor coefficients $I_{3,i}^{[d+],st}$ in~(\ref{toobtain}),
have been reduced to scalar 2-,3-, and 4-point integrals in generic dimension $d$ by means of
(\ref{A511}) and~(\ref{A312}).
{The 2-point functions cancel here.}
Further we need the identity
\bea  \label{tuzeroa}
{s\choose s}_5 {0st\choose 0st}_5 = {0s\choose 0s}_5 {st\choose st}_5 -  {ts\choose 0s}_5^2,
\eea
and in order to get rid of the vector indices in the $2$-point functions, we need the relation
\bea
\left[{ts\choose 0s}_5 {ust\choose jst}_5-{ts\choose js}_5 {ust\choose 0st}_5\right]{s\choose s}_5
=\left[{ts\choose 0s}_5{us\choose js}_5-{ts\choose js}_5{us\choose 0s}_5\right]{st\choose st}_5,
\label{tuzero}
\eea
which shows that after cancellation of ${st\choose st}_5$, Eq.~(\ref{tuzero}) is antisymmetric in
$t$ and $u$ such that it can be effectively considered to vanish after summation over $t$ and $u$.
This
allows finally to introduce $I_{3}^{[d+],st}$ according to~(\ref{A301}) into~\eqref{squarebr}.

There is a further subtlety concerning~(\ref{squarebr}).
The ultraviolet (UV) divergency
\bea\label{eq-uv2}
I_{3,\mathrm{UV}}^{[d+],st} = -\frac{1}{2 \eps},
\eea
when combined with $\frac{d-2}{2}=1-\eps$, yields a constant finite contribution $\frac{1}{2}$.\footnote{See the discussion after~(\ref{A301}).}
Since, however,
\bea
\sum_{t=1}^{5}{ts\choose is}_5=0,
\label{zerosum}
\eea
this term does \emph{not} contribute and we can put $d=4$.
In that case~(\ref{squarebr}) reads
\bea
T^s_{ij}={s\choose i}_5 I_{4,j }^{[d+]^2,s}+{s\choose j}_5 I_{4,i }^{[d+]^2,s} ,
\label{bra2}
\eea
to be compared with~(\ref{bra1}).
 According to our general scheme, each $q_i$
generates a factor ${s\choose i}_5$, the further factor being a higher-dimensional
integral with index (indices) being the same as in the remaining chords.
In fact,
(\ref{bra1}) is a vector coefficient so that no additional index is available
and thus the  higher-dimensional integral cannot carry an index.

We just mention that, due to~(\ref{zerosum}) and~(\ref{A511}), also  the integral $I_{4,i }^{[d+]^2,s}$
is UV and infrared (IR-) finite.
Applying~(\ref{cancel}) in~(\ref{starter3}), we obtain products
${s\choose 0}_5 I_{4,i }^{[d+]^2,s}$, for which we can write, using
~(\ref{Zauberei1}),~(\ref{Zauberei3}),~(\ref{Zauberei2})  and~(\ref{A401}) and setting $d \rightarrow d+2$:
\bea
{s\choose 0}_5 I_{4,i }^{[d+]^2,s}={0s\choose 0i}_5 I_{4 }^{[d+],s}-
{s\choose i}_5 (d-1) I_{4 }^{[d+]^2,s}.
\label{I4idx}
\eea
Collecting all the contributions, our final result for the tensor of rank $R=3$ can be written as follows:
\begin{eqnarray}
I_{4}^{\mu\, \nu\, \lambda}&&= \sum_{i,j,k=1}^{5} \, q_i^{\mu}\, q_j^{\nu} \, q_k^{\lambda}
E_{ijk}+\sum_{k=1}^5 g^{[\mu \nu} q_k^{\lambda]} E_{00k},
\label{Exyz0}
\end{eqnarray}
with
\bea
\label{Exyz1}
E_{00j} &\equiv&   \sum_{s=1}^5 E_{00j}^s
\nn\\
&=& \sum_{s=1}^5 \frac{1}{{0\choose 0}_5}  \left[\frac{1}{2} {0s\choose 0j}_5 I_4^{[d+],s}- \frac{d-1}{3} {s\choose j}_5 I_4^{[d+]^2,s} \right]  ,
\\
E_{ijk} &\equiv&   \sum_{s=1}^5 E_{ijk}^s
\nn\\
&=&-  \sum_{s=1}^5\frac{1}{{0\choose 0}_5}  \left\{ \left[{0j\choose sk}_5 I_{4,i}^{[d+]^2,s}+
(i \leftrightarrow j)\right]+{0s\choose 0k}_5 {\nu}_{ij} I_{4,ij}^{[d+]^2,s} \right\}.
\label{Exyz2}
\eea
In ~(\ref{Exyz1}), we can put $d=4$ because, similar to the discussion by means of~(\ref{zerosum}), it is
\bea
\sum_{s=1}^5 {s\choose j}_5=0.
\label{SumZero}
\eea
Another possibility to argue uses that, due to~(\ref{A401}), $l=2$, the $d-1$ cancels.

The rank $R=3$ tensors were also treated in~\cite{Diakonidis:2008ij}.
We proved there successfully the cancellation of  $1/()_5$, although the corresponding formulae  were quite a bit longer than here: see Eqs. (3.41)--(3.42), (3.30)--(3.33), (3.40) in~\cite{Diakonidis:2008ij}. For the rank $R>3$, however, the tensor reduction would become really
awkward with the older approach.

\subsection{\label{appT4}Reduction of integrals with rank $R=4$}

For the tensor integral of  rank $R=4$, Eq. ~(\ref{starter}) reads:
\bea
{0\choose 0}_5 I_5^{\mu \nu \lambda \rho}  =
\left[{0\choose 0}_5 I_5^{\mu \nu \lambda}-\sum_{s=1}^{5} {s\choose 0}_5
I_4^{\mu \nu \lambda,s } \right] Q_0^{\rho} - \sum_{s=1}^{5}I_4^{\mu \nu \lambda,s }
{\bar{Q}}_s^{0,\rho}.
\label{starter4}
\eea
Here $I_5^{\mu \nu \lambda}$ is given in~(\ref{Exyz0}) to~(\ref{Exyz2}), $I_4^{\mu \nu \lambda,s}$
in~(\ref{tensor3}), taken at  $n=4$. In a similar manner we decompose the square
bracket in~(\ref{starter4}):
\begin{eqnarray}\label{eq-t5mnl}
T_{5}^{\mu\, \nu\, \lambda}&=& \sum_{s=1}^5 T_{5}^{\mu\, \nu\, \lambda,s},
\\
T_{5}^{\mu\, \nu\, \lambda,s}&=& \sum_{i,j,k=1}^{5} \, q_i^{\mu}\, q_j^{\nu} \, q_k^{\lambda}
T_{ijk}^s+\sum_{i=1}^5 g^{[\mu \nu} q_i^{\lambda]} T_{00i}^s,
\label{T3}
\end{eqnarray}
according to which:
\bea
T_{00i}^s&=&\left[{0\choose 0}_5 E_{00i}^s-\frac{1}{2}{s\choose 0}_5 I_{4,i}^{[d+]^2,s}\right]\nn\\
&=& \frac{1}{2} {s\choose i}_5 I_4^{[d+]^2,s}.
\label{T00k}
\eea
Obviously, the tensor coefficient $E_{00k}^s$ has been completely eliminated - as observed before in~\eqref{CancEi}.
As in~(\ref{Exyz0}), we now write
\begin{eqnarray}
I_{4}^{\mu\, \nu\, \lambda \rho}&&= \sum_{i,j,k,l=1}^{5}  q_i^{\mu} q_j^{\nu}  q_k^{\lambda} q_l^{\rho}
E_{ijkl}+\sum_{i,j=1}^5 g^{[\mu \nu} q_i^{\lambda} q_j^{\rho]} E_{00ij}+ g^{[\mu \nu} g^{\lambda \rho]}
E_{0000}.
\label{Ewxyz0}
\end{eqnarray}
Proceeding as before, using~(\ref{cancel}), from~(\ref{T00k}) we obtain:
\bea\label{Ew0000a}
E_{0000} &=& \sum_{s=1}^5 E_{0000}^s,
\\
E_{0000}^s&=& \frac{1}{4} \frac{{s\choose 0}_5}{{0\choose 0}_5} I_4^{[d+]^2,s}.
\label{Ew0000}
\eea
We remark that since~(\ref{T00k}) is summed over $s$, all (constant UV-) divergent contributions
from $I_4^{[d+]^2,s}$ can be dropped;
see also the discussion at the end of sect.~\ref{appT3}.

In the next step we calculate $T_{ijk}^s$:
\begin{align}\label{eq-tsijk}
T_{ijk}^s
= &{0\choose 0}_5 E_{ijk}^s+{s\choose 0}_5 {\nu}_{ij}{\nu}_{ijk}I_{4,ijk}^{[d+]^3,s}
\notag
\\
= &\frac{1}{{s\choose s}_5} \left\{  {s\choose s}_5\left[-{0i\choose sk}_5 I_{4,j }^{[d+]^2,s}-{0j\choose sk}_5 I_{4,i }^{[d+]^2,s}-
{0s\choose 0k}_5  {\nu}_{ij}I_{4,ij}^{[d+]^2,s}\right] 
\right.
\notag \\
& +~\left. {s\choose 0}_5  \left[{is\choose ks}_5 I_{4,j }^{[d+]^2,s}+{js\choose ks}_5
I_{4,i }^{[d+]^2,s}-{0s\choose ks}_5  {\nu}_{ij}I_{4,ij}^{[d+]^2,s}+\sum_{t=1}^5
{ts\choose ks}_5  {\nu}_{ij}I_{3,ij}^{[d+]^2,st} \right]
\right\}
\notag \\
= &\left\{
{s\choose i}_5 {0s\choose ks}_5 I_{4,j }^{[d+]^2,s}+{s\choose j}_5 {0s\choose ks}_5 I_{4,i }^{[d+]^2,s}-
{s\choose k}_5 {0s\choose 0s}_5 {\nu}_{ij}I_{4,ij }^{[d+]^2,s}
\right.
\notag \\
&+~ \left. {s\choose k}_5 \sum_{t=1}^5{ts\choose 0s}_5 {\nu}_{ij}I_{3,ij }^{[d+]^2,st}
\right\}\frac{1}{{s\choose s}_5},
\end{align}
where again~(\ref{Zauberei1}),~(\ref{Zauberei3}) and~(\ref{Zauberei2})  have been applied
and ${\nu}_{ijk}I_{4,ijk}^{[d+]^3,s}$ has been replaced by means of~(\ref{A533}).
Again we observe that the complete tensor of lower rank (here $E_{ijk}^s$) cancels.
 After
further lengthy manipulations and subsequent symmetrization,
the following analogue of~(\ref{bra1}) and~(\ref{bra2}) can be verified:
\bea
T^s_{ijk}=-
 \left\{
 {s\choose i}_5 {\nu}_{jk}I_{4,jk }^{[d+]^3,s}+{s\choose j}_5 {\nu}_{ik}I_{4,ik }^{[d+]^3,s}
+{s\choose k}_5 {\nu}_{ij}I_{4,ij }^{[d+]^3,s}
\right\}.
\label{bra3}
\eea
Using again~(\ref{cancel}), we can immediately write down the pure spatial components:
\bea
E_{ijkl} &\equiv&  \sum_{s=1}^5 E_{ijkl}^s
\nn\\
&=&  \sum_{s=1}^5\frac{1}{{0\choose 0}_5}  \left\{ \left[
{0k\choose sl}_5 {\nu}_{ij}I_{4,ij }^{[d+]^3,s}+(i \leftrightarrow k)+(j \leftrightarrow k)\right]
+{0s\choose 0l}_5{n}_{ijk} I_{4,ijk}^{[d+]^3,s} \right\}.
\label{Ewxyz3}
\eea
For the mixed terms $E^s_{00ij}$, i.e. those containing the metric tensor, we have contributions
from different origins.
From $T_{00k}^s$ (see~(\ref{T00k})) we get
\bea\label{eq-non}
-\frac{1}{2}\sum_{k,l=1}^5 g^{[\mu \nu} q_k^{\lambda]} q_l^{\rho}{0k\choose sl}_5 I_{4 }^{[d+]^2,s}.
\eea
From~(\ref{bra3}),  we get
\bea\label{eq-non2}
&&-\sum_{i,j,k,l=1}^{5} q_i^{\mu} q_j^{\nu} q_k^{\lambda} q_l^{\rho}{s\choose 0}_5\left\{
\frac{{i\choose l}_5}{{\left( \right)}_5} {\nu}_{jk}I_{4,jk }^{[d+]^3,s}+
\frac{{j\choose l}_5}{{\left( \right)}_5} {\nu}_{ik}I_{4,ik }^{[d+]^3,s}+
\frac{{k\choose l}_5}{{\left( \right)}_5} {\nu}_{ij}I_{4,ij }^{[d+]^3,s}
\right\}.\nn \\
=&&-\frac{1}{2}{s\choose 0}_5
\left\{g^{\mu \rho}\sum_{j,k=1}^5 q_j^{\nu} q_k^{\lambda}{\nu}_{jk}I_{4,jk }^{[d+]^3,s}+
       g^{\nu \rho}\sum_{i,k=1}^5 q_i^{\mu} q_k^{\lambda}{\nu}_{ik}I_{4,ik }^{[d+]^3,s}+
    g^{\lambda\rho}\sum_{i,j=1}^5 q_i^{\mu} q_j^{\nu}    {\nu}_{ij}I_{4,ij }^{[d+]^3,s} \right\}.\nn \\
\eea
Finally, there is a contribution from the second term of~(\ref{tensor3}):
\bea\label{eq-n1}
-\frac{1}{2} \sum_{i,l=1}^{5} (\, g^{\mu \nu}  \, q_i^{\lambda} \,+ g^{\mu \lambda}  \, q_i^{\nu} \,+
    \, g^{\nu \lambda}  \, q_i^{\mu} \, ) q_l^{\rho} {0s\choose 0l}_5 I_{4,i}^{[d+]^2,s} .
\eea
Collecting these contributions without symmetrization we have:
\bea\label{eq-n2}
E_{00ij} &=&  \sum_{s=1}^5 E_{00ij}^s,
\\
E^s_{00ij}&=&-\frac{1}{4 {0\choose 0}_5} \left\{{0i\choose sj}_5 I_{4 }^{[d+]^2,s}+
{0s\choose 0j}_5 I_{4,i }^{[d+]^2,s}+{s\choose 0}_5{\nu}_{ij}I_{4,ij }^{[d+]^3,s} \right\} .
\label{E00ij}
\eea
A general comment is in order at this place:
The only UV divergent term in~(\ref{E00ij})
is $I_{4 }^{[d+]^2,s}$, which comes from~(\ref{T00k}).
We see, however, that due to
(\ref{SumZero}) this (constant) term does not contribute when $T^s_{00k}$ is summed over $s$.
Thus, the UV divergent  part can be dropped in $I_{4 }^{[d+]^2,s}$
and as a consequence it does also not appear in~(\ref{E00ij}).
This, after all, is only an
expression of the fact that the original tensor integral under consideration is finite.

\subsection{\label{appT5}Reduction of integrals with rank $R=5$}
For the tensor integral of rank $R=5$, Eq.~(\ref{starter}) reads:
\bea
{0\choose 0}_5 I_5^{\mu \nu \lambda \rho \sigma}  =
\left[{0\choose 0}_5 I_5^{\mu \nu \lambda \rho}-\sum_{s=1}^{5} {s\choose 0}_5
I_4^{\mu \nu \lambda \rho,s } \right] Q_0^{\sigma} - \sum_{s=1}^{5}I_4^{\mu \nu \lambda \rho,s }
{\bar{Q}}_s^{0,\sigma}.
\label{starter5}
\eea
Writing the square bracket in~(\ref{starter5}) as
\begin{eqnarray}
\label{s5a}
T_{5}^{\mu\, \nu\, \lambda \rho}
&=& \sum_{s=1}^5  T_{5}^{\mu\, \nu\, \lambda \rho,s},
\\
T_{5}^{\mu\, \nu\, \lambda \rho,s}
&=&
\sum_{i,j,k,l=1}^{5}  q_i^{\mu} q_j^{\nu}  q_k^{\lambda} q_l^{\rho}
T_{ijkl}^s+\sum_{i,j=1}^5 g^{[\mu \nu} q_i^{\lambda} q_j^{\rho]} T_{00ij}^s+ g^{[\mu \nu} g^{\lambda \rho]}
T_{0000}^s,
\label{T4}
\end{eqnarray}
we, first of all, observe that the $g^{[\mu \nu} g^{\lambda \rho]}$ term vanishes.
Indeed,
from~(\ref{Ew0000}) and ~(\ref{tensor5}) we have:
\bea
T^s_{0000}={0\choose 0}_5  E_{0000}^s-\frac{1}{4}{s\choose 0}_5 I_4^{[d+]^2,s} =0.
\label{rank5cancel}
\eea
Again we have a situation like in~(\ref{rank3cancel}): There is no vector
$q_i$, and no ${s\choose i}_5$ is produced. Thus, no inverse Gram determinant appears
since this term vanishes.
  The next term, $T^s_{00ij}$, is calculated similarly as is scetched in
sect.~\ref{appT3} with a result generalizing~(\ref{T00k}):
\bea
T_{00ij}^s &=& \left[{0\choose 0}_5 E_{00ij}^s+\frac{1}{2}{s\choose 0}_5
{\nu}_{ij} I_{4,ij}^{[d+]^3,s}\right]
\nl
&=&- \frac{d}{8}\left\{ {s\choose i}_5 I_{4,j}^{[d+]^3,s}+
{s\choose j}_5 I_{4,i}^{[d+]^3,s} \right\} .
\label{T00ij}
\eea
In contrary to the discussion at the end of sect.  {\ref{appT4}, summing ~(\ref{T00ij}) over $s$,
the UV divergence of the integrals $I_{4,i}^{[d+]^3,s}$ does not drop out since
$I_{4,i}^{[d+]^3,s}=0$ for $s=i$.  The corresponding divergence cancels in this case
against a divergence coming from the last term of~(\ref{tensor4}).

In analogy to~(\ref{bra1}),~(\ref{bra2}) and~(\ref{bra3}) we also have
\bea
T^s_{ijkl}=
  {s\choose i}_5 {n}_{jkl}I_{4,jkl}^{[d+]^4,s}+{s\choose j}_5 {n}_{ikl}I_{4,ikl}^{[d+]^4,s}
+ {s\choose k}_5 {n}_{ijl}I_{4,ijl}^{[d+]^4,s}+{s\choose l}_5 {n}_{ijk}I_{4,ijk}^{[d+]^4,s} .
\label{bra4}
\eea
It is interesting to note that a second chain of tensor coefficients has developed for
the square bracket tensor $T^{\mu_1  \dots \mu_{R-1}}$~(\ref{tensorT})
which follows the same rule when proceeding to
higher ranks, namely~(\ref{T00k}) and~(\ref{T00ij}) to be compared with the chain
(\ref{bra1}),~(\ref{bra2}),~(\ref{bra3}) and~(\ref{bra4}).

The complete tensor of rank $R=5$~\eqref{tensor5} now reads
\bea\label{compl5a}
I_{5}^{\mu\, \nu\, \lambda \rho \sigma} &=&  \sum_{s=1}^5 I_{5}^{\mu\, \nu\, \lambda \rho \sigma,s} ,
\\
I_{5}^{\mu\, \nu\, \lambda \rho \sigma,s}
&=& \sum_{i,j,k,l,m=1}^{5}  q_i^{\mu} q_j^{\nu}  q_k^{\lambda} q_l^{\rho} q_m^{\sigma} E_{ijklm}^s+
\sum_{i,j,k=1}^5 g^{[\mu \nu} q_i^{\lambda} q_j^{\rho} q_k^{\sigma]} E_{00ijk}^s
+
\sum_{i=1}^5
~g^{[\mu \nu} g^{\lambda \rho} q_i^{\sigma]} E_{0000i}^s .
\nl
\label{compl5}
\eea
Using~(\ref{barQ}) and~(\ref{cancel}), we obtain for the pure spatial part
\bea
E_{ijklm}^s &=&
-\frac{1}{{0\choose 0}_5}  \left\{ \left[
{0l\choose sm}_5 {n}_{ijk}I_{4,ijk}^{[d+]^4,s}+(i \leftrightarrow l)+(j \leftrightarrow l)+
(k \leftrightarrow l)\right]
+{0s\choose 0m}_5{n}_{ijkl} I_{4,ijkl}^{[d+]^4,s} \right\}.
\nl
\label{Ewxyz5}
\eea
Next we consider again the mixed terms and begin with $E_{00ijk}$.
From~(\ref{T00ij}) we have
\bea
\frac{1}{2}\sum_{i,j,k=1}^5 g^{[\mu \nu} q_i^{\lambda} q_j^{\rho]}  q_k^{\sigma}
\left[ {0i\choose sk}_5 I_{4,j}^{[d+]^3,s}+{0j\choose sk}_5 I_{4,i}^{[d+]^3,s} \right].
\label{Mix51}
\eea
From~(\ref{bra4}) we get:
\bea
&&\sum_{i,j,k,l.m=1}^{5} q_i^{\mu} q_j^{\nu} q_k^{\lambda} q_l^{\rho} q_m^{\sigma}
{s\choose 0}_5\left\{
\frac{{i\choose m}_5}{{\left( \right)}_5} {n}_{jkl}I_{4,jkl}^{[d+]^4,s}+
\frac{{j\choose m}_5}{{\left( \right)}_5} {n}_{ikl}I_{4,ikl}^{[d+]^4,s}
\right. \nn \\
&&\left. ~~~~~~~~~~~~~~~~~~~~~~~~~~~~~~~~~~~~~~~~~~~~~~~
+ \frac{{k\choose m}_5}{{\left( \right)}_5} {n}_{ijl}I_{4,ijl}^{[d+]^4,s}+
\frac{{l\choose m}_5}{{\left( \right)}_5} {n}_{ijk}I_{4,ijk}^{[d+]^4,s}
\right\}\nn \\
=&&\frac{1}{2}{s\choose 0}_5
\left\{g^{\mu \sigma}\sum_{j,k,l=1}^5 q_j^{\nu} q_k^{\lambda} q_l^{\rho}{n}_{jkl}I_{4,jkl}^{[d+]^4,s}+
       g^{\nu \sigma}\sum_{i,k,l=1}^5 q_i^{\mu} q_k^{\lambda} q_l^{\rho}{n}_{ikl}I_{4,ikl}^{[d+]^4,s}
~\right. \nn \\
&&\left. ~~~~~~~~~~~~~~~~~~~~
 +   g^{\lambda\sigma}\sum_{i,j,l=1}^5 q_i^{\mu} q_j^{\nu} q_l^{\rho}    {n}_{ijl}I_{4,ijl}^{[d+]^4,s}+
 g^{\rho\sigma}\sum_{i,j,k=1}^5 q_i^{\mu} q_j^{\nu} q_k^{\lambda }{n}_{ijk}I_{4,ijk}^{[d+]^4,s}
 \right\}.
\label{Mix52}
\eea
There is a $4$-point contribution from the second term of~(\ref{tensor4}):
\bea
\frac{1}{2} \sum_{i,j,k=1}^{5}  g^{[\mu \nu}  \, q_i^{\lambda} \, q_j^{\rho]} q_k^{\sigma}
{0s\choose 0k}_5 {\nu}_{ij} I_{4,ij}^{[d+]^3,s} .
\label{Mix53}
\eea
In~(\ref{Mix51}) and ~(\ref{Mix53}) we have the tensor structure
$g^{[\mu \nu} q_i^{\lambda} q_j^{\rho]}  q_k^{\sigma}$,
and in~(\ref{compl5}) the structure $g^{[\mu \nu} q_i^{\lambda} q_j^{\rho} q_k^{\sigma]}$.
In order to identify them, we can make
use of the fact that the tensor $I_5^{\mu \nu \lambda \rho \sigma}$ is symmetric in all indices
due to which we have only to count the number of terms in the structures to be compared:
e.g.
(\ref{G2V3}) contains ten terms,  while~(\ref{G2V2}) contains six terms.
 Thus, replacing
(\ref{G2V2}), multiplied by $q_k^{\sigma}$,
by~(\ref{G2V3}) we have to introduce a
factor taking care of the ratio
of the numbers of terms in each of them.
Similarly this applies for~(\ref{Mix52}).
In this way we obtain
\bea
E^s_{00ijk}=\frac{1}{2 {0\choose 0}_5} \left\{
\frac{3}{5}\left[\frac{d}{4}{0i\choose sk}_5 I_{4,j}^{[d+]^3,s}+\frac{d}{4}{0j\choose sk}_5 I_{4,i}^{[d+]^3,s}+
{0s\choose 0k}_5 {\nu}_{ij}I_{4,ij}^{[d+]^3,s}\right]+
\frac{2}{5} {s\choose 0}_5{n}_{ijk}I_{4,ijk}^{[d+]^4,s} \right\} .
\nl
\label{E00ijk}
\eea
This concludes the $E_{00ijk}$ and
finally we have to collect the contributions to $E_{0000i}$. They come from
the last term of~(\ref{tensor4}) and~(\ref{T00ij}) with the result
\bea
E^s_{0000i}=-\frac{1}{4 {0\choose 0}_5}\frac{1}{5}  \left\{
{0s\choose 0i}_5 I_{4}^{[d+]^2,s}+d
{s\choose 0}_5 I_{4,i}^{[d+]^3,s} \right\}.
\label{E0000i}
\eea
}
We will not proceed here further, but by now it may be evident to the reader how to treat tensors of higher rank.
One remark, however, is in order: identifying the $I_{4,i \cdots}^{[d+]^l}$ in
\eqref{tensor1} -~\eqref{tensor5} as $4$-point tensor coefficients, the above tensor coefficients
of the $5$-point functions should be equivalent to (6.17) -(6.21) of~\cite{Denner:2005nn}.
Nevertheless, working with higher dimensional $4$-point functions and in particular
using the \emph{algebra of the signed minors} appears advantageous to us.

\section{\label{4togeneric}Calculation of higher-dimensional $4$-point functions}
In the foregoing sects. the $5$-point tensor coefficients $I_{5,i\cdots}^{[d+]^{R-r}}$ have been
rewritten in terms of $4$-point tensor coefficients.
The factor $1/()_5$ has been completely avoided.
In detail we have:
\begin{itemize} 
 \item Tensors with $R=2$:
\\The tensor coefficients
$E_{00}, E_{ij}$ are expressed by $I_{4}^{[d+],s},  I_{4,i}^{[d+],s} $.
 \item Tensors with $R=3$:
\\ The tensor coefficients
$E_{00k}, E_{ijk}$ are expressed by $I_{4}^{[d+],s}, I_{4}^{[d+]^2,s}, I_{4,i}^{[d+]^2,s}, I_{4,ij}^{[d+]^2,s} $.
 \item Tensors with $R=4$:
\\ The tensor coefficients
$E_{0000}, E_{00ij}, E_{ijkl}$ are expressed by $I_{4}^{[d+]^2,s}, I_{4,i}^{[d+]^2,s}, I_{4,ij}^{[d+]^3,s},  I_{4,ijk}^{[d+]^3,s}$.
 \item Tensors with $R=5$:
\\ The tensor coefficients
$E_{0000i}, E_{00ijk}, E_{ijklm}$ are expressed by $I_{4}^{[d+]^2,s}$, $I_{4,i}^{[d+]^3,s}$, $I_{4,ij}^{[d+]^3,s}$, $I_{4,ijk}^{[d+]^4,s}$, $I_{4,ijkl}^{[d+]^4,s}$.
\end{itemize}

It is our goal to find a representation of these integrals which is suited
for the most problematic cases occurring in practical calculations,
namely for
vanishing sub-Gram determinants $()_4={s\choose s}_5 $.
In the numerics we will make use of opensource programs for the calculation of few \emph{master integrals}, chosen here to be the scalar 1-point to 4-point functions in generic dimension $d=4-\eps$,
in standard
notation the integrals $A_0$, $B_0$, $C_0$ and $D_0$.
They are available from e.g. the LoopTools/FF package~\cite{Hahn:1998yk,vanOldenborgh:1990yc} or from the QCDloop/FF package~\cite{Ellis:2007qk,vanOldenborgh:1990yc}.
For this purpose,
we have to reduce dimension and indices of the above integrals.
This may be done
by  recurrence relations~\eqref{eq:RR1} and~\eqref{eq:RR2},
given in detail in app.~\ref{App} . In each
recursion step an inverse power of $()_4$ is generated, which causes  numerical
problems  for small $()_4$ although the original integrals $ I_{4,i\cdots}^{[d+]^{l},s}$
are finite and well-behaved there.

We proceed in two steps.
In  subsect.~\ref{highdim4}, an intermediate step, we manage to write the integrals
in the form:
\bea\label{eq-i4andz}
I_{4,ij\cdots k}^{[d+]^l,s} \sim \frac{{0s\choose ks}_5}{{s\choose s}_5}  \left[I_{4,ij\cdots}^{[d+]^{l-1},s}-Z_{4,ij\cdots}^{[d+]^{l-1},s}\right]+R_{4,ij\cdots k}.
\eea
Here $Z_{4,ij\cdots}^{[d+]^{l-1},s}$ is constructed such that
in the limit $()_4 \rightarrow 0$
it has the same value as $I_{4,ij\cdots}^{[d+]^{l-1},s}$, i.e.
the first term, the difference quotient
$\left[I_{4,ij\cdots}^{[d+]^{l-1},s}-Z_{4,ij\cdots}^{[d+]^{l-1},s}\right] / {s\choose s}_5$,
 stays finite in this limit.
Further, dimension and indices
are reduced.
The remainder $R_{4,ij\cdots k}$  does no contain an inverse $()_4$.
In a second step, in subsect.~\ref{DiffQuo}, we will eliminate the inverse $()_4$
in the difference quotient.

\subsection{\label{highdim4}%
{Difference quotients with $1/()_4$}}
In this subsect.  we derive optimized, compact expressions, where the appearance of
possible singular $1/()_4$-terms is reduced as much as possible.
We will treat the singular behaviour using the fact that the integrals are exactly known
in the limit ${\left( \right)}_n \rightarrow 0$.
In fact, if ${\left( \right)}_n=0$, due to~\eqref{eq:RR2}
the $n$-point integrals degenerate to integrals with scratched propagators:
\bea
\lim_{()_n \rightarrow 0} I_{n,i\cdots}^{(d)} = \sum_{t=1}^n \frac{{t\choose 0}_n}{{0\choose 0}_n}~~ {\bf t^-}I_{n,i\cdots}^{(d)}.
\label{ItoZ0}
\eea
Accordingly we define objects
 which  converge  in the limit $()_4 \rightarrow 0$  to the corresponding
tensor coefficients, taken in that limit:
\bea\label{z4d}
 Z_4^{(d),s}&=&\sum_{t=1}^5 \frac{{ts\choose 0s}_5}{{0s\choose 0s}_5} I_3^{(d),st},
\\
Z_{4,i}^{(d),s}&=& 
\sum_{t=1,t \ne i}^5 \frac{{ts\choose 0s}_5}{{0s\choose 0s}_5} I_{3,i}^{(d),st}+
\frac{{is\choose 0s}_5}{{0s\choose 0s}_5} I_4^{(d),s} ,
\label{Z4id}
\\
{\nu}_{ij}Z_{4,ij}^{(d),s}&=&  
\sum_{t=1,t \ne i,j}^5 \frac{{ts\choose 0s}_5}{{0s\choose 0s}_5} {\nu}_{ij}I_{3,ij}^{(d),st}+
\frac{{is\choose 0s}_5}{{0s\choose 0s}_5} I_{4,j}^{(d),s}+
\frac{{js\choose 0s}_5}{{0s\choose 0s}_5} I_{4,i}^{(d),s} ,
\label{Z4ij}
\\\label{Z4ijk}
{\nu}_{ij}{\nu}_{ijk} Z_{4,ijk}^{(d),s}&=&  
\sum_{t=1,t \ne i,j,k}^5 \frac{{ts\choose 0s}_5}{{0s\choose 0s}_5}
{\nu}_{ij}{\nu}_{ijk} I_{3,ijk}^{(d),st}
\nn \\
&&+~
\frac{{ks\choose 0s}_5}{{0s\choose 0s}_5} {\nu}_{ij}I_{4,ij}^{(d),s}
+
\frac{{js\choose 0s}_5}{{0s\choose 0s}_5} {\nu}_{ik}I_{4,ik}^{(d),s}
+
\frac{{is\choose 0s}_5}{{0s\choose 0s}_5} {\nu}_{jk}I_{4,jk}^{(d),s} .
\eea
Eq.~(\ref{A401}) reads in this notation
\bea
I_4^{(d+2),s}
&=&{0s\choose 0s}_5 \frac{1}{{s\choose s}_5}\left[I_4^{(d),s}
-
\sum_{t=1}^5 \frac{{ts\choose 0s}_5}{{0s\choose 0s}_5} I_3^{(d),st} \right]\frac{1}{d-3}
\nl
&\equiv&
{0s\choose 0s}_5 \frac{1}{{s\choose s}_5}\left[I_4^{(d),s}-Z_4^{(d),s}
\right]\frac{1}{d-3} ,
\label{I4d+}
\eea
such that indeed for ${s\choose s}_5 \rightarrow 0$ the $I_4^{(d+2),s}$ remains finite.
We need this relation with $d=4-2 \eps$ and $d=[d+]=6-2\eps$  for tensors of rank $R=2,3$  and rank $R=3,4,5$, respectively.

The next integral is $I_{4,i}^{(d),s}$.
The recursion of integrals with one  index, $I_{4,i}^{(d),s}$, is~\eqref{A511}.
To rewrite~\eqref{A511} in a similar manner as~\eqref{I4d+}, we evaluate the
right hand side of~\eqref{A511}, replacing $I_4$ by $Z_4$ :
\bea\label{i4deqa}
\frac{{0s\choose is}_5}{{s\choose s}_5} Z_4^{(d),s}
 - \sum_{t=1}^5 \frac{{ts\choose is}_5}{{s\choose s}_5} I_3^{(d),st}
&=&
\sum_{t=1}^5 \frac{1}{{s\choose s}_5} \left[ {0s\choose is}_5 \frac{{ts\choose 0s}_5}{{0s\choose 0s}_5}
-{ts\choose is}_5\right] I_3^{(d),st}
\nl
&=& -~\frac{1}{{0s\choose 0s}_5}\sum_{t=1}^5 {0st\choose 0si}_5 I_3^{(d),st},
\eea
where the latter eqn.   is due to
\bea\label{i4deq}
{0s\choose is}_5 {ts\choose 0s}_5 - {0s\choose 0s}_5 {ts\choose is}_5
= -{s\choose s}_5 {0st\choose 0si}_5.
\eea
The factor ${1} / {s\choose s}_5$ has cancelled and we obtain the analogue to~\eqref{I4d+}:
\bea
I_{4,i}^{(d+2),s} 
&=&
-\frac{{0s\choose is}_5}{{s\choose s}_5} \left[I_4^{(d),s}-Z_4^{(d),s}\right]+
\frac{1}{{0s\choose 0s}_5}\sum_{t=1}^5 {0st\choose 0si}_5 I_3^{(d),st}
 \nn \\
&=&
\frac{1}{{0s\choose 0s}_5}\left[-{0s\choose is}_5  (d-3) I_4^{(d+2),s}
+\sum_{t=1}^5 {0st\choose 0si}_5 I_3^{(d),st}\right].
\label{I4id+2}
\eea
In the first line of~(\ref{I4id+2}) we have introduced a difference quotient
which can in the next step be replaced, due to~(\ref{I4d+}), by $I_4^{(d+2),s}$, i.e.
an integral of the same dimension as the original integral on the left hand side.
In fact the  second line of~(\ref{I4id+2})
is already our final result for this type of tensor coefficient.
We need this result  for the 5-point tensors of
rank $R=2$
{with $d=4-2 \eps$ (generic dimension),  for $R=3, 4$ with $d=\left[d+\right]$, and for  $R=5$ with $d=\left[d+\right]^2$.
Eqn.~\eqref{I4id+2} demonstrates our principle as described above.}

In the tensor integrals of higher rank more complicated difference quotients appear and
 will be dealt with in the next subsect.~\ref{DiffQuo}.
The procedure of calculation is the same as before.
In order  to obtain, e.g., ${\nu}_{ij} I_{4,ij}^{\left[d+\right]^2,s}$,
we calculate the right hand side of~\eqref{A522} (for $l=2$), replacing $I_{4,i}^{\left[d+\right]}$ by
$Z_{4,i}^{\left[d+\right]}$. Using again relation~\eqref{i4deq} we find
\bea
{\nu}_{ij} I_{4,ij}^{[d+]^2,s}=-\frac{{0s\choose js}_5}{{s\choose s}_5} \left[I_{4,i}^{[d+],s}-Z_{4,i}^{[d+],s}\right]+
\frac{1}{{0s\choose 0s}_5}\left[{0si\choose 0sj}_5 I_4^{[d+],s}+\sum_{t=1,t \ne i}^5 {0st\choose 0sj}_5 I_{3,i}^{[d+],st}\right]. \nn \\
\label{d3}
\eea
Next we obtain from~(\ref{A533})
\bea
{\nu}_{ij}{\nu}_{ijk} I_{4,ijk}^{[d+]^3,s}=
&&-\frac{{0s\choose ks}_5}{{s\choose s}_5} {\nu}_{ij}
\left[I_{4,ij}^{[d+]^2,s}-Z_{4,ij}^{[d+]^2,s}\right] \nn \\
&&+\frac{1}{{0s\choose 0s}_5}\left[{0si\choose 0sk}_5 I_{4,j}^{[d+]^2,s}+
                                {0sj\choose 0sk}_5 I_{4,i}^{[d+]^2,s}
+\sum_{t=1,t \ne i,j}^5 {0st\choose 0sk}_5 {\nu}_{ij} I_{3,ij}^{[d+]^2,st}\right], \nn \\
\label{I4ijkd+3}
\eea
and with~(\ref{A555})
\bea
{n}_{ijkl}I_{4,ijkl}^{[d+]^4,s}
&=&-\frac{{0s\choose ls}_5}{{s\choose s}_5} {\nu}_{ij} {\nu}_{ijk}
\left[I_{4,ijk}^{[d+]^3,s}-Z_{4,ijk}^{[d+]^3,s}\right]
+\frac{1}{{0s\choose 0s}_5}\left[{0sk\choose 0sl}_5 {\nu}_{ij}I_{4,ij}^{[d+]^3,s}+ \right.
\nn \\
&&\left.  {0sj\choose 0sl}_5 {\nu}_{ik}I_{4,ik}^{[d+]^3,s}+
{0si\choose 0sl}_5 {\nu}_{jk}I_{4,jk}^{[d+]^3,s}
+\sum_{t=1,t \ne i,j,k}^5 {0st\choose 0sl}_5 {\nu}_{ij} {\nu}_{ijk}I_{3,ijk}^{[d+]^3,st}\right]. \nn \\
\label{I4ijkld+4}
\eea
We now have collected all contributions to higher-dimensional integrals
with an ${s\choose s}_5$ in the denominator in
such a way that also the numerator vanishes for ${s\choose s}_5=0$,
see~(\ref{ItoZ0}). These results are only a rewriting of the
recurrence relations, but they make the finiteness of the integrals at  ${s\choose s}_5=0$
manifest.
They will be a starting point to find a final representation, which
is truly optimal for kinematical points around $()_4=0$.
In the second line of
(\ref{I4id+2})
we observe that
there are no explicit inverse Gram determinants anymore.
In the following we will show that this also holds for integrals
with any number of indices.

\subsection{\label{DiffQuo}%
{Reduction of the difference quotients}}
In app.~\ref{App} we reproduce a list of the
recurrence relations needed for the evaluation
of the $5$-point functions. In fact, since all tensor coefficients
of the $5$-point functions have been reduced to higher-dimensional $4$-point functions,
we need only the recursions for the latter.
 When applying these formulae to $5$-point functions,
we have to identify ${\left(  \right)_4}={s\choose s}_5$ and $I_{4}^{[d+]}=I_{5}^{[d+],s}$, etc.
In the present sect.  we will drop the index $s$ in the Gram determinant and in the upper indices of
the integrals.

We now discuss the
 higher-dimensional $4$-point functions needed
for the different tensor ranks of the $5$-point functions.
For the tensor of rank $R=2$~(\ref{Exy}) we need $I_4^{[d+]}$ and $I_{4,i}^{[d+]}$  given
in~(\ref{I4d+}) and~(\ref{I4id+2}).
In the spirit of our approach they are already
in the final form.
For the tensor of rank $R=3$,~(\ref{Exyz1}) and~(\ref{Exyz2}), we further need
$I_4^{[d+]^2}$, $I_{4,i}^{[d+]^2}$ and ${\nu}_{ij} I_{4,ij}^{[d+]^2}$.
These are given in
(\ref{I4d+}),~(\ref{I4id+2}) and~(\ref{d3}).
In fact, the first two are already in the
final form, while in the last one a new difference quotient appears.

Our general approach to cancel Gram determinants is, first of all, to use the recurrence
relations "backward", i.e. to express a $4$-point function of dimension $d$ by one of
dimension $d+2$, multiplied by a Gram determinant, plus a sum over $3$-point functions.
The factorized Gram determinant can be cancelled and for the collected sum over
$3$-point functions the algebra of Cayley determinants allows to combine them such that
again the Gram determinant factorizes and can be cancelled.

With the notation
$\left( \right) \equiv {\left( \right)}_4$ we obtain
\bea
\frac{{0\choose 0}}{\left( \right)} \left[I_{4,i}^{[d+]}-Z_{4,i}^{[d+]}\right]=
-(d-2)\left[\frac{{0\choose i}}{{0\choose 0}}(d-1)I_4^{[d+]^2}-\frac{1}{{0\choose 0}}
\sum_{t=1}^4 {0t\choose 0i}I_3^{[d+],t}\right] ,
\label{DQ1}
\eea
and from~(\ref{d3})
\bea
{\nu}_{ij} I_{4,ij}^{[d+]^2}= && ~~\frac{{0\choose i}}{{0\choose 0}}\frac{{0\choose j}}{{0\choose 0}}
(d-2)(d-1)I_4^{[d+]^2}+\frac{{0i\choose 0j}}{{0\choose 0}}I_{4}^{[d+]}\nn \\
&&-\frac{{0\choose j}}{{0\choose 0}}\frac{d-2}{{0\choose 0}}\sum_{t=1}^4 {0t\choose 0i}I_3^{[d+],t} ~~~+
\frac{1}{{0\choose 0}}\sum_{t=1}^4 {0t\choose 0j}I_{3,i}^{[d+],t} .
\label{want1}
\eea
This is the form we wanted to obtain.
The higher-dimensional $3$-point functions can
be calculated by means of the recurrence relations given in app.~\ref{sub3}, reducing them to
scalar functions in generic dimension.
If $\left( \right)$ is not small, the same applies for the higher-dimensional
$4$-point functions, in particular ~(\ref{A401});
otherwise
 we will use~(\ref{z4d}) by setting up an expansion in the small
Gram determinant.
This will be done in subsect.~\ref{Gram}.

It is worth mentioning that we can deal with $3$-point functions like $I_{3,i}^{(d),t}$
in ~(\ref{want1}) in the same manner as we dealt with the $4$-point functions:
\bea
I_{3,i}^{[d+],t}=-\frac{{0t\choose it}}{{0t\choose 0t}}(d-2) I_{3}^{[d+],t}+\frac{1}{{0t\choose 0t}}
\sum_{u=1}^4 {0tu\choose 0ti}I_2^{tu} ,
\label{indices}
\eea
to be compared with~\eqref{I4id+2}.
This allows to handle the $3$-point functions in case ${t\choose t}=0$, for which
case~(\ref{A301}) does not work - expanding in small ${t\choose t}$ if needed by the
use of~(\ref{ItoZ0}).
The ${0t\choose 0t}$, however, vanishes for an infrared $3$-point
function  - thus we have to assume here that ${t\choose t}$ and ${0t\choose 0t}$ don't vanish
simultaneously in order to be able to apply at least one of them; this is also implicitly assumed for the case of the $4$-point functions.

Exploiting this approach systematically, it can be achieved in general
that the indices are carried, like in~\eqref{want1} and~\eqref{indices}, only by the Cayley determinants,
multiplied by scalar integrals in higher dimension. This property might become
useful for further analytical evaluation of the original Feynman diagrams, in performing partial sums over indices explicitly where needed.
We point out that due to the powers of  $d$ in front of higher-dimensional integrals we have
to take into account finite rational contributions arising from the divergencies of the integrals;
see
app.~\ref{Bpp} for a list of examples.

For the tensors of rank $R=4$(see~(\ref{Ew0000}),~(\ref{Ewxyz3}) and~(\ref{E00ij}))
we further need $I_{4,ij}^{[d+]^3}$ and ${\nu}_{ij}{\nu}_{ijk} I_{4,ijk}^{[d+]^3}$.
The tensor with two indices was treated in~(\ref{want1}) and we have only to shift
the dimension: $d \rightarrow d+2$.

Much more involved is now the calculation of $I_{4,ijk}^{[d+]^3}$.  The crucial point of
our approach is to obtain here an expression for the following difference quotient with
the envisaged properties:
\bea
\frac{{0\choose 0}}{\left( \right)} {\nu}_{ij} \left[I_{4,ij}^{[d+]^2}-Z_{4,ij}^{[d+]^2}\right]=&&
d\frac{{0\choose i}}{{0\choose 0}}\frac{{0\choose j}}{{0\choose 0}}(d-3)(d+1)I_4^{[d+]^3}+
(d-3)\frac{1}{{0\choose 0}}{0i\choose 0j}I_4^{[d+]^2}  + 2 {\nu}_{ij}I_{4,ij}^{[d+]^3} \nn \\
&&+\frac{1}{\left( \right)}
\left\{
d\frac{{0\choose i}}{{0\choose 0}}\frac{{0\choose j}}{{0\choose 0}}(d-3)
\sum_{t=1}^4 {t\choose 0}I_3^{[d+]^2,t}+\frac{1}{{0\choose 0}}{0i\choose 0j}\sum_{t=1}^4 {t\choose 0}I_3^{[d+],t} \right. \nn \\
&&- \left. (d-2)\frac{{0\choose j}}{{0\choose 0}} \sum_{t=1}^4 {0t\choose 0i}I_3^{[d+],t}+
\sum_{t=1}^4 {0t\choose 0j}I_{3,i}^{[d+],t} \right. \nn \\
&&-\sum_{t=1}^4 {t\choose 0}{\nu}_{ij}  I_{3,ij}^{[d+]^2,t}
- \sum_{t=1}^4 {t\choose i}I_{3,j}^{[d+]^2,t}-
\left. \sum_{t=1}^4 {t\choose j}I_{3,i}^{[d+]^2,t} \right\} .
\label{DQ3}
\eea
By construction, the $4$-point functions have no explicit inverse Gram determinant anymore. It is interesting
that the formerly calculated $I_{4,ij}^{[d+]^3}$ (see~(\ref{want1})) enters here as a whole.
The remaining task is now to show that in the sum of $3$-point functions in~(\ref{DQ3}) a
Gram determinant $\left( \right)$ factorizes and thus cancels its overall factor $1/\left( \right)$. Indeed this is so.
After a tremendeous amount of cancellations one gets  the result:
\bea
\frac{{0\choose 0}}{\left( \right)} {\nu}_{ij} \left[I_{4,ij}^{[d+]^2}-Z_{4,ij}^{[d+]^2}\right]&&=
\frac{{0\choose i}}{{0\choose 0}}\frac{{0\choose j}}{{0\choose 0}}(d-1)d(d+1)I_4^{[d+]^3}+
(d-1)\frac{1}{{0\choose 0}}{0i\choose 0j}I_4^{[d+]^2}
  \nn \\
&&-~\frac{(d-1)d}{{0\choose 0}}\frac{{0\choose j}}{{0\choose 0}}\sum_{t=1}^4 {0t\choose 0i}I_{3}^{[d+]^2,t}+
\frac{d-1}{{0\choose 0}}\sum_{t=1}^4 {0t\choose 0j}I_{3,i}^{[d+]^2,t}.
\label{DQ3simp}
\eea
The ${\nu}_{ij}  I_{4,ij}^{[d+]^3}$ from~(\ref{want1}) ($d \rightarrow d+2$)
has now been explicitly inserted since it has the same structure as the final result.
Adding all contributions, using~(\ref{I4id+2}), we finally have
\bea
{\nu}_{ij}{\nu}_{ijk} I_{4,ijk}^{[d+]^3}=&&
-\frac{{0\choose i}}{{0\choose 0}}
\frac{{0\choose j}}{{0\choose 0}}\frac{{0\choose k}}{{0\choose 0}}(d-1)d(d+1)I_4^{[d+]^3}
-\frac{{0i\choose 0j}{0\choose k}+{0i\choose 0k}{0\choose j}+{0j\choose 0k}{0\choose i}}
{{0\choose 0}^2}(d-1)I_4^{[d+]^2} \nn \\
&&+\frac{{0\choose j}}{{0\choose 0}}\frac{{0\choose k}}{{0\choose 0}}\frac{(d-1)d}{{0\choose 0}}
\sum_{t=1}^4 {0t\choose 0i}I_{3}^{[d+]^2,t}
-\frac{{0\choose k}}{{0\choose 0}}\frac{d-1}{{0\choose 0}}
\sum_{t=1}^4 {0t\choose 0j}I_{3,i}^{[d+]^2,t}\nn \\
&&+\sum_{t=1}^4 \frac{{0i\choose 0k}{0t\choose 0j}+
{0j\choose 0k}{0t\choose 0i}}{{0\choose 0}^2}I_{3}^{[d+],t}
+\frac{1}{{0\choose 0}}
\sum_{t=1,t \ne i,j}^4 {0t\choose 0k} {\nu}_{ij} I_{3,ij}^{[d+]^2,t}.
\label{fulld3}
\eea
It is obvious where the various contributions come from: Those being proportional to $-{0\choose k}/{0\choose 0}$
come from~(\ref{DQ3simp}), while all the others come from the second part of~(\ref{I4ijkd+3}).
We could indeed list these contributions separately without inserting the second part explicitly.
However, in this way the coefficient of $I_4^{[d+]^2}$ gets contributions from both terms which
combine to make the resulting coefficient explicitly symmetric in all indices.
The symmetry of the $3$-point functions is not so easily seen. Neverheless, numerically, it might
be faster not to combine these terms but just to retain what had been derived.

Also here, as discussed in~(\ref{indices}), we can replace the  tensor-
$3$-point functions:
\bea
{\nu}_{ij}I_{3,ij}^{[d+]^2,t}=&&~~~ \frac{{0t\choose it}}{{0t\choose 0t}}\frac{{0t\choose jt}}{{0t\choose 0t}}
(d-1)d~I_3^{[d+]^2,t}~~~~~~+\frac{1}{{0t\choose 0t}}{0ti\choose 0tj}I_{3}^{[d+],t} \nn \\
&&-\frac{{0t\choose jt}}{{0t\choose 0t}}\frac{d-1}{{0t\choose 0t}}\sum_{u=1}^4 {0tu\choose 0ti}I_2^{[d+],tu}
+\frac{1}{{0t\choose 0t}}\sum_{u=1}^4 {0tu\choose 0tj}I_{2,i}^{[d+],tu}.
\label{indiyes}
\eea

At this point, we observe a simple rule
of how to obtain~(\ref{fulld3}): replace in~(\ref{want1}) $d \rightarrow d+2$ and multiply with
$ -(d-1)\frac{{0\choose k}}{{0\choose 0}}$, where $(d-1)$ is to be chosen such that all factors $(d+i)$
increase by steps of 1 (see also~(\ref{want1})  ). In this manner we increase simulaneously
the dimension and the number of indices. Then one has to add the second part of~(\ref{I4ijkd+3}).

Finally, for the tensor of  rank $R=5$ we need $I_{4,ijkl}^{[d+]^4}$. Due to the above  rule
we need not again perform a complicated calculation, rather we apply the rule
as proceeding from~(\ref{want1}) to~(\ref{fulld3}), increasing simultaneously
dimension and number of indices:
we have to shift in~(\ref{fulld3}) $d \rightarrow d+2$, muliply with $-d\frac{{0\choose l}}{{0\choose 0}}$
and  add the second part of~(\ref{I4ijkld+4}).
We obtain:
\bea
&&{\nu}_{ij}{\nu}_{ijk} {\nu}_{ijkl} I_{4,ijkl}^{[d+]^4}=
\frac{{0\choose i}}{{0\choose 0}}
\frac{{0\choose j}}{{0\choose 0}}\frac{{0\choose k}}{{0\choose 0}}\frac{{0\choose l}}{{0\choose 0}}
d(d+1)(d+2)(d+3)I_4^{[d+]^4} \nn \\
&&+\frac{{0i\choose 0j}{0\choose k}{0\choose l}+{0i\choose 0k}{0\choose j}{0\choose l}+{0j\choose 0k}{0\choose i}{0\choose l}+
{0i\choose 0l}{0\choose j}{0\choose k}+{0j\choose 0l}{0\choose i}{0\choose k}+{0k\choose 0l}{0\choose i}{0\choose j}}
{{0\choose 0}^3}  d(d+1)I_4^{[d+]^3}\nn \\
&&+\frac{{0i\choose 0l}{0j\choose 0k}+{0j\choose 0l}{0i\choose 0k}+{0k\choose 0l}{0i\choose 0j}}
{{0\choose 0}^2}I_4^{[d+]^2} \nn \\
&&-\frac{{0\choose j}}{{0\choose 0}}\frac{{0\choose k}}{{0\choose 0}}\frac{{0\choose l}}{{0\choose 0}}
\frac{d(d+1)(d+2)}{{0\choose 0}}\sum_{t=1}^4 {0t\choose 0i}I_{3}^{[d+]^3,t}
+\frac{{0\choose k}}{{0\choose 0}}\frac{{0\choose l}}{{0\choose 0}}\frac{d(d+1)}{{0\choose 0}}
\sum_{t=1}^4 {0t\choose 0j}I_{3,i}^{[d+]^3,t}\nn \\
&&-\frac{d}{{0\choose 0}^3}\sum_{t=1}^4
\left[{0i\choose 0k}{0t\choose 0j}+{0j\choose 0k}{0t\choose 0i}\right]{0\choose l}I_{3}^{[d+]^2,t}\nn \\
&&-\frac{d}{{0\choose 0}^3}\sum_{t=1}^4 \left[{0j\choose 0l}{0t\choose 0i}{0\choose k}+
{0i\choose 0l}{0t\choose 0j}{0\choose k}+{0k\choose 0l}{0t\choose 0i}{0\choose j}\right]I_{3}^{[d+]^2,t} \nn \\
&&+\frac{1}{{0\choose 0}^2}\sum_{t=1}^4 \left[{0j\choose 0l}{0t\choose 0k}I_{3,i}^{[d+]^2,t}+
{0i\choose 0l}{0t\choose 0k}I_{3,j}^{[d+]^2,t}+{0k\choose 0l}{0t\choose 0j}I_{3,i}^{[d+]^2,t}\right] \nn \\
&&-\frac{{0\choose l}}{{0\choose 0}}\frac{d}{{0\choose 0}}\sum_{t=1}^4{0t\choose 0k}{\nu}_{ij}I_{3,ij}^{[d+]^3,t}
+\frac{1}{{0\choose 0}}
\sum_{t=1,t \ne i,j}^4 {0t\choose 0l} {\nu}_{ij} {\nu}_{ijk} I_{3,ijk}^{[d+]^3,t}.
\label{fulld4}
\eea
Again we, first of all, mention that the $3$-point function $I_{3,ijk}^{[d+]^3,t}$
appearing here can as well be calculated like~(\ref{indices}) and~(\ref{indiyes}),
essentially by taking over the $4$-point result, shifting $d \rightarrow d+1$:
\bea
{\nu}_{ij}{\nu}_{ijk} I_{3,ijk}^{[d+]^3,t}=&&
-\frac{{0t\choose it}}{{0t\choose 0t}}
\frac{{0t\choose jt}}{{0t\choose 0t}}\frac{{0t\choose kt}}{{0t\choose 0t}}d(d+1)(d+2)I_3^{[d+]^3,t}
-\frac{{0ti\choose 0tj}{0t\choose kt}+{0ti\choose 0tk}{0t\choose jt}+{0tj\choose 0tk}{0t\choose it}}
{{0t\choose 0t}^2}dI_3^{[d+]^2,t} \nn \\
&&+\frac{{0t\choose jt}}{{0t\choose 0t}}\frac{{0t\choose kt}}{{0t\choose 0t}}\frac{d(d+1)}{{0t\choose 0t}}
\sum_{u=1}^4 {0tu\choose 0ti}I_{2}^{[d+]^2,tu}
-\frac{{0t\choose kt}}{{0t\choose 0t}}\frac{d}{{0t\choose 0t}}
\sum_{u=1}^4 {0tu\choose 0tj}I_{2,i}^{[d+]^2,tu}\nn \\
&&+\sum_{u=1}^4 \frac{{0ti\choose 0tk}{0tu\choose 0tj}+
{0tj\choose 0tk}{0tu\choose 0ti}}{{0t\choose 0t}^2}I_{2}^{[d+],tu}
+\frac{1}{{0t\choose 0t}}
\sum_{u=1,t \ne i,j}^4 {0tu\choose 0tk} {\nu}_{ij} I_{2,ij}^{[d+]^2,tu}.
\label{fullt3}
\eea

In fact, having a closer look at our $4$-point tensor coefficients, we observe that
apart from the higher
dimensional $4$-point functions, all other terms are $3$-point tensor coefficients,
occasionally of higher tensor rank.
Eq.~\ref{fulld4} has been obtained  by an educated guess.
Indeed, a step by step derivation would have been extremely tedious if one
would have had the courage at all to try the calculation.
Of course one needs a verification by numerical checks:
among others we found for non-exceptional  Gram determinants an agreement
with LoopTools of typically more than ten decimals;
see sect.~\ref{Num} for some details.
Concerning these non-exceptional Gram determinants, we just mention that we
evaluate~(\ref{want1}),~(\ref{fulld3}) and~(\ref{fulld4}) by means of the recurrence relations
of app.~\ref{App}. For further details see Sec.~\ref{LaGra}.

\subsection{\label{Gram}%
Expansion of $I_4^{[d+]^L}$ for small Gram determinants}
The tensor coefficients in~(\ref{tensor1}) to~(\ref{tensor4}), in particular~\eqref{want1},
(\ref{fulld3}) and~(\ref{fulld4}),
have been expressed in terms of $4$-point functions in higher dimensions:
$I_4^{[d+]},I_4^{[d+]^2},I_4^{[d+]^3}$ and $I_4^{[d+]^4}$.
In our approach only these integrals can cause problems for small Gram determinants and
therefore, finding a special approach for their calculation, will finalize the problem
of calculating the $4$-poin tensor coefficients. We start from
the fact that for exactly vanishing Gram determinants~(\ref{z4d}) yields a finite value for $4$-point functions of any dimension and, taking into account higher orders, we set up an infinite
series in terms of powers of the Gram determinant:\footnote{The series is not a Taylor series {because the expansion coefficients are not the derivatives of} $I_4^{[d+]^L}$ at $()=0$.}
\bea
I_4^{[d+]^L}=\sum_{j=0}^{\infty} r^j ~~ I_{4,j}^L,
~~~L=1,\cdots,4,
\label{wish}
\eea
where the coefficients $I_{4,j}^L$ have to be determined and
\bea\label{rr}
r = \frac{{\left( \right)}}{{0\choose 0}} .
\eea
We use the observation that the recurrence relation with shift of dimension~(\ref{eq:RR2}) (see also~\ref{A401}) for the scalar higher-dimensional $4$-point
functions can be written as follows:
\bea
\label{Z4d1l}
I_4^{[d+]^l}&=&Z_4^{[d+]^l}+\frac{{\left( \right)}}{{0\choose 0}}\left[(2 l +1) -2 \eps \right]I_4^{[d+]^{(l+1)}},
~~~l=1,\cdots
\eea
where we re-wrote~(\ref{z4d}) as follows:
\bea
Z_4^{[d+]^l}=\frac{1}{{0\choose 0}}\sum_{t=1}^4 {t\choose 0} I_3^{[d+]^l,t},
~~~l=1,\cdots
\label{Z4dla}
\eea
From now on we assume $l>0$, so that the integrals are infrared finite and the leading singularity in $\eps$ is at most of the order $1/\eps$.
Integrals in generic dimension ($l=0$) will be discussed at the end of the section.
We treat the finite and divergent parts separately:
\bea
\label{Z4dl}
{I_4^{[d+]^l} }&=& {F_4^{[d+]^l} + \frac{D_4^{[d+]^l}}{\eps} + {\cal O}(\eps^2) ,}
\\
\label{Z4dlb}
{Z_4^{[d+]^l}} &=& {Z_{4F}^l + \frac{Z_{4D}^l}{\eps} + {\cal O}(\eps^2)}.
\eea
The first few iterations of~\eqref{Z4d1l} are
\bea
\label{Z4d3l}
I_4^{[d+]^l}
&=&
Z_4^{[d+]^l}+\frac{{\left( \right)}}{{0\choose 0}}
c_{l+1} 
I_4^{[d+]^{(l+1)}}
\\
&=&
Z_4^{[d+]^l}
+
\frac{{\left( \right)}}{{0\choose 0}}
c_{l+1} 
\left\{
Z_4^{[d+]^{l+1}}+\frac{{\left( \right)}}{{0\choose 0}}
c_{l+2} 
I_4^{[d+]^{(l+2)}}
\right\}
\nl
&=&
Z_4^{[d+]^l}
+
\frac{{\left( \right)}}{{0\choose 0}}
c_{l+1} 
Z_4^{[d+]^{l+1}}
+
\frac{{\left( \right)^2}}{{0\choose 0}^2}
c_{l+1} c_{l+2}
\left\{
Z_4^{[d+]^{l+2}}+\frac{{\left( \right)}}{{0\choose 0}}
c_{l+3} 
I_4^{[d+]^{(l+3)}}
\right\}
\nl
&=& \cdots ~=~
Z_4^{[d+]^l}
+
\sum_{i=1}^{\infty} \frac{{\left( \right)^i}}{{0\choose 0}^i} \left[ \prod_{j=1}^{i} c_{l+j}\right]  ~~Z_4^{[d+]^{l+i}} ,
\eea
with
\bea
c_{l+j} &=& 2(l+j)-1-2\eps .
\eea
The 4-point functions are expressed in terms of an infinite power series in $\left( \right)/{0\choose 0}$ with higher-dimensional 3-point functions in the expansion coeffients.

For the finite and divergent part {of~(\ref{Z4dl})} we get to lowest order
\bea
\label{finite}
F_4^{[d+]^l}&=&Z_{4F}^l+\frac{{\left( \right)}}{{0\choose 0}} \left[(2 l +1) F_4^{[d+]^{(l+1)}}
-2  D_4^{[d+]^{(l+1)}}  \right] ,
\\
\label{divergent}
D_4^{[d+]^l}&=&Z_{4D}^l+\frac{{\left( \right)}}{{0\choose 0}}\left[(2 l +1)D_4^{[d+]^{(l+1)}}\right].
\eea
Now the second part in ~(\ref{finite}) is proportional to ${\left( \right)}$ and
can be considered as a correction term for small Gram determinants, where also a
proper approximation for $F_4^{[d+]^{(l+1)}}$ has to be chosen.
{There are two simple choices.
One has just to set in~\eqref{finite}  $F_4^{[d+]^{(l+1)}}=Z_{4F}^{l+1}$ and neglect the unknown higher order terms, or one selects a close kinematical point with $()=0$ and uses $F_4^{[d+]^{(l+1)}}=Z_{4F}^{l+1}|_{()=0}$.
We come back to these two alternatives  later.
}

To calculate higher
order corrections, we perform now iterations.
Defining the correction term as
\bea
{\delta Z_{4F,i}^l=\frac{{\left( \right)}}{{0\choose 0}}\left[(2 l +1) Z_{4F,i}^{(l+1)}-2~ D_4^{[d+]^{(l+1)}} \right], ~~~i=0,1,2 \cdots ,}
\label{Correction}
\eea
the iterative scheme then reads:
\bea
{Z_{4F,i}^l=Z_{4F}^l+\delta Z_{4F,(i-1)}^l, ~~~ i=1,2, \cdots}
\label{iteration}
\eea
The index $i$ counts the highest power of ${\left( \right)}_4$  and the series $Z_{4F,i}^l$
is supposed to converge for growing $i$ towards $F_4^{[d+]^l}$.
As a condition of applicability of the iteration we can obviously use
\bea
\frac{\delta Z_{4F,i}^l}{Z_{4F}^l}  \sim
 \frac{{\left( \right)}}{{0\choose 0}} \times \mathrm{scale} \ll 1,
\label{CondRec}
\eea
where the scale has dimension of a squared mass.
It is worth to perform the first few steps in the iteration explicitly:
\bea\label{eq-fewsteps}
Z_{4F,i}^l&=&Z_{4F}^l+\frac{{\left( \right)}}{{0\choose 0}}\left[(2l+1)Z_{4F,(i-1)}^{(l+1)}-2 D_4^{[d+]^{(l+1)}}\right]
\\ \nn
&=&Z_{4F}^l+\frac{{\left( \right)}}{{0\choose 0}} (2l+1) \left\{ Z_{4F}^{(l+1)}+
\frac{{\left( \right)}}{{0\choose 0}} \left[(2l+3) Z_{4F,(i-1)}^{(l+2)}-2 D_4^{[d+]^{(l+2)}}\right]\right\}-2\frac{{\left( \right)}}{{0\choose 0}} D_4^{[d+]^{(l+1)}}
\\ \nn
&=& F_4^{[d+]^l} + \mathcal{O}(r^{i}) .
\eea
Performing $i$ steps in the iteration, we have
\bea
Z_{4F,{i}}^{L}&=&\sum_{j=0}^{i-1} a_j^L r^j Z_{4F}^{(L+j)}+a_i^L r^i Z_{4F,0}^{(L+i)}
            -2 \sum_{j=0}^{i-1} a_j^L r^{j+1} D_4^{[d+]^{(L+j+1)}},
\label{finitesum}
\eea
where { $r$ is given in~\eqref{rr} }
 and
\bea\label{eq-aj}
a_j^L=2^j \frac{\Gamma(L+j+\frac{1}{2})}{\Gamma(L+\frac{1}{2})}.
\eea
The $\Gamma(z)$ is the Euler Gamma function.
In~(\ref{finitesum}) we have to define yet $Z_{4F,0}^{(L+i)}$.
Strictly speaking for $Z_{4F,0}^{(L+i)}=F_4^{[d+]^{L+i}}$ we have $Z_{4F,i}^{(l)}=F_4^{[d+]^{l}}$
($i=1,2 \cdots$)
but in order to evaluate~(\ref{finitesum}) we have to choose an appropriate approximation.

Taking $i \rightarrow \infty$, however, and assuming convergence of the series, we have
\bea
{F_4^{[d+]^L}=\sum_{j=0}^{\infty} a_j^L r^j Z_{4F}^{(L+j)}-2\sum_{j=0}^{\infty} a_j^L r^{j+1} D_4^{[d+]^{(L+j+1)}},}
\label{infinitesum}
\eea
i.e. in this limit the term with $Z_{4F,0}^{(L+i)}$ drops out. The choice of an approximation
for $Z_{4F,0}^{(L+i)}$ thus can influence only the first few partial sums of~\eqref{infinitesum}.
In fact, as will be seen in Sec.~\ref{Num} in an example, the convergence is quite good for
moderate $r$ and after a few steps the result is not very much dependent on the approximant
of $Z_{4F,0}^{(L+i)}$.

It remains to deal with the last term in~(\ref{finitesum}), i.e. $D_4^{[d+]^{(L+j+1)}}$.
As in~(\ref{finitesum}),
we define a partial sum
\bea
Z_{4D,{i}}^{L}=&&\sum_{j=0}^{i} a_j^L r^j Z_{4D}^{(L+j)} ,
\label{Divfinitesum}
\eea
and get corresponding to~(\ref{infinitesum})
\bea
D_4^{[d+]^L}=\sum_{j=0}^{\infty} a_j^L r^j Z_{4D}^{(L+j)}.
\label{Dinfinitesum}
\eea
In order to avoid in~(\ref{finitesum}) higher order terms coming from $D_4^{[d+]^{(L+j+1)}}$,
i.e. higher than those contained in the finite part, we re-write
\bea
\sum_{j=0}^{i-1} a_j^L r^{j+1} D_4^{[d+]^{(L+j+1)}}= \sum_{j=0}^{i-1} a_j^L r^{j+1}Z_{4D,i-1-j}^{(L+j+1)}
+{\cal O}(r^{(i+1)}) .
\label{Div4}
\eea
E.g. for $j=i-1$ on the right hand side of~(\ref{Div4}) there contributes a term $r^{i}Z_{4D,0}^{(L+i)}$, where $i$ is the highest power of $r$ which occurs also in the finite part of~(\ref{finitesum}).
Some algebra yields
\bea\label{somealg}
\sum_{j=0}^{i-1} a_j^L r^{j+1}Z_{4D,i-1-j}^{(L+j+1)}=\frac{1}{2} \sum_{j=0}^{i} a_j^L  b_j^L r^{j}Z_{4D}^{(L+j)} ,
\eea
with coefficients
\bea\label{digamma}
b_j^L
&=& \sum_{k=L+1}^{L+j}  \frac{2}{2k-1}
\nl
&=&
\psi(L+j+\frac{1}{2})-\psi(L+\frac{1}{2}) ,
\eea
such that as final expression in terms of an infinite sum, we can write the solution of~\eqref{wish}:
\bea
I_4^{[d+]^L}=\sum_{j=0}^{\infty} a_j^L r^j \left[Z_{4}^{(L+j)}-b_j^L Z_{4D}^{(L+j)}\right],~~~~
L=0, \cdots 4.
\label{final}
\eea
The $\psi(z)$ is the logarithmic derivative of the Gamma function (digamma function).

It is interesting to note that~(\ref{final}) is also valid for $L=0$. Possible additional infrared
divergent terms are then contained for $j=0$ in $Z_{4}^{(0)}$. Comparing~(\ref{final})
with~(\ref{wish}), we see that our goal is achieved.

{Having now obtained the compact expression~(\ref{final}),
it remains to discuss how to calculate the $Z_{4}^l$.
Quite naturally one first calculates the needed
$I_3^{[d+]^l,t}$ by means of recurrence~(\ref{A301}) and sums  over $t$ according to~(\ref{Z4dla}).
This is possible since in general for $\left( \right)_4=0$, ${t\choose t}_4 \ne 0$
for $t=0,\cdots,4$.
Thus in order to evaluate~\eqref{final}, what is needed at the end, are
the $I_3^t, ~t=1, \cdots, 4$ , $I_2^{t,u}, ~t,u=1, \cdots, 4, t \ne u$
and $4$ $1$-point funcions for the kinematical point
under consideration, i.e. we need $14$ master integrals.
Applying~(\ref{A301}) in order to get the finite parts of $I_3^{[d+]^l,t}$, one also has to calculate the divergent parts of the higher-dimensional $2$- and $3$-point functions $I_2^{[d+]^{l-1},tu}$ and
$I_3^{[d+]^{l-1},t}$.
These have been discussed in app.~\ref{Bpp},
but only for low values of $l$.
They are needed for quite large $l$ ($l \sim 10$ and larger), and for larger $l$
the analytic expressions blow up considerably.
Further, for $l > 6$ the
analytic cancellation of the occurring Gram determinants is hard to perform.
So, we preferred to work numerically.
Amazingly, the situation
is different for the $2$-point functions. Without any problem we can produce with Mathematica
any higher divergences with recurrence ~(\ref{A211}), cancelling thereby the ${tu\choose tu}$ Gram
determinants.
For details see the discussion in sect.~\ref{sub4}.
To remain numerically as accurate as possible we use these analytic expressions when
calculating the divergences of the $3$-point functions numerically by recursion, starting
with $D_3^{[d+]}(t)=-\frac{1}{2}$, see~(\ref{D3t}).
As mentioned above, this has to be done for $l=1,\cdots ,l_{max}-1$.}
As a result, in a practical calculation one has the objects
\bea
{Z_{4F,{i}}^{L},~~ i=0, \dots ,l_{max}-L,}
\label{I4L}
\eea
{which are a sequence of approximations for the finite part of the integrals $I_4^{[d+]^L}$,
i.e. $F_4^{[d+]^L}$.} In app.~\ref{Num} we give an example for their numerical evaluation.

At this point we stress that formulae {\eqref{want1},~\eqref{fulld3} and~\eqref{fulld4}}
are free of indexed inegrals $I_{4,i \cdots}^{[d+]^l}$ and thus enable a new access to the
calculation of tensor $4$-point functions. Relation~\eqref{final} was already obained
in~\cite{Fleischer:2003rm}, see (36) there: for $n$-point functions ($n=1,2,3,4$)
of arbitrary dimension \emph{with generic indices} $\nu_i=1$ this series was
derived in similar manner as above and a general scheme was developed of how
to find analyic continuations to kinematical domains, where this series does not converge.
E.g., if ${0\choose 0}_4$ is small, $r$ in~\eqref{rr} is large. For this case
\cite{Fleischer:2003rm} contains the description of how to
modify the procedure of solving the recursion such that the expansion parameter is small,
i.e. how to obtain from the recursion relation a series in $\frac{1}{r}$, see eqn. (23) ibid.
Beyond that, from this series
the $d$-dimensional $n$-point functions are obtained iteratively in terms of multiple
hypergeometric series with ratios of different signed minors as arguments.
For the 4-point function, e.g.,  the generalized hypergeometric functions
$_2F_1$, Appell function $F_1$ and the Lauricella-Saran funcion $F_S$ appear, see e.g.~(98) of~\cite{Fleischer:2003rm}.
Transformation formulae of the generalized
hypergeometric functions  allow to extend their applicability to different domains
of the phase space.
In particular in our situation
we have to deal with integrals of dimension $d \ge 6 - 2 \eps$. The hypergeometric functions in
\cite{Fleischer:2003rm}  are also expressed in terms of 1-dimensional integrals and inspection
shows that for the large dimensions these are particularly well suited for numerical
evaluation (see eqns.~(78) and~(96)).
We leave this for further study.

%

{Another} attempt to perform the described series of approximations was undertaken in~\cite{Giele:2004ub}; see Eq.~(5) there.
A specific example was studied, namely forward light-by-light scattering through a massless fermion loop.
The approach was then not further followed.

\section{\label{LaGra}
Symmetrized recurrence relations
}
So far in the former sect.  we were concerned with the evaluation of the $4$-point tensor coefficients for small Gram determinants.
If, however, the Gram determinant is not small there are other ways of doing the reduction.
In fact the ''standard'' Passarino-Veltman~\cite{Passarino:1978jh} reduction is one possibility.
This is, however, not a unique procedure.

While in~\cite{Diakonidis:2009fx} a systematic application of recursion relations of
type ~(\ref{tensor5general}) was performed for all tensor $n$-point functions, here we
take a different point of view, namely to arrange the tensor coefficients in~(\ref{tensor1}) to
(\ref{tensor4}) for the $4$-point functions in such a way that a possible analytic
simplification of the tensor as a whole is achieved.

We begin with $I_4^{\mu\nu}$ defined in~\eqref{tensor2}, containing as most complicated object $I_{4,ij}^{[d+]^2}$.
This integral, represented also by~\eqref{want1}, may be reduced by~\eqref{eq:RR1},
\bea
{\nu}_{ij} I_{4,ij}^{[d+]^2}= && \frac{{0\choose i}{0\choose j}}{{\left( \right)}^2}I_{4}+
\frac{{i\choose j}}{{\left( \right)}}I_{4}^{[d+]}+R_{3,ij}^{[d+]^2},
\label{wantx}
\eea
with
\bea
R_{3,ij}^{[d+]^2}&&=-\frac{{0\choose j}}{{\left( \right)}}\sum_{t=1}^4
\frac{{t\choose i}}{{\left( \right)}}I_{3}^{t}+
\sum_{t=1}^4\frac{{t\choose j}}{{\left( \right)}}I_{3,i}^{[d+],t}\nn \\
&&=-\sum_{t=1}^4\frac{{0\choose i} {t\choose j}
+{0\choose j}{t\choose i} -{t\choose i}{t\choose j}
\frac{{t\choose 0}}{{t\choose t}}}{{\left( \right)}^2}I_3^{t}
+\frac{1}{{\left( \right)}}\sum_{t,u=1}^4\frac{{t\choose j}}{{t\choose t}}
{ut\choose it}I_2^{tu}.
\label{wanty}
\eea
The first observation of interest here is the symmetry in the indices $i,j$. Only the last
term is not obviously symmetric. As was mentioned earlier, the symmetry is
in general seen only after summation over $s,t$, which we can exemplify here.
We use the relation
\bea\label{eq-wa-4}
{t\choose j}{ut\choose it}={t\choose i}{ut\choose jt}+{t\choose t}{tu\choose ji}.
\eea
Inserting the left hand side into~(\ref{wanty}), the first term
on the right hand side has just exchanged indices
$i,j$. In the second contribution ${t\choose t}$ cancels due to which the sum over $s,t$
vanishes since ${tu\choose ji}$ is antisymmetric in $t,u$.

In~(\ref{wantx}) there remains as higher-dimensional integral $I_{4}^{[d+]}$, which is evaluated
according to~(\ref{A401}). Often it is as well used as ``master integral'' since it is UV and
IR finite. Having a look at~(\ref{tensor2}) we see that the second amplitude of the rank $R=2$
tensor is also just $I_{4}^{[d+]}$. This allows the following way of writing for this tensor.
Similarly as in~\cite{Diakonidis:2009fx} we introduce \footnote{$G^{\mu \nu}$ here differs
from the definition (24) in~\cite{Diakonidis:2009fx}  by a factor of 2.
}
\bea\label{eq-wa-5}
G^{\mu \nu}&=&g^{\mu \nu}-2 \sum_{i,j=1}^4 q_i^{\mu} q_j^{\nu}
\frac{{i\choose j}}{{\left( \right)}}
\nl
&=&\frac{8 v^{\mu} v^{\nu}}{{\left( \right)}} ,
\label{Gg}
\eea
with
\bea\label{eq-wa-6}
v^{\mu}={\eps}^{\mu \lambda \rho \sigma} (q_1-q_4)_{\lambda}(q_2-q_4)_{\rho}(q_3-q_4)_{\sigma}
,
\eea
and $v^2=\frac{1}{8}{\left( \right)}$. This allows to drop $I_{4}^{[d+]}$ in~(\ref{wantx})
and to replace $g^{\mu \nu}$ in~(\ref{tensor2}) by $G^{\mu \nu}$:
\begin{eqnarray}\label{eq-i4alt}
I_4^{\mu\nu}
&=&
\sum_{i,j=1}^{4} \, q_i^{\mu}\, q_j^{\nu} \,
\left[
\frac{{0\choose i}{0\choose j}}{{\left( \right)}^2}I_{4}
-\sum_{t=1}^4\frac{{0\choose i} {t\choose j}
+{0\choose j}{t\choose i} -{t\choose i}{t\choose j}
\frac{{t\choose 0}}{{t\choose t}}}{{\left( \right)}^2}I_3^{t}
+\frac{1}{{\left( \right)}}\sum_{t,u=1}^4\frac{{t\choose j}}{{t\choose t}}
{ut\choose it}I_2^{tu}
\right]
\nl
&&-~ \frac{1}{2}
   \, G^{\mu \nu}  \, I_{4}^{[d+]} .
\end{eqnarray}
This representation  may become
advantageous in the analytic evaluation of diagrams since $G^{\mu \nu}$, contracted with
a (proper difference of) chord(s), vanishes. Remember that any external momentum may be
written as a sum of chords. For the tensor of rank $R=2$ this corresponds to (20) of
\cite{Diakonidis:2009fx}.

We will derive here the corresponding relations for the higher tensors as well.

Proceeding to $I_4^{\mu\nu\lambda}$,~(\ref{tensor3}), we ,first of all, need in the second term on the right hand side
$I_{4,i}^{[d+]^2}$, (\eqref{A511} with $l=2$):
\bea
I_{4,i}^{[d+]^2}=-\frac{{0\choose i}}{{\left( \right)}}I_{4}^{[d+]}+
\sum_{t=1}^4\frac{{t\choose i}}{{\left( \right)}}I_{3}^{[d+],t},
\label{I447}
\eea
where again $I_{4}^{[d+]}$ appears and $I_{3}^{[d+],t}$ is given in~(\ref{A301}).
From the recursions of subsect.~\ref{sub2} we obtain
\bea
&&{\nu}_{ij}{\nu}_{ijk} I_{4,ijk}^{[d+]^3}=-\frac{{0\choose i}{0\choose j}{0\choose k}}
{{\left( \right)}^3}I_{4}+\left\{
\frac{{i\choose j}}{{\left( \right)}} I_{4,k}^{[d+]^2}+
(j \leftrightarrow k) + (i \leftrightarrow k) \right\}+R_{3,ijk}^{[d+]^3},
\label{fullx}
\eea
where  we have introduced an abbreviation for the remaining
$3$- and $2$-point functions:
\bea
R_{3,ijk}^{[d+]^3} =&&
-\frac{{0\choose k}}{{\left( \right)}}R_{3,ij}^{[d+]^2}-
\sum_{t=1}^4\frac{{t\choose k}}{{\left( \right)}}\frac{{0t\choose jt}}{{t\choose t}}I_{3,i}^{[d+],t}-
\sum_{t=1}^4 \frac{{t\choose i}{t\choose j}{t\choose k}}{{\left( \right)}^2{t\choose t}} I_{3}^{[d+],t}
+\sum_{t,u=1}^4\frac{{t\choose k}{ut\choose jt}}{{\left( \right)}{t\choose t}}
 I_{2,i}^{[d+],tu} \nn \\
=&&\sum_{t=1}^4\frac{{0\choose i}{0\choose j}{t\choose k}+{0\choose i}{t\choose j}{0\choose k}+
{t\choose i}{0\choose j}{0\choose k}} {{\left( \right)}^3}I_{3}^{t}
\nn \\
-&&\sum_{t=1}^4\frac{{0\choose i}{t\choose j}{t\choose k}+{t\choose i}{0\choose j}{t\choose k}+
{t\choose i}{t\choose j}{0\choose k}-{t\choose i}{t\choose j}{t\choose k}\frac{{t\choose 0}}{{t\choose t}}}{{\left( \right)}^3} \frac{{t\choose 0}}{{t\choose t}}I_{3}^{t}
-\sum_{t=1}^4 \frac{{t\choose i}{t\choose j}{t\choose k}}{{\left( \right)}^2{t\choose t}} I_{3}^{[d+],t}
\nn \\
-&&\sum_{t,u=1}^4\frac{\left[{0\choose j}{t\choose k}+{0\choose k}{t\choose j}-{t\choose j}{t\choose k}
\frac{{t\choose 0}}{{t\choose t}}\right]}{{\left( \right)}^2}\frac{{ut\choose it}}{{t\choose t}}I_{2}^{tu}
+\sum_{t,u=1}^4\frac{{t\choose k}{ut\choose jt}}{{\left( \right)}{t\choose t}}
 I_{2,i}^{[d+],tu} .
\label{fullxr}
\eea
In~(\ref{fullx}), the $\frac{{i\choose j}}{{\left( \right)}}I_{4}^{[d+]}$ of~(\ref{wantx}) together
with part of the last contribution in~(\ref{A322}) has now been absorbed in $\frac{{i\choose j}}{{\left( \right)}} I_{4,k}^{[d+]^2}$: in~(\ref{A533}) only two terms of this type appear explicitely,
i.e. this form has no obvious symmetry. Further, in~(\ref{fullx}) and~(\ref{fullxr})
only the  $4$- and $3$- point functions are explicitly symmetric in the indices $i,j,k$. In
order to demonstrate symmetry also for the $2$-point functions one would have to reduce
also $I_{2,i}^{[d+],tu}$, which is given in app.~\ref{App}.

We may now combine the results and simplify $I_4^{\mu\nu\lambda}$ correspondingly:
The $I_{4,k}^{[d+]^2}$  in~(\ref{fullx})
can be combined with the second part of~(\ref{tensor3}), i.e. it can be dropped in~(\ref{fullx}),
and in~(\ref{tensor3}) the $g^{\mu \nu}$ must then be replaced by $G^{\mu \nu}$.
(\ref{fullx}) has also been identified with a complete reduction of~(\ref{fulld3}).

For  $I_4^{\mu\nu\lambda\rho}$~(\ref{tensor4}), we  start from~(\ref{A555}):
\bea
{n}_{ijkl} I_{4,ijkl}^{[d+]^4}
&=&
\frac{{0\choose i}{0\choose j}{0\choose k}{0\choose l}}
{{\left( \right)}^4}I_{4}-\frac{{0\choose l}}{{\left( \right)}}\left\{
\frac{{i\choose j}}{{\left( \right)}} I_{4,k}^{[d+]^2}+
(j \leftrightarrow k) + (i \leftrightarrow k) +R_{3,ijk}^{[d+]^3}\right\}
\nn \\
&&
+\left\{
\frac{{i\choose l}}{{\left( \right)}}\frac{{j\choose k}}{{\left( \right)}}+
\frac{{j\choose l}}{{\left( \right)}}\frac{{i\choose k}}{{\left( \right)}}+
\frac{{k\choose l}}{{\left( \right)}}\frac{{i\choose j}}{{\left( \right)}} \right\}I_4^{[d+]^2}
 \\
&&
-\left\{
\frac{{i\choose l}}{{\left( \right)}}\frac{{0\choose k}}{{\left( \right)}}I_{4,j}^{[d+]^2}+
\frac{{j\choose l}}{{\left( \right)}}\frac{{0\choose k}}{{\left( \right)}}I_{4,i}^{[d+]^2}+
\frac{{k\choose l}}{{\left( \right)}}\frac{{0\choose j}}{{\left( \right)}}I_{4,i}^{[d+]^2}\right\}
\nn \\\nn
&&
+\frac{{i\choose l}}{{\left( \right)}}\sum_{t=1}^4 \frac{{t\choose k}}{{\left( \right)}}I_{3,j}^{[d+]^2,t}
+\frac{{j\choose l}}{{\left( \right)}}\sum_{t=1}^4 \frac{{t\choose k}}{{\left( \right)}}I_{3,i}^{[d+]^2,t}
+\frac{{k\choose l}}{{\left( \right)}}\sum_{t=1}^4 \frac{{t\choose j}}{{\left( \right)}}I_{3,i}^{[d+]^2,t}+
\sum_{t=1}^4 \frac{{t\choose l}}{{\left( \right)}}n_{ijk}I_{3,ijk}^{[d+]^3,t}.
\label{fully}
\eea
As can be  seen, the ${n}_{ijkl} I_{4,ijkl}^{[d+]^4}$ can be expressed in a form which mainly contains
terms which also occur in
\bea
{\nu}_{ij} I_{4,ij}^{[d+]^3}=-\frac{{0\choose j}}{{\left( \right)}}I_{4,i}^{[d+]^2}+
\frac{{i\choose j}}{{\left( \right)}}I_4^{[d+]^2}+ \sum_{t=1}^4 \frac{{t\choose j}}{{\left( \right)}}I_{3,i}^{[d+]^2,t},
\label{I451}
\eea
see~(\ref{A522}).
The  $n_{ijk}I_{3,ijk}^{[d+]^3,t}$
can be written similarly like~(\ref{fullx}):
\bea
&&{\nu}_{ij}{\nu}_{ijk} I_{3,ijk}^{[d+]^3,t}=-\frac{{0t\choose it}{0t\choose jt}{0t\choose kt}}
{{t\choose t}^3}I_{3}^t+\left\{ \left[
\frac{{i\choose j}}{{\left( \right)}}- \frac{{t\choose i}{t\choose j}}{{\left( \right)}{t\choose t}}\right]I_{3,k}^{[d+]^2,t}+
(j \leftrightarrow k) + (i \leftrightarrow k) \right\}+R_{2,ijk}^{[d+]^3,t}, \nn \\
\label{fullx3}
\eea
where $R_{2,ijk}^{[d+]^3,t}$ collects the remaining $2$- and $1$-point functions and is obtained
from~(\ref{fullxr}) as follows: all $3$-point functions are replaced by $2$-point functions
($3 \rightarrow 2$) and the $2$-point functions are replaced by $1$-point functions ($2 \rightarrow 1$).
All summation indices $u$ must be replaced by $v$ and  summation indices $t$ must be replaced by $u$.
Finally in all determinants and integrals columns, lines and propagators $t$ must be scratched - like
in~(\ref{fullx3}), i.e.
\bea\label{fullx23}
R_{2,ijk}^{[d+]^3,t} =
-\frac{{0t\choose kt}}{{t\choose t}}R_{2,ij}^{[d+]^2,t}-
\sum_{u=1}^4\frac{{ut\choose kt}}{{t\choose t}}\frac{{0tu\choose jtu}}{{tu\choose tu}}I_{2,i}^{[d+],tu}-
\sum_{u=1}^4 \frac{{ut\choose it}{ut\choose jt}{ut\choose kt}}{{t\choose t}^2{tu\choose tu}} I_{2}^{[d+],tu}
+\sum_{u,v=1}^4\frac{{ut\choose kt}{vtu\choose jtu}}{{t\choose t}{tu\choose tu}}
 I_{1,i}^{[d+],tuv} , \nn \\
\eea
and
\bea\label{fullx22}
R_{2,ij}^{[d+]^2,t}=-\frac{{0t\choose jt}}{{t\choose t}}\sum_{u=1}^4
\frac{{ut\choose it}}{{t\choose t}}I_{2}^{tu}+
\sum_{u=1}^4\frac{{ut\choose jt}}{{t\choose t}}I_{2,i}^{[d+],tu}.
\eea
To indicate how the various tensor components may be combined to simplify the result,
we only count here the number of contributions of a certain type. The second term
in~\eqref{tensor4} contains, according to~\eqref{G2V2}, $6$ terms of type~\eqref{I451}.
Each type of the $3$ terms in~\eqref{I451} is also contained in the other tensor
components in~\eqref{tensor4}:
\begin{itemize}
 \item $\frac{{0\choose j}}{{\left( \right)}}I_{4,i}^{[d+]^2}$ occurs $6$ \rm{times} in~\eqref{fully},
 \item $\frac{{i\choose j}}{{\left( \right)}}I_4^{[d+]^2}$ occurs $3$ \rm{times} in~\eqref{fully}
       and occurs $3$ \rm{times} in the $3^{rd}$ term of~\eqref{tensor4},
 \item $\sum_{t=1}^4 \frac{{t\choose j}}{{\left( \right)}}I_{3,i}^{[d+]^2,t}$ occurs $3$ \rm{times} in~\eqref{fully} and occurs $3$ \rm{times} in~\eqref{fullx3}.
\end{itemize}
Thus, rewriting~\eqref{Gg} as
\bea\label{eq23}
\sum_{i,j=1}^4 q_i^{\mu} q_j^{\nu}
\frac{{i\choose j}}{{\left( \right)}}=\frac{1}{2} \left( g^{\mu \nu}-G^{\mu \nu} \right),
\eea
one can convince oneself that $g^{\mu \nu}$ cancels and it remains $G^{\mu \nu}$, which after
contraction with a chord drops out.

As we have seen, the above treatment of tensor coefficients requires at least
one step of iteration of the recursion relation, because of which it is applicable only for
non-vanishing Gram determinants. For these, however, it is very useful when we consider
$4$-point functions obtained by scratching one line of a $5$-point function. In this
case it is indeed possible to perform a cancellation of terms, which above still have
the factor $G^{\mu \nu}$. Also for this case, which we dealt with in sect.~\ref{LargeGrams},
the above presentation of the $4$-point tensor coefficients can be applied
- only scratching of one propagator has to be taken into account.

Beyond that the present approach results in reductions which make the
symmetry in the indices $i,j,k,l$ more transparent and as a consequence yield
certain blocks which can be calculated separately and combined to yield the complete
tensor coefficients.

\section{\label{Simplify}Analytic simplifications for contractions of tensors with chords}
Before a numerical program for calculations is set up, it turns out to be
advantageous to simplify Feynman diagrams analytically. A standard example
is the following.
If in the numerator of a Feynman integral a scalar
product $q_i \cdot k$ of a chord and an integration momentum occurs,
this {product} is usually expressed in terms of the difference of two
scalar propagators which can be cancelled against propagators in the
denominator. Already in~\cite{Diakonidis:2009fx} an alternative was
indicated, making use of the fact that the contraction of a vector
of the type~(\ref{Qs}) with a chord yields a simple expression:
\bea
q_i \cdot Q_0 =\sum_{j=1}^{n-1} q_i q_j \frac{{0\choose j}_n}{{\left(\right)}_n}=-\frac{1}{2}\left(
 Y_{in}-Y_{nn} \right), ~~~i=1, \dots , n-1,
\label{Scalar1}
\eea
and
\bea
q_i \cdot Q_s =\sum_{j=1}^{n-1} q_i q_j \frac{{s\choose j}_n}{{\left( \right)}_n}=\frac{1}{2}\left(
{\delta}_{is}-{\delta}_{ns}\right), ~~~i=1, \dots , n-1, ~~~s=1, \dots n.
\label{Scalar2}
\eea
In~(\ref{Scalar1}) and~(\ref{Scalar2}) $q_n=0$ is assumed since only in this case
\bea\label{Scalar2a}
q_i \cdot q_j =\frac{1}{2} \left[Y_{ij}-Y_{in}-Y_{nj}+Y_{nn} \right],
\eea
which is needed for their derivations.
Thus, if the reduction relation~(\ref{tensor5general}) for $I_5^{\mu_1  \dots \mu_{R-1} \mu}$ is contracted
with a $q_{i,\mu}$, we can advantageously apply~(\ref{Scalar1}) and~(\ref{Scalar2}) with $n=5$.

In case one considers a process with $5$ external legs, one can choose from the very beginning
$q_5=0$ in the tensor integrals. If, however, the $5$-point tensor is obtained by
reducing a $6$-point tensor,
cases with $q_5 \ne 0$ will occur.
In order to be able to
apply~(\ref{Scalar1}) and~(\ref{Scalar2}), it is recommended to perform a shift of the
integration momentum like $k \rightarrow k+ q_5$, i.e. $q_i \rightarrow q_i -q_5$. Such
a shift is not a problem at all, nevertheless it is interesting to see how this
shift can be implemented in the formalism.
The scalar integrals and the signed minors
are invariant under the shift.
We exemplify this for $I_n^{\mu\nu}$, writing
\bea
-\sum_{i,j=1}^n (q_i-q_n)^{\mu}(q_j-q_n)^{\nu} {\nu}_{ij}I_{n,ij}^{[d+]^2}=
&&-\sum_{i,j=1}^n q_i^{\mu}q_j^{\nu} {\nu}_{ij}I_{n,ij}^{[d+]^2}+
q_n^{\mu}\sum_{i,j=1}^n q_j^{\nu} {\nu}_{ij}I_{n,ij}^{[d+]^2}
\nn \\
&&
+~
q_n^{\nu}\sum_{i,i=1}^n q_i^{\nu} {\nu}_{ij}I_{n,ij}^{[d+]^2}
-q_n^{\mu}q_n^{\nu} \sum_{i,j=1}^n {\nu}_{ij}I_{n,ij}^{[d+]^2} ,
\label{Inijd21}\eea
with
\bea
{\nu}_{ij} I_{n,ij}^{[d+]^2}&=&-\frac{{0\choose j}_n}{\left(  \right)_n} I_{n,i}^{[d+]} +
 \sum_{t=1,t \ne i}^{n} \frac{{t\choose j}_n}{\left(  \right)_n} I_{n-1,i}^{[d+],t} +
 \frac{{i\choose j}_n}{\left(  \right)_n} I_{n}^{[d+]}.
\label{Inijd2}
\eea
The sums
\bea\label{eq-wa-83r}
\sum_{j}^n {\nu}_{ij}I_{n,ij}^{[d+]^2}&=&-I_{n,i}^{[d+]},
\\\label{eq-wa-83q}
\sum_{i=1}^n {\nu}_{ij}I_{n,ij}^{[d+]^2}&=&-I_{n,j}^{[d+]} ,
\\\label{eq-wa-83s}
\sum_{i,j=1}^n {\nu}_{ij}I_{n,ij}^{[d+]^2} &=&I_n
\eea
can be obtained by making use of
\bea\label{eq-wa-8}
\sum_{i=1}^n {0\choose i}_n &=& {\left( \right)}_n ,
\\\label{eq-wa-9}
\sum_{i=1}^n {s\choose i}_n &=& 0.
\eea
Thus one has
\bea\label{eq-wa-90}
\sum_{i,j=1}^n q_i^{\mu}q_j^{\nu} {\nu}_{ij}I_{n,ij}^{[d+]^2}=\sum_{i,j=1}^{n-1} {q^{'}}_i^{\mu}{q^{'}}_i^{\nu} {\nu}_{ij}I_{n,ij}^{[d+]^2}-q_n^{\mu}\sum_{i=1}^n q_i^{\nu}I_{n,i}^{[d+]}-
q_n^{\nu}\sum_{i=1}^n q_i^{\mu} I_{n,i}^{[d+]}-q_n^{\mu}q_n^{\nu}I_n ,
\eea
using the abbreviation ${q^{'}}_i^{\mu}=q_i^{\mu}-q_n^{\mu}$,
and with
\bea\label{eq-wa-80}
-\sum_{i=1}^n q_i^{\mu}I_{n,i}^{[d+]}=-\sum_{i=1}^{n-1}{q^{'}}_i^{\mu}I_{n,i}^{[d+]}+q_n^{\mu} I_n,
\eea
derived in the same manner, with~(\ref{tensor1}) and~(\ref{tensor2})
the standard result of shifting the integration momentum is obtained.
The point is that the extra contributions obtained by the shift contain only integrals which
were needed already in the unshifted integral so that the shift does not require the calculation
of any new integral, see~(\ref{Inijd2}).

Assume now again that we are dealing with the $5$-point tensor and have $q_5=0$. The above trick
to avoid an increase of the tensor rank can be applied to~(\ref{starter}) as well:
contracting with $q_{i,\mu} $ the first term yields the contribution $q_i \cdot Q_0$
given in~(\ref{Scalar1}), for the second term we have
to find a formula for the scalar product $q_i \cdot {\bar{Q}}_s^{0}$. Indeed,
\bea\label{eq-wa-81}
q_i {\bar{Q}}_s^{0}&=&\sum_{j=1}^{4} q_i q_j {0s\choose 0j}_5
\nl
&=& \frac{1}{2}
\left[{0\choose 0}_5 \left({\delta_{is}}-{\delta_{5s}}\right)+{s\choose 0}_5 \left(Y_{i5}-Y_{55}\right)\right].
\eea
For the first term in~(\ref{starter}) also another possibility exists, provided a
contraction with a further vector is available. In such a case the first term
on the right hand side of
(\ref{cancel}), which shows up explicitly in the tensor components
$E_{ij \dots}$, yields a double-sum like
\bea\label{eq-wa-82}
\sum_{i,j=1}^{4} (q_a \cdot q_i) (q_b \cdot q_j) {0i\choose sj}_5=\frac{1}{2}q_a \cdot q_b{s\choose 0}_5
+\frac{1}{4}{\left(\right)}_5\left(Y_{b5}-Y_{55} \right)\left({\delta}_{as}-{\delta}_{5s}\right).
\eea
Further sums are obtained if the $4$-point tensors are contracted. These are all of the
type ${is\choose js}_5$, i.e. with line $s$ scratched. We just list a few of them:
\bea\label{eq-wa-83}
\sum_{j=1}^{4}q_a \cdot q_j  {0s\choose js}_5 &&=-\frac{1}{2} \left[
{s\choose 0}_5\left({\delta}_{as}-{\delta}_{5s}\right)+
{s\choose s}_5\left(Y_{a5}-Y_{55} \right)\right], \\\label{eq-wa-83a}
\sum_{i,j=1}^{4} q_i \cdot q_j {0s\choose is}_5{0s\choose js}_5 &&=\frac{1}{2} {s\choose s}_5 \left[
{0s\choose 0s}_5+Y_{55}{s\choose s}_5+2 {s\choose 0}_5{\delta}_{5s}\right],
 \\\label{eq-wa-83v}
\sum_{i,j=1}^{4} q_i \cdot q_j {is\choose js}_5 &&=\frac{3}{2} {s\choose s}_5,  \\\label{eq-wa-83b}
\sum_{i,j=1}^{4} (q_a \cdot q_i) (q_b \cdot q_j) {is\choose js}_5 &&=\frac{1}{2}q_a \cdot q_b{s\choose s}_5
-\frac{1}{4}{\left(\right)}_5\left({\delta}_{ab}{\delta}_{as}+{\delta}_{5s}\right),
\eea
and for $3$-point functions
\bea\label{eq-wa-84}
\sum_{j=1}^{4}q_a \cdot q_j  {ts\choose js}_5
&&=\frac{1}{2} \left\{{s\choose s}_5\left[
(1-{\delta}_{as}){\delta}_{at}-(1-{\delta}_{5s}){\delta}_{5t}\right]\right.
\nn \\
&&\left. ~~~~~~-{t\choose a}_5
(1-{\delta}_{at}){\delta}_{as}+{t\choose 5}_5(1-{\delta}_{5t}){\delta}_{5s} \right\} , \\\label{eq-wa-851}
\sum_{i,j=1}^{4} q_i \cdot q_j{ts\choose is}_5 {ts\choose js}_5 &&=\frac{1}{2} {s\choose s}_5
{st\choose st}_5 ,
\\\label{eq-wa-852}
\sum_{i,j=1}^{4} q_i \cdot q_j{ts\choose is}_5 {0s\choose js}_5 &&=\frac{1}{2} {s\choose s}_5
\left\{{0s\choose ts}_5-{s\choose s}_5(1-{\delta}_{5s}){\delta}_{5t}+{t\choose 5}_5 (1-{\delta}_{5t}){\delta}_{5s}\right\} ,
\\\label{eq-wa-854}
\sum_{i,j=1}^{4} q_i \cdot q_j{ist\choose jst}_5 &&={st\choose st}_5 .
\eea
Even a quadrupel sum appears:
\bea\label{eq-wa-85}
\sum_{i,j,k,l=1}^{4}(q_i \cdot q_j)(q_k \cdot q_l){0i\choose sl}_5{ts\choose js}_5{ts\choose ks}_5
=\frac{1}{4} {s\choose 0}_5{s\choose s}_5{st\choose st}_5.
\eea
In fact, there are many more such sums. Our conclusion here is that in every scalar, which is obtained by
contraction with chords the appearing sums can be evaluated analytically in order to yield
compact expressions. This is due to the fact that the indices of the chords
$i,j, c\dots $ are  carried by signed minors while the integrals don't necessarily
carry  indices anymore.
\section{\label{conclude}Conclusions}
We have developed a new approach to reduce tensorial one-loop $n$-point Feynman integrals  based
on an algebraic method elaborated in earlier papers.
The approach is worked out
up to 6-point tensors with rank $R \leq 6$
and a rule is found how to extend the method to higher ranks.
The first step was to reduce $5$-point tensors up to rank $5$ to $4$-point tensor
coefficients given in terms of higher-dimensional, indexed $4$-point functions.
The latter are expressed in terms of higher-dimensional integrals $I_4^{[d+]^L}$, $L=1, \cdots ,4$,
plus $3$-point tensor coefficients in~\eqref{I4id+2},~\eqref{want1},~\eqref{fulld3} and~\eqref{fulld4}.
So far no Gram determinants $\left( \right)_4$ occur.
Inverse powers of $\left( \right)_4$ do occur if the integrals $I_4^{[d+]^L}$ are reduced to
standard $A_0,B_0,C_0,D_0$ functions in generic dimension. For small $\left( \right)_4$
this is avoided by using the expansion~\eqref{final} in positive powers of $\left( \right)_4$.
Application of Pad\'{e}-approximants based on the $\eps$-algorithm
to this expansion allows to calculate the
 $I_4^{[d+]^L}$ in a simple manner to such a precision
that the complete phase space is covered with high numerical precision.
{A numerical opensource code of the formulae derived in this article is under development.}

As a matter of fact,~\eqref{final} is a special case of the general
method developed in~\cite{Fleischer:2003rm}. Apart from this special series expansion,
the integrals $I_4^{[d+]^L}$ are expressed in~\cite{Fleischer:2003rm}
in terms of multiple hypergeometric functions $_2F_1$, Appell function $F_1$ and Lauricella-Saran function
$F_S$.
In this context, it is a crucial property of eqns.~\eqref{I4id+2},~\eqref{want1},~\eqref{fulld3} and~\eqref{fulld4} to be free of integrals $I_{4,i\cdots}^{[d+]^L}$ with indices larger than one.
Using the notion of special functions  allows a variety of options to adjust to various kinematical situations, and it might be interesting to explore their potential for a further improvement of numerical programs.


\section*{Acknowledgements}
We would like thank  F. Campanario, Th. Diakonidis, B. Tausk and V. Yundin for useful discussions and  V. Yundin for a careful reading of the manuscript.
{We are grateful to A. Denner for some clarifying discussion.}
J.F. thanks DESY for kind hospitality.
Work supported in part by Sonderforschungsbereich/Trans\-re\-gio SFB/TRR 9 of DFG
``Com\-pu\-ter\-ge\-st\"utz\-te Theoretische Teil\-chen\-phy\-sik"
and by the European Community's Marie-Curie Research Trai\-ning Network
MRTN-CT-2006-035505
``HEPTOOLS''.

\appendix

\section{\label{App}Recursion relations for the reduction of higher-dimensional 4-point functions}
In this app. we quote explicitly the needed recurrence relations for the reduction
of the 4-point functions. In~\cite{Fleischer:1999hq} three recurrence relations are
given: one reducing simultaneously index and dimension, one reducing only the dimension
and another one reducing only the index. In the present work only the first two relations
are used, i.e.~(\ref{eq:RR1}) and~(\ref{eq:RR2}).
Special cases of~(\ref{eq:RR2}) are~(\ref{A401}),~(\ref{A301}) and
(\ref{A211}), i.e. these are the ones used to reduce the dimension of the scalar
4-, 3- and 2-point functions. All the other ones are special cases of~(\ref{eq:RR1}) and
reduce the tensor indices. Reducing $4$-point functions, $3$- and $2$-point functions are
generated, for which we also have to give the corresponding recurrence relations.

\subsection{\label{sub2}Reduction of 4-point integrals}
To learn about practical applications of the recurrence relations it is useful to
investigate the most complicated one under consideration in some detail.
Recall the definition given after~(\ref{eq:Inij}), \\$[d+]^l=4+2 l -2 \eps=d+2 l$.
With~\eqref{eq:RR1},
\begin{eqnarray}
\label{A544}
{\nu}_{ijkl} I_{4,ijkl}^{[d+]^4}&=&-\frac{{0\choose l}_4}{\left(  \right)_4} I_{4,ijk}^{[d+]^3} +
 \sum_{t=1,t \ne i,j,k}^{4} \frac{{t\choose l}_4}{\left(  \right)_4} I_{3,ijk}^{[d+]^3,t} +
 \frac{{i\choose l}_4}{\left(  \right)_4} I_{4,jk}^{[d+]^3}+
 \frac{{j\choose l}_4}{\left(  \right)_4} I_{4,ik}^{[d+]^3}+
 \frac{{k\choose l}_4}{\left(  \right)_4} I_{4,ij}^{[d+]^3}.
\end{eqnarray}
The integral contributes to the rank $R=5$ tensor coefficient $E_{ijklm}$, see~\eqref{I4ijkld+4}.
Strictly speaking~(\ref{A544}) is in this form valid only if all indices $i,j,k$ are different.
In case that some or all indices are equal there is no repetition of the same terms on the right
hand side
and the question is how to take into account this property in a general manner. Let us
recall that in fact in the original integral~(\ref{tensor5}) we have to deal with
${n}_{ijkl} I_{4,ijkl}^{[d+]^4}$, where ${n}_{ijkl}={\nu}_{ij}{\nu}_{ijk}{\nu}_{ijkl}$.
Thus it is recommended to multiply~(\ref{A544}) with ${\nu}_{ij}{\nu}_{ijk}$.
Let us further introduce
for the sum of the last three terms in ~(\ref{A544}) the notation $[ijk]^{(l)}$ (with repetition)
and $[ijk]^{(l)}_{red}$ (without repetition). Then we have the following useful relation:
\begin{eqnarray}
{\nu}_{ij}{\nu}_{ijk} [ijk]^{(l)}_{red}&&=[ijk]^{(l)}+
{\delta}_{jk}\frac{{i\choose l}_4}{\left(  \right)_4} I_{4,jk}^{[d+]^3}+
{\delta}_{ik}\frac{{j\choose l}_4}{\left(  \right)_4} I_{4,ik}^{[d+]^3}+
{\delta}_{ij}\frac{{k\choose l}_4}{\left(  \right)_4} I_{4,ij}^{[d+]^3} \nn\\
&&={\nu}_{jk}\frac{{i\choose l}_4}{\left(  \right)_4} I_{4,jk}^{[d+]^3}+
{\nu}_{ik}\frac{{j\choose l}_4}{\left(  \right)_4} I_{4,ik}^{[d+]^3}+
{\nu}_{ij}\frac{{k\choose l}_4}{\left(  \right)_4} I_{4,ij}^{[d+]^3}.
\label{observation}
\end{eqnarray}
Thus, with ${n}_{ijk}={\nu}_{ij}{\nu}_{ijk}$~(\ref{A544}) reads
\begin{eqnarray}
\label{A555}
{n}_{ijkl} I_{4,ijkl}^{[d+]^4}=&&-\frac{{0\choose l}_4}{\left(  \right)_4} {n}_{ijk}I_{4,ijk}^{[d+]^3}\nn\\
&& +
 \sum_{t=1,t \ne i,j,k}^{4} \frac{{t\choose l}_4}{\left(  \right)_4} {n}_{ijk}I_{3,ijk}^{[d+]^3,t} +
 \frac{{i\choose l}_4}{\left(  \right)_4}{\nu}_{jk} I_{4,jk}^{[d+]^3}+
 \frac{{j\choose l}_4}{\left(  \right)_4}{\nu}_{ik} I_{4,ik}^{[d+]^3}+
 \frac{{k\choose l}_4}{\left(  \right)_4}{\nu}_{ij} I_{4,ij}^{[d+]^3}.
\end{eqnarray}
Correspondingly we have
\begin{eqnarray}
\label{A533}
{n}_{ijk} I_{4,ijk}^{[d+]^l}&=&-\frac{{0\choose k}_4}{\left(  \right)_4}{\nu}_{ij} I_{4,ij}^{[d+]^{l-1}} +
 \sum_{t=1,t \ne i,j}^{4} \frac{{t\choose k}_4}{\left(  \right)_4} {\nu}_{ij}I_{3,ij}^{[d+]^{l-1},t} +
 \frac{{i\choose k}_4}{\left(  \right)_4} I_{4,j}^{[d+]^{l-1}}+
 \frac{{j\choose k}_4}{\left(  \right)_4} I_{4,i}^{[d+]^{l-1}}, \\
\label{A522}
{\nu}_{ij} I_{4,ij}^{[d+]^l}&=&-\frac{{0\choose j}_4}{\left(  \right)_4} I_{4,i}^{[d+]^{l-1}} +
 \sum_{t=1,t \ne i}^{4} \frac{{t\choose j}_4}{\left(  \right)_4} I_{3,i}^{[d+]^{l-1},t} +
 \frac{{i\choose j}_4}{\left(  \right)_4} I_{4}^{[d+]^{l-1}} ,
\end{eqnarray}
and directly from~\eqref{eq:RR1} and~\eqref{eq:RR2} we obtain:
\begin{eqnarray}
\label{A511}
I_{4,i}^{[d+]^l}&=&-\frac{{0\choose i}_4}{\left(  \right)_4} I_{4}^{[d+]^{l-1}} +
 \sum_{t=1}^{4} \frac{{t\choose i}_4}{\left(  \right)_4} I_{3}^{[d+]^{l-1},t},
\\
\label{A401}
I_{4}^{[d+]^l}&=&\left[\frac{{0\choose 0}_4}{\left(  \right)_4}I_{4}^{[d+]^{l-1}}-
\sum_{t=1}^{4} \frac{{t\choose 0}_4} {\left(  \right)_4} I_{3}^{[d+]^{l-1},t}  \right]
\frac{1}{d+2 l-5}.
\end{eqnarray}
In~(\ref{A401}) we can put $d=4$ for $l=1$ since $I_{4}^{[d+]}$ is UV and IR finite.
Concerning the factors ${n}_{ijkl\cdots}$ and  ${\nu}_{ijkl\cdots}$ we see that the
recursions work as if there were no such factors at all: each recursion eliminates one
of the factors ${\nu}_{ij\cdots}$. This property continues
to be valid for $3$- and $2$-point functions.
\subsection{\label{sub3}Reduction of 3-point integrals}
\begin{eqnarray}
\label{A433}
{n}_{ijk} I_{3,ijk}^{[d+]^3,t}&=&-\frac{{0t\choose kt}_4} {{t\choose t}_4}{\nu}_{ij}I_{3,ij}^{[d+]^2,t} +
\sum_{u=1,u \ne i,j}^{4} \frac{{ut \choose kt}_4} {{t\choose t}_4} {\nu}_{ij}I_{2,ij}^{[d+]^2,tu} +
\frac{{it\choose kt}_4} {{t\choose t}_4}I_{3,j}^{[d+]^2,t}+
\frac{{jt\choose kt}_4} {{t\choose t}_4}I_{3,i}^{[d+]^2,t}, \\
\label{A322}
{\nu}_{ij}I_{3,ij}^{[d+]^l,t}&=&-\frac{{0t\choose jt}_4}{{t\choose t}_4}I_{3,i}^{[d+]^{l-1},t}+
 \sum_{u=1,u \ne t, i}^{4} \frac{{ut\choose jt}_4}{{t\choose t}_4}
I_{2,i}^{[d+]^{l-1},tu}+\frac{{it\choose jt}_4}{{t\choose t}_4}I_{3}^{[d+]^{l-1},t},\\
\label{A312}
I_{3,i}^{[d+]^l,t}&=&-\frac{{0t\choose it}_4}{{t\choose t}_4}I_{3}^{[d+]^{l-1},t}+
 \sum_{u=1,u \ne t}^{4} \frac{{ut\choose it}_4}{{t\choose t}_4}
I_{2}^{[d+]^{l-1},tu}, \\
 I_{3  }^{[d+]^l,t}&=& \left[ \frac{{0t\choose 0t}_4}{{t\choose t}_4}  I_{3  }^{[d+]^{l-1},t}-
\sum_{u=1, u \ne t}^4 \frac{ {ut\choose 0t}_4}{{t\choose t}_4} I_{2  }^{[d+]^{l-1},tu} \right]\frac{1}{d+2 l-4} ,
\label{A301}
\end{eqnarray}
Of special interest is the case $l=1$:
the $I_{3}^{[d+],t}$ is IR finite but UV infinite.
The $I_{3 }^{t}$ is in any case UV finite, however, if it is IR divergent then the coefficient ${0t\choose 0t}_4$ is zero.
Thus for the 3-point function we can put $d=4$. On the other hand,
the UV divergence of $I_{3}^{[d+],t}$, coming from $I_{2,UV}^{tu}={1}/{\eps}$,
results in $I_{3,\mathrm{UV} }^{[d+],t}=-{1}/{(2 \eps)}$. Thus we have to
keep the factor ${1}/{(d-2)}$ in~(\ref{A301}) when multipying $I_{2  }^{tu}$.
\subsection{\label{sub4}Reduction of 2-point integrals}
For the $2$-point functions, surprisingly enough, a number of peculiarities occur. Let
us begin with the scalar integrals in arbitrary dimension. The recurrence relation
quite generally reads
\begin{eqnarray}
I_{2}^{[d+]^l,tu}&=&\left\{\frac{{0tu\choose 0tu}_4}{{tu\choose tu}_4}I_{2}^{[d+]^{l-1},tu}-
 \sum_{v=1,v \ne  t, u}^{4} \frac{{vtu\choose 0tu}_4}{{tu\choose tu}_4}
I_{1}^{[d+]^{l-1},tuv} \right\}\frac{1}{d+2l-3} .
\label{A211}
\end{eqnarray}
The ${tu\choose tu}_4=-2 (q_i-q_{i'})^2 \equiv -2 q^2, ~i,i' \ne t,u$  ($i' \ne i$)
is independent of masses and is in particular the argument of the $2$-point function,
i.e. $I_2=I_2(m_1,m_2,q^2)$. Quite often $q^2=0$, which has to be considered separately.
This situation occurs, e.g., in our case of calculating the higher-dimensional
$4$-point functions where the corresponding $2$-point
functions are generated by the application of the recurrence relations for the $4$-point
functions. A more physical case, e.g., occurs
if we consider radiation of a photon from an internal massive line. For $q^2 \ne 0$,
nevertheless, we can apply~(\ref{A211}). In that case it is first of all worth
to investigate the $1$-point functions. These depend only on one mass $m$ and can be
expressed for arbitrary dimension as
\bea
I_1^{[d+]^l}=(-1)^l ~I_1^d~ \frac{(2 m^2)^l}{d(d+2) \cdots (d+2l -2)} ,
\label{A10}
\eea
with
\bea
I_1^d=-\frac{\Gamma(1-\frac{d}{2})}{(m^2)^{1-\frac{d}{2}}} \sim A_0(m^2)+\frac{m^2}{\eps}.
\label{A11}
\eea
The question now arises how to implement these expressions in recurrence ~(\ref{A211}).
Here the following observation is useful: in ~(\ref{A211}),
performing the sum over $v$, there are only contributions if all indices $t,u,v$ are different. In the
$I_1$-functions the propagators with indices $t,u,v$ are scratched. Let us call the remaining
propagator $w$. Since all indices are running from $1$ to $4$ and all are different, theire sum is $10$.
Thus $w=10-t-u-v$ and the mass in $I_1$,~(\ref{A10}), can be expressed as
$m^2=\frac{1}{2}Y_{10-t-u-v,10-t-u-v}$, where the $Y_{ij}$ is defined in~\eqref{gram}. This property has in particular been used to
calculate the divergent parts of the $2$-point functions $I_{2}^{[d+]^l,tu}$ for $t,u=1, \cdots ,4$
as it was needed in Sec.~\ref{Gram}.

Making use of
\bea\label{eq-wa-87}
(m_i^2-m_{i'}^2) I_2^{(d)}(m_i,m_{i'},q^2=0)=I_1^{(d)}(m_i)-I_1^{(d)}(m_{i'}) ,
\eea
we obtain the following relation of special interest:
\begin{eqnarray}
I_{2}^{[d+],tu}=&&\frac{1}{2} \left\{
\left[q^2-2(m_i^2+m_{i'}^2) \right]I_2^{tu}(q^2)+
(m_i^2-m_{i'}^2)^2 \frac{I_2^{tu}(q^2)-I_2^{tu}(0)}{q^2} \right. \nn \\
&&\left. -\left[I_1^{tui}+I_1^{tui'}\right]
\right\} \frac{1}{d-1},
\label{A201}
\end{eqnarray}
where $q^2=(q_i-q_{i'})^2$. The second term is obviously UV finite
and vanishing for equal masses. Also for $q^2=0$ this expression makes sense, the difference quotient
resulting in the  $DB0$-function in LoopTools notation~\cite{Hahn:1998yk}.
In general, however, and in particular for high dimensions one should start from~(\ref{ItoZ0})
in case of $q^2=0$. For convenience we may drop indices $t,u$ in $I_{2}^{[d+],tu}$, i.e. we
work in this case with $2$-point functions only:
\bea\label{eq-wa-88}
I_2^{[d+]^l}(m_i,m_{i'},q^2=0)&&=Z_2^{[d+]^l}(m_i,m_{i'},q^2=0) \nn \\
&&=
\sum_{t=1}^2 \frac{{t\choose 0}_2}{{0\choose 0}_2}I_1^{[d+]^l,t}(m_i,m_{i'},q^2=0),
\eea
and with $q^2=0$:
\bea\label{eq-wa-89}
\frac{{t\choose 0}_2}{{0\choose 0}_2}=(-1)^t\frac{1}{m_i^2-m_{i'}^2}
,~~m_i \neq m_{i'} ,
\eea
such that
\bea
I_2^{[d+]^l}(m_i,m_{i'},q^2=0)&=& \frac{1}{m_i^2-m_{i'}^2} I_1^{[d+]^l}(m_i) + \frac{1}{m_{i'}^2-m_i^2} I_1^{[d+]^l}(m_{i'}).
\label{A2000}
\eea
For the case $m_i = m_{i'}$ and $q^2=0$ the ratio
${{t\choose 0}_2}/{{0\choose 0}_2} = {0}/{0}$ does not yield correct results \footnote{In fact the result depends on the order
of taking the limits $m_i- m_{i'} \to 0$ or $q^2 \to 0$.}
 and we have to choose a different approach, based on relation  (29) of~\cite{Fleischer:1999hq}
(see also~\cite{Davydychev:1991va},~\cite{Tarasov:1996br} ):
\bea
\sum_{j=1}^n {\nu}_{j} {\bf j}^+ I_n^{(d+2)}=-I_n^{(d)}.
\label{DavTa}
\eea
With
\bea\label{eq-wa-92}
I_2^{(d)}(m,m,q^2=0) = I_{1,1}^{(d)},
\eea
the eqnn.~(\ref{DavTa}) yields
\bea
I_{1,1}^{(d)}&=& -I_1^{(d-2)}
\nl
&=&\frac{d-2}{2 m^2} ~I_1^{(d)}.
\label{A1xy}
\eea
The latter eqn. is  obtained from~(\ref{A11}), and finally
\bea
I_2^{(d)}(m,m,q^2=0)=\frac{d-2}{2 m^2} ~I_1^{(d)}.
\label{A2d}
\eea
For the vector integral $I_2^{\mu}$ the recursion reads
\begin{eqnarray}
I_{2,i}^{[d+]^l,tu}&=&-\frac{{0tu\choose itu}_4}{{tu\choose tu}_4}I_{2}^{[d+]^{l-1},tu}+
 \sum_{v=1,v \ne  t, u}^{4} \frac{{vtu\choose itu}_4}{{tu\choose tu}_4}
I_{1}^{[d+]^{l-1},tuv} ,
\label{A21l}
\end{eqnarray}
and similarly to~(\ref{A201}):
\begin{eqnarray}
I_{2,i}^{[d+],tu}&=&-\frac{1}{2} I_2^{tu}(q^2)+
                      \frac{1}{2} (m_i^2-m_{i'}^2) \frac{I_2^{tu}(q^2)-I_2^{tu}(0)}{q^2},~~
i,i' \ne t,u,
\label{A2nn}
\end{eqnarray}
which seems particularly useful for numerical evaluations even if $q^2=0$, yielding the
function $DB0(0,m_1^2,m_2^2)$ of LoopTools.
To clarify the notation we mention that the integral $I_{2,j}^{[d+],tu}$ occurs with
the indices $j=i$ and $j=i'$ ($i \ne i'$), corresponding to the chords $q_i$ and $q_{i'}$ and the
masses $m_i$ and $m_{i'}$ of propagators $c_i$ and $c_{i'}$, respectively
\footnote{For the comparison wih LoopTools. e.g.,  we put $q_2=0,~ m_{i'}=m_1^{LT}$ and
$q_1=-p, m_i=m_2^{LT}$.}.

For $m_i=m_{i'}=m$ and $q^2=0$ we have
\bea
I_{2,i}^{(d+2)}(m,m,q^2=0)=-\frac{1}{2} I_2^{(d)}(m,m,q^2=0)=-\frac{d-2}{4 m^2} ~I_1^{(d)}(m).
\label{I2i}
\eea
Further we need the tensor coefficients
\bea
{\nu}_{ij}I_{2,ij}^{[d+]^l,tu}&=&-\frac{{0tu\choose jtu}_4}{{tu\choose tu}_4}I_{2,i}^{[d+]^{l-1},tu}+
 \sum_{v=1,v \ne t,u, i}^{4} \frac{{vtu\choose jtu}_4}{{tu\choose tu}_4}
I_{1,i}^{[d+]^{l-1},tuv}+\frac{{itu\choose jtu}_4}{{tu\choose tu}_4}I_{2}^{[d+]^{l-1},tu},
~l=1,2 .
 \nn \\
\label{tensorijd2}
\eea
For $q^2 \ne 0$ all the occurring integrals on the right-hand side of~(\ref{tensorijd2})
are given above. For $q^2=0$ and $m_i \ne m_{i'}$ we now have
\bea
{\nu}_{ij}I_{2,ij}^{[d+]^2}(q^2=0)=&&{\nu}_{ij}Z_{2,ij}^{[d+]^2}(q^2=0) \nn \\
=&&\frac{{t\choose 0}_2}{{0\choose 0}_2} {\nu}_{ij}I_{1,ij}^{[d+]^2,t}+
\frac{{i\choose 0}_2}{{0\choose 0}_2} I_{2,j}^{[d+]^2}(q^2=0)+\frac{{j\choose 0}_2}{{0\choose 0}_2} I_{2,i}^{[d+]^2}(q^2=0),~~ t \ne i,j.\nn \\
\label{unif}
\eea
Note that for $i \ne j $ the first term does not exist and for $i=j$ we have ${\nu}_{ii}=2$.
Thus we have for $i \ne j $:
\bea
I_{2,ij}^{[d+]^2}(q^2=0)=\frac{{i\choose 0}_2}{{0\choose 0}_2} I_{2,j}^{[d+]^2}(q^2=0)+\frac{{j\choose 0}_2}{{0\choose 0}_2} I_{2,i}^{[d+]^2}(q^2=0) ,
\label{Tijne}
\eea
and for $i=j$ we have
\bea
I_{2,ii}^{[d+]^2}(q^2=0)=\frac{{i\choose 0}_2}{{0\choose 0}_2} I_{2,i}^{[d+]^2}(q^2=0)+
\frac{{t\choose 0}_2}{{0\choose 0}_2} \frac{1}{2} I_1^d(m_i),~~~t \ne i,
\label{Tijeq}
\eea
where due to ~(\ref{DavTa})
\bea\label{eq-wa-923}
I_{1,11}^{[d+]^2}=\frac{1}{2}I_{1}^{d} ,
\eea
with $I_{1}^{d}$ given in~(\ref{A11}). To evaluate~(\ref{Tijne}) and~(\ref{Tijeq}) we need
\bea\label{eq-wa-924}
I_{2,i}^{[d+]^2}(q^2=0)=-\frac{1}{2}I_{2}^{[d+]}(q^2=0)+\frac{1}{2}(m_i^2-m_{i'}^2)
\frac{\partial I_2^{[d+]}}{\partial q^2} (q^2=0)
\eea
with
\bea\label{eq-wa-925}
\frac{\partial I_2^{[d+]}}{\partial q^2}(q^2=0)=\frac{1}{2}\left\{I_2-2\left(m_i^2+m_{i'}^2\right)
\frac{\partial I_2}{\partial q^2}+\frac{1}{2}\left(m_i^2-m_{i'}^2\right)^2
\frac{{\partial}^2 I_2}{\partial (q^2)^2}\right\}(q^2=0)\frac{1}{d-1},
\eea
i.e. the second derivative $\frac{{\partial}^2 I_2}{\partial (q^2)^2}(q^2=0)$ is needed in addition
to the function DB0 used also in LoopTools.

Last but not least we have to deal with $q^2=0$ and $m_i=m_{i'}=m$, where we avoid the
appearance of a ratio $\frac{0}{0}$ as follows:~(\ref{DavTa}) reads in this case
\bea\label{eq-wa-926}
\sum_{j=1}^2 {\nu}_{ij} I_{2,ij}^{[d+]^2}&=&-I_{2,i}^{[d+]}
\nl
&=&\frac{d-2}{4 m^2} ~I_1^{d}(m),
\eea
the latter eqn.   being obtained from~(\ref{I2i}). Assuming $q=0$, all integrals $I_{2,ij}^{[d+]^2}$ are equal
and we have~\cite{Fleischer:2003rm}
\bea\label{eq-wa-927}
I_{2,ij}^{[d+]^2}=\frac{1}{3}\frac{d-2}{4 m^2} ~I_1^{d}(m).
\eea
\section{\label{Bpp}Divergent parts of higher-dimensional integrals}
In our final results ~(\ref{want1}),~(\ref{fulld3}) and~(\ref{fulld4}) we
obtained contributions of certain higher-dimensional integrals muliplied with polynomials
in $d=4-2 \eps$ such that the ${1}/{\eps}$ parts of the UV divergent 4- and 3-point integrals
combine with the $\eps$-powers of the polynomials to yield finite contributions.
In a numerical approach these contributions have to be explicitly calculated, and for that
purpose we list the infinite parts of those integrals and of scalar $2$-point functions appearing in the reductions.
As described before, for the
calculation of corrections for small Gram determinants, we need 3-point integrals of
even higher dimensions, which are,
however, too complicated to be listed -- apart from being difficult to obtain.
They have been calculated iteratively in the numerics, see sect.~\ref{Gram} for details.

The higher-dimensional integrals are in general UV divergent
and we write them in the form
\bea\label{eq-wa-98}
I_n^{(d)}=F_n^{(d)}+\frac{1}{\eps}D_n^{(d)} + {\cal O}(\eps^2), ~~~d=[d+]^l.
\label{Decomp}
\eea
The terms $D_n^{(d)}$ are obtained from the recurrence relations (see app.  A), starting
from lower dimensions.
We mention that, whenever in the following list two of the occurring indices
$i,j,k,t,u,v$ are equal, the corresponding $D's$ vanish.

For 1-point functions, it is
\bea\label{eq-wa-929}
D_1^{[d+]^l,t,u,v}=(-1)^l \frac{1}{2 \cdot 4 \cdots 2(l+1)} Y_{10-t-u-v,10-t-u-v}^{l+1} .
\eea
For $i,j \ne t,u$ we have for 2-point functions:
\bea\label{eq-wa-95}
D_2^{t,u}~~~~~~~~=&&~~~1, \nn \\
D_2^{[d+],t,u}~~=&&-\frac{1}{6}\left[Y_{i i}+Y_{j j}+Y_{i j}\right],  \nn \\
D_2^{[d+]^2,t,u}=&&\frac{1}{120}\left[3 Y_{i i}^2+Y_{i i} Y_{j j}+3 Y_{j j}^2 +3 Y_{i j}
\left( Y_{i i}+Y_{j j} \right)+2Y_{i j}^2\right],  \nn \\
D_2^{[d+]^3,t,u}=&&-\frac{1}{1680}\left[5 Y_{i i}^3+Y_{i i}^2 Y_{j j}+Y_{i i} Y_{j j}^2+5 Y_{j j}^3+
Y_{i j} \left(5 Y_{i i}^2+ 3 Y_{i i} Y_{j j}+ 5 Y_{j j}^2 \right) \right.\nn \\
&&\left. +~~~~~~~~~~~4 Y_{i j}^2 \left( Y_{i i}+Y_{j j} \right)+2Y_{i j}^3 \right], \nn \\
D_2^{[d+]^4,t,u}=&&\frac{1}{120960}\left[35 Y_{i i}^4+ 5 Y_{i i}^3 Y_{j j}+3 Y_{i i}^2 Y_{j j}^2+
5 Y_{i i} Y_{j j}^3+35 Y_{j j}^4 \right. \nn \\
&&\left. +~~~~~~~~~~~5
Y_{i j} \left(7 Y_{i i}^3+ 3 Y_{i i}^2 Y_{j j}+3 Y_{i i} Y_{j j}^2+7 Y_{j j}^3 \right) \right. \nn \\
&&\left. +~~~~~~~~~~~
6 Y_{i j}^2 \left(5 Y_{i i}^2 +4 Y_{i i}Y_{j j}+5 Y_{j j}^2 \right)+20Y_{i j}^3 \left(Y_{i i}+Y_{j j}
\right)+8 Y_{i j}^4 \right], \nn \\
D_{2,i}^{[d+]^2,t,u}=&&\frac{1}{24}\left[3 Y_{i i}+2 Y_{i j} +Y_{j j}\right].
\eea
For $i,j,k \ne t$ we have for 3- and 4-point functions:
\bea
D_3^{[d+],t}~=&&-\frac{1}{2}, \nn \\
D_3^{[d+]^2,t}=&&~~\frac{1}{24} \left[Y_{i i}+Y_{i j}+Y_{ i k}+Y_{j j}+Y_{j k}+Y_{ k k}\right],  \nn \\
D_3^{[d+]^3,t}=&&-\frac{1}{720} \left\{3\left[Y_{i i}\left(Y_{i i}+Y_{i j}+Y_{ i k}\right)+Y_{j j}
\left(Y_{j i}+Y_{j j}+Y_{j k}\right)+Y_{ k k}\left(Y_{k i}+Y_{k j}+Y_{ k k}\right)\right] \right.\nn \\
&&\left.+ ~~~~~~~~~
2\left[Y_{i j}\left(Y_{i j}+Y_{j k}\right)+Y_{i k}\left(Y_{i j}+Y_{i k}\right)+
Y_{j k}\left(Y_{i k}+Y_{j k}\right)\right] \right.\nn \\
&&\left.+~~~~~~~~~~~
\left[Y_{i i}\left(Y_{j j}+Y_{j k}\right)+Y_{j j}\left(Y_{i k}+Y_{k k}\right)+
Y_{k k}\left(Y_{i i}+Y_{i j}\right)\right]\right\}, \nn \\
D_{3,i}^{[d+]^2,t}=&&~~~~\frac{1}{6},  \nn \\
D_{3,i}^{[d+]^3,t}=&&-\frac{1}{120}\left\{3~~Y_{i i}+2(Y_{i j}+Y_{ i k})+Y_{j j}+Y_{j k}+
Y_{ k k}\right\},~~j,k \ne i,t ~~~~~~ (j \ne k), \nn \\
D_{3,ij}^{[d+]^3,t}=&&-\frac{1}{24} ,
\label{D3t}
\eea
and
\bea\label{eq-wa-940}
D_4^{[d+]}~=&&~~~~~~0,\nn \\
D_4^{[d+]^2}=&&~~~~~~\frac{1}{6},\nn \\
D_4^{[d+]^3}=&&-\frac{1}{120}\left[Y_{1 1}+Y_{1 2}+Y_{ 1 3}+Y_{1 4}+Y_{2 2}+Y_{ 2 3}+
Y_{2 4}+Y_{3 3}+Y_{ 3 4}+Y_{4 4}\right],\nn \\
D_4^{[d+]^4}=&&\frac{1}{5040} \left\{3 \left[
              Y_{1 1}(Y_{1 1} + Y_{1 2} + Y_{1 3} + Y_{1 4}) +
              Y_{2 2}(Y_{1 2} + Y_{2 2} + Y_{2 3} + Y_{2 4}) ~+\right. \right. \nn \\
&&\left. \left.~~~~~~~~~~~~~      Y_{3 3}(Y_{1 3} + Y_{2 3} + Y_{3 3} + Y_{3 4}) +
              Y_{4 4}(Y_{1 4} + Y_{2 4} + Y_{3 4} + Y_{4 4})\right] +\right. \nn \\
&&\left.~~~~~~~~~~ 2\left[    Y_{1 2}(Y_{1 2} + Y_{1 3} + Y_{1 4}) +
             Y_{1 3}(Y_{1 3} + Y_{1 4} + Y_{2 3}) +
              Y_{1 4}(Y_{1 4} + Y_{2 4} + Y_{3 4}) ~~+\right. \right.\nn \\
&&\left.\left.~~~~~~~~~~~~~~               Y_{2 3}(Y_{1 2} + Y_{2 3} + Y_{2 4}) +
              Y_{2 4}(Y_{1 2} + Y_{2 4} + Y_{3 4}) +
              Y_{3 4}(Y_{1 3} + Y_{2 3} + Y_{3 4})\right]  +\right.\nn \\
&&\left.~~~~~~~~~~~ Y_{1 1}(Y_{2 2} + Y_{2 3} + Y_{2 4} + Y_{3 3} + Y_{3 4}) +
              Y_{2 2}(Y_{1 3} + Y_{1 4} + Y_{3 3} + Y_{3 4} + Y_{4 4}) ~~~+\right.\nn \\
&&\left.~~~~~~~~~~~Y_{3 3}(Y_{1 2} + Y_{1 4} + Y_{2 4} + Y_{4 4}) ~~~~~~~~~~+
              Y_{4 4}(Y_{1 1} + Y_{1 2} + Y_{1 3} + Y_{2 3}) +\right.\nn \\
&&\left.~~~~~~~~~~~Y_{1 2} Y_{3 4} + Y_{1 3} Y_{2 4} + Y_{1 4}  Y_{2 3}
\right\}, \nn \\
D_{4,i}^{[d+]^3}=&&-\frac{1}{24}, \nn \\
D_{4,ij}^{[d+]^3}=&&~~~~~~0, \nn \\
D_{4,ijk}^{[d+]^4}=&&~~~~~~0.
\eea

\section{\label{Num}A numerical example}
\newcommand{\snu}{s_ {\bar{\nu} u}} 
\newcommand{\ted}{t_{ed}} 
\newcommand{\tedcr}{t_{ed,\mathrm{crit}}} 
\newcommand{\smnu}{s_ {\mu \bar{\nu}   u}} 
\newcommand{\tem}{t_ {\bar{e} \mu}} 
In order to demonstrate the use of our small Gram determinant expansion, we reproduce the numerics for the topology shown in Fig.~\ref{fig-4-and-6-point} (b), which arises from the on-shell six-point topology of Fig.~\ref{fig-4-and-6-point} (a).
The example is taken from~\cite{Denner:THHH2009}, and  first results obtained with  our approach were reported in~\cite{Fleischer:2010mq}.
In LoopTools~\cite{Hahn:1998yk} conventions, the tensor coefficients $D_{ijl}$ are defined as follows:
\bea\label{dmunula}
D_{\mu\nu\lambda}
&=&
\sum_{i,j,l=1}^3 K_{i\mu}K_{j\nu}K_{l\lambda} D_{ijl}
+ \sum_{i=1}^3 (g_{\mu\nu}K_{i\lambda} + g_{\nu\lambda}K_{i\mu}  +g_{\lambda\mu}K_{i\nu}) D_{00i}.
\eea
For our conventions see ~(\ref{tensor3}).
The external momenta are assumed to be incoming: $p_{1}=p_{e^+}, p_{2}=p_{\mu^-}, p_{3}=p_{{\bar \nu}_{\mu}}+p_{u}, p_{4}=p_{e^-}+p_{{\bar d}}$.
The inverse propagators are here $c_j=  [(k-q_j)^2-m_j^2]$, and in LoopTools conventions  $c_j = [(k+K_{j-1})^2-m_j^2]$.
The  $K_i$ are the internal momenta, expressible by the $p_i$: $K_1=p_1, K_2=K_1+p_2, K_3=K_2+p_3, K_4=0$.
Then, with $p_i^2=s_i, (p_i+p_j)^2=s_{ij}$, we set $s_{12}=\tem, s_{23}=\smnu ,   s_3=\snu , s_4=\ted.$
The corresponding tensor integrals are, in  LoopTools~\cite{Hahn:1998yk} notation:
\bea\label{n1}
\mathrm{D0i}(\mathrm{id},0,0,s_{\bar{\nu}u},t_{ed},\tem,\smnu,0,M_Z^2,0,0). 
\eea
The Gram determinant is:
\bea\label{n2}
()_4 ~=~ \varDelta^{(3)} &=& - \mathrm{Det}\left( 2 K_iK_j \right)
\nl
 &=&
-~2   \tem [\smnu^2 + \snu \ted - \smnu (\snu + \ted - \tem)],
 \eea
and it vanishes if
\bea\label{n3}
\ted \to \tedcr =
\frac{ \smnu ( \smnu -  \snu +  \tem)}{ \smnu -  \snu} .
\eea
In terms of a dimensionless scaling parameter $x$,
\bea\label{n4}
\ted &=& (1+x) \tedcr,
\eea
the Gram determinant becomes
\bea\label{n5}
()_4 =- 2 ~x~ \smnu \tem (\smnu - \snu + \tem) .
\label{FoxGra}
\eea
Following~\cite{Denner:THHH2009}, we have chosen
\bea\label{n7}
\smnu 
&=& 2 \times 10^4 \mathrm{GeV}^2,
\nl
\snu 
&=&1 \times 10^4 \mathrm{GeV}^2,
\nl
\tem 
&=& -4 \times 10^4 \mathrm{GeV}^2,
\eea
and get $ \tedcr = -6 \times 10^4 \mathrm{GeV}^2$.
For $x$=1, the Gram determinant becomes
 $()_4 =- 4.8 \times 10^{13}$ GeV$^{3}$.

We also need the modified Cayley determinant:
\begin{eqnarray}\label{cay1}
{0\choose 0}_4
&=&
  \begin{vmatrix}
2 M_Z^2     & M_Z^2 & M_Z^2-\smnu & M_Z^2
\\
M_Z^2       & 0     & -\snu       & M_Z^2
\\
M_Z^2-\smnu & -\snu & 0           & -\ted
\\
M_Z^2       & -\tem & -\ted       & 0
  \end{vmatrix}
\nl\nl
&=&
 \smnu^2 \tem^2 +
2~ M_Z^2 \tem [-2 \snu \ted + \smnu (\snu + \ted - \tem)]
\nl
&&+~
 M_Z^4 \left[\snu^2 + (\ted - \tem)^2 - 2 \snu (\ted +\tem)\right] .
\end{eqnarray}
From~\eqref{Correction} we see that  a \emph{small-Gram determinant expansion} will be applicable when the following dimensionless parameter becomes small:
\bea\label{n6}
R &=& \frac{()_4}{{0\choose 0}_4}~\times~ s,
\eea
where $s$ is a typical scale of the process.
We  choose here $s=\smnu$.
Obviously,
due to~\eqref{n5}, $x$ must also be small.
Indeed, for e.g. $x=0.01$ we have $R=-0.064$.\footnote{The dependence of $R(x)$ on $x$ is shown in Fig. 1 of
\cite{riemann-corfutalk-2010aa}.}

\begin{figure}[t]
\begin{center}
\subfloat[][]{\includegraphics[width=.45\textwidth]{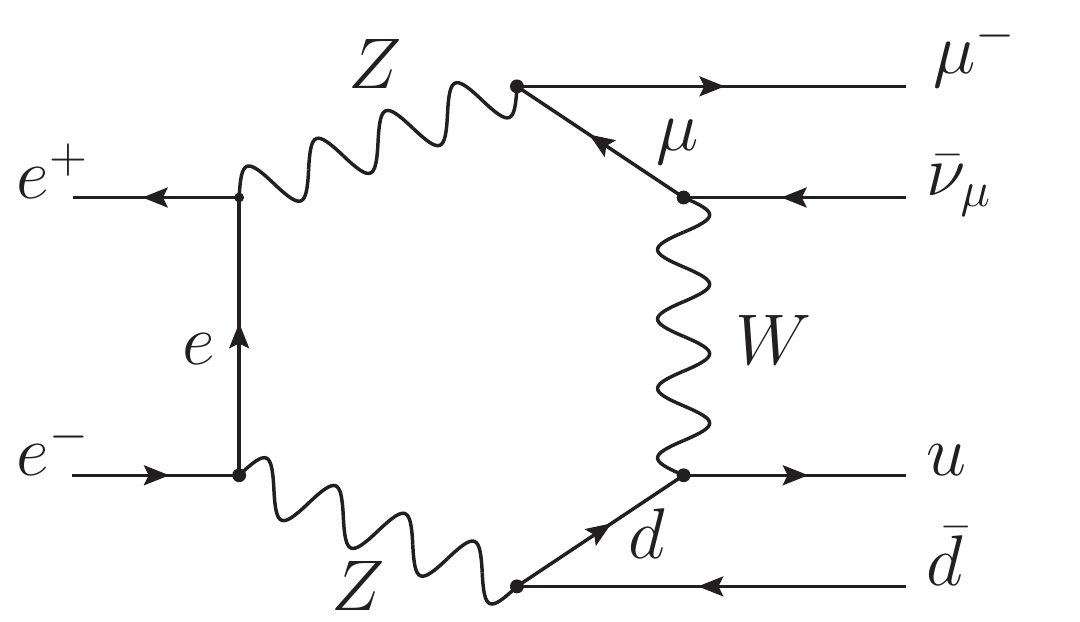}}
\hspace*{0.5cm}
\subfloat[][]{\includegraphics[width=.45\textwidth]{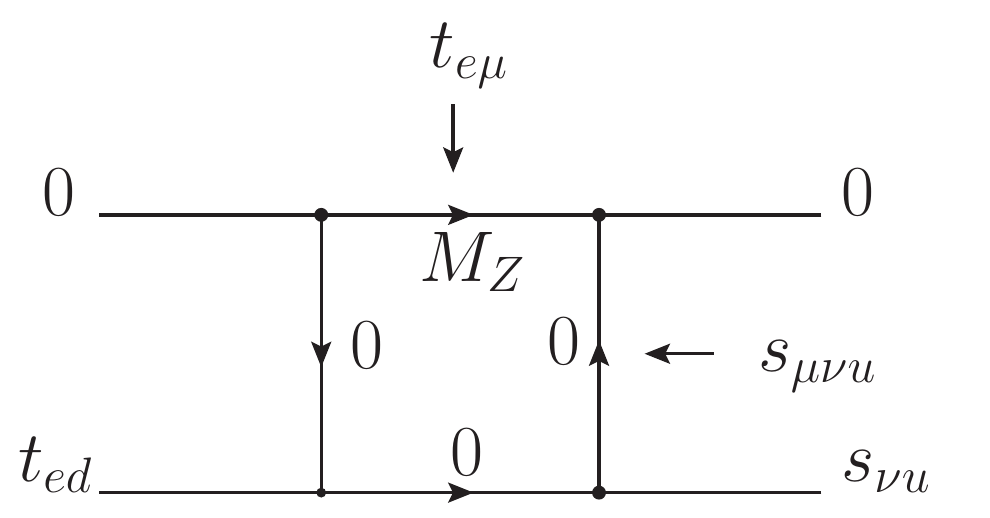}}
\end{center}
\caption[(a) A six-point topology; (b) a four-point topology derived from (a).]{%
\emph{(a) A six-point topology; (b) a four-point topology derived from (a).}
}
\label{fig-4-and-6-point}
\end{figure}

\begin{figure}[t]
 \begin{center}
 \subfloat[][]{\includegraphics[width=.7\textwidth]{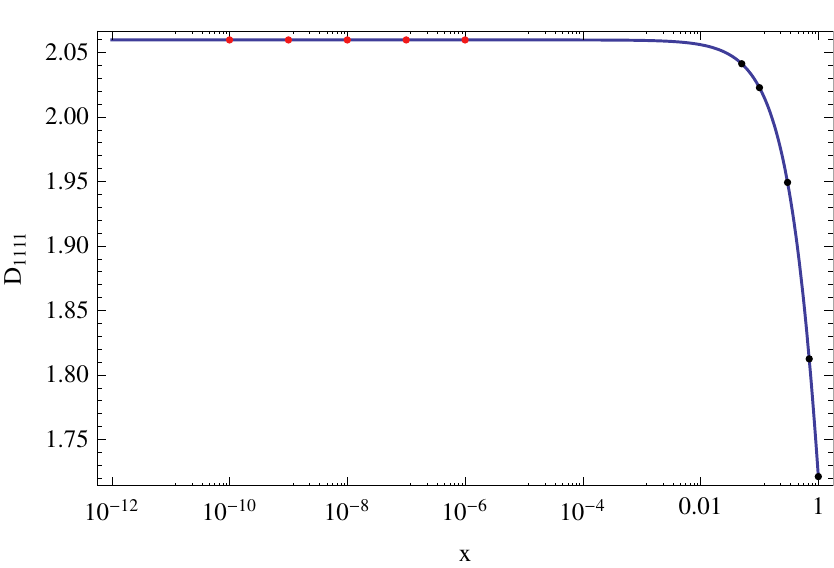}}
 \end{center}
 \caption[The tensor coefficient $D_{1111}(x)$.]{%
\emph{The tensor coefficient $D_{1111}(x)$; the data points for a least
 square fit are shown. Input:  grey (red) dots originate from the expansion at
 small $x$,  black dots from the values at larger $x$.}
 }
 \label{fig-D1111}
 \end{figure}

In Fig.~\ref{fig-D1111} we show  the tensor coefficient $D_{1111}(x)$, and in  Tabs.
\ref{tab-d1111}--\ref{tab-d111}  the tensor coefficients $D_{1111}$ and  $D_{111}$ are tabulated for the region of interest of $x$.
Because we assume in our  formulae $q_1=0$, and in LoopTools $K_4=0$ is assumed, one has to care about specific correspondences, in particular:
\bea\label{n6a}
D_{1111} &=& n_{2222} \times I_ {4, 2222}^{[d +]^4},
\\\label{n6b}
D_{111} &=& n_{222}  \times I_ {4, 222}^{[d +]^3}.
\eea
The integral $I_ {4, 2222}^{[d +]^4}$ contributes also to the rank $R=5$ tensor coefficients $E_{2222m}$, see~\eqref{Ewxyz5}, and  $I_ {4, 222}^{[d +]^3}$ to the rank $R=4$ tensor coefficients $E_{222l}$, see~\eqref{Ewxyz3}.

The numerics have been performed with Mathematica v.7.0, using LoopTools v.2.5
for the scalar  $4$-, $3$- and $2$-point functions in generic dimension.
Of course the first step to be done is to verify~\eqref{fulld4},~\eqref{fulld3}, and~\eqref{want1}
by comparing their numerical evaluation with results from LoopTools v.2.5 \footnote{In fact in some
cases we found a slightly better agreement with LoopTools v.2.5 than with LoopTools v.2.4.}. This verification
is naturally only possible for non-vanishing Gram determinants, i.e. in this case all (higher-dimensional) integrals have been calculated by means of the recurrence relations given in app.~\ref{App}.
For $x\in (0.1,1.0)$ we find in
general an agreement to more than 10 decimals: in Tabs.~\ref{tab-d1111}--\ref{tab-d111}
these cases are marked by $I_{4,2222}^{[d+]^4}$ and $I_{4,222}^{[d+]^3}$, respectively.
In the next step we have to evaluate~\eqref{fulld4},~\eqref{fulld3}, and~\eqref{want1} for
small Gram determinants.
If $l=l_{max}$ is the highest available index for which the $Z_{4F}^l$ are calculated, we see
that the upper limit $i=i_{max}$ in~\eqref{finitesum} can be at most  $i_{max}=l_{max}-L$.
Since the integral of highest dimension is $I_4^{[d+]^4}$, we take the value $L=4$
as reference value.

 Tabs.~\ref{tab-d1111}--\ref{tab-d111} have been produced with a varying number of
correction terms, specified by $i_{max}=l_{max}-4$, dependent on the size of $\left( \right)_4$,
specified in terms of the parameter $x$ (see~\eqref{FoxGra}). It is clear that for
very small Gram determinants a few terms only have to be taken into account. Higher terms
don't change anything beyond the shown accuracy.

Due to our discussion following~\eqref{I4L} we have a certain choice for the value
of $Z_{4F,0}^{(L+i)}$ for $i=i_{max}$, see~\eqref{finitesum}. The choices are
$\left( \right)_4=0$ and $\left( \right)_4 \ne 0$, taken at the kinematics under
consideration. Due to~\eqref{FoxGra} the two options are parametrized
here by $x$, specified in~\eqref{n4}. In the tables we mark the approximations correspondingly
by $0$ and $x$. We specify the expansions as [exp p,$i_{max}$], where $Z_{4F,0}^{(L+i_{max})}$ in~\eqref{finitesum} is calculated for $p=0$ and $p=x$, respectively.

\begin{table}[h] 
\footnotesize
\centering
\begin{tabular}{|l|r@{.}l|r@{.}l|
}
\hline
$x$ &  \multicolumn{2}{|c|}{ $\EuFrak{Re}$ $D_{1111}$} & \multicolumn{2}{|c|}{$\EuFrak{Im}$  $D_{1111}$ }
\\
\hline
$0.$ \hfill  [exp 0,0]     & 2&05969289730~E-10 & 1&55594910118~E-10
\\\hline
$10^{-8}$ \hfill  [exp x,2]& 2&05969289342~E-10 & 1&55594909187~E-10
\\
 \hfill  [exp 0,2]& 2&05969289349~E-10 & 1&55594909187~E-10
\\\hline
$10^{-4}$  \hfill  [exp x,5]& 2&05965609497~E-10 & 1&55585605343~E-10
\\
 \hfill  [exp 0,5]& 2&05965609495~E-10 & 1&55585605343~E-10
 \\
\hline
$0.001$ \hfill  [exp 0,6]  & 2&05932484380~E-10 & 1&55501912433~E-10
\\
\hfill  [exp x,6] & 2&05932484381~E-10 & 1&55501912433~E-10
\\
\cline{2-5}
      \hfill      $I_{4,2222}^{[d+]^4}$       & 2&02292295240~E-10 & 1&54974785467~E-10
\\
      \hfill   $D_{1111}$&     2&01707671668~E-10 &   1&62587142251~E-10
\\
\hline
$0.005$ \hfill  [exp 0,6]  & 2&05786054801~E-10 & 1&55131031024
~E-10
\\
 \hfill  [pade 0,3]& 2&05785198947~E-10 & 1&55131031003~E-10
\\
\cline{2-5}
\hfill  [exp x,6] & 2&05786364440~E-10 & 1&55131031024~E-10
\\
\hfill [pade x,3] & 2&05785199805~E-10 & 1&55131030706~E-10
\\
\cline{2-5}
    \hfill   $I_{4,2222}^{[d+]^4}$            & 2&05778894114~E-10 & 1&55135794453~E-10
\\
 \hfill   $D_{1111}$& 2&05779811490~E-10 & 1&55136343923~E-10
\\\hline
$0.01$ \hfill  [exp 0,6]   & 2&05703298143~E-10 & 1&54669910676~E-10
\\
 \hfill  [pade 0,3]   & 2&05600940065~E-10 & 1&54669907784~E-10
\\
\cline{2-5}
\hfill  [exp 0,10] & 2&05600964693~E-10 & 1&54669910676~E-10
\\
\hfill [pade 0,5] & 2&05600955381~E-10 & 1&~54669910676E-10
\\
\cline{2-5}
\hfill  [exp x,10] & 2&05600963675~E-10 & 1&54669910676~E-10
\\
\hfill [pade x,5] & 2&05600955381~E-10 & 1&54669910676~E-10
\\
\cline{2-5}
   \hfill     $I_{4,2222}^{[d+]^4}$            & 2&05600013702~E-10 & 1&54670651917~E-10
\\
 \hfill   $D_{1111}$
& 2&05600239280~E-10 & 1&54670771210~E-10
\\\hline %
$0.05$ \hfill  [exp 0,6]   & 4&83822963052~E-09 & 1&51077429118~E-10
\\
 \hfill  [pade 0,3]   & 2&01518061131
~E-10 & 1&50591643209~E-10
\\
\cline{2-5}
  \hfill [exp 0,20] &  2&04218962072~E-10  & 1&51077424143~E-10
\\
  \hfill [pade 0,10] & 2&04122727654~E-10  & 1&51077424149~E-10
\\
\cline{2-5}
  \hfill [exp x,20] &  2&04190274030~E-10  & 1&51077424143~E-10
\\
  \hfill [pade x,10] & 2&04122727971~E-10  & 1&51077423985~E-10
\\
\cline{2-5}
  \hfill   $I_{4,2222}^{[d+]^4}$               & 2&04122726387~E-10 & 1&51077422901~E-10
\\
 \hfill   $D_{1111}$
& 2&04122726601~E-10 & 1&51077423320~E-10
\\\hline %
 $0.1$
\hfill  [exp 0,26] & 2&20215264409~E-08 & 1&46815247004~E-10
\\
\hfill [pade 0,13] & 2&01749674352~E-10 & 1&46681287362~E-10
\\
\cline{2-5}
\hfill  [exp x,26] & 2&08190721550~E-08 & 1&46815247004~E-10
\\
\hfill [pade x,13] & 2&03995221326~E-10 & 1&46785977364~E-10
\\
\cline{2-5}
    \hfill  $I_{4,2222}^{[d+]^4}$               & 2&02269485177~E-10 & 1&46815247061~E-10
\\
\hfill   $D_{1111}$ & 2&02269485217~E-10 & 1&46815247051~E-10
\\ \hline
   $1.$
\hfill  $I_{4,2222}^{[d+]^4}$& 1&72115440143~E-10 & 9&74550747662~E-11
\\
 \hfill  $D_{1111}$& 1&72115440148~E-10 & 9&74550747662~E-11
\\
\hline
\end{tabular}
\caption[Numerical values for the tensor coefficient $D_{1111}$.]{%
\emph{Numerical values for the tensor coefficient $D_{1111}$.
Values marked by $D_{1111}$ are evaluated with LoopTools, the $I_{4,2222}^{[d+]^4}$ corresponds to~\eqref{fulld4}.
The labels [exp 0,2n] and  [pade 0,n] denote iteration $2n$ and Pad\'{e} approximant $[n,n]$ when the small Gram determinant expansion starts at $x=0$, and [exp x,2n] and  [pade x,n] are the corresponding numbers for an expansion starting at $x$.}}
\label{tab-d1111}
\end{table}

\begin{table}[t] 
{
\footnotesize
\centering
\begin{tabular}{|l|r@{.}l|r@{.}l|}
\hline
$x$ &  \multicolumn{2}{|c|}{ $\EuFrak{Re}$ $D_{111}$} & \multicolumn{2}{|c|}{$\EuFrak{Im}$  $D_{111}$ }
\\
\hline
%
$0$ \hfill  [exp 0,0]& --3&15407250453~E-10 & --3&31837792634~E-10
\\\hline
$10^{-8}$
\hfill [exp x,1]& --3&15407250057   ~E-10 & --3&31837790700~E-10
\\
\hfill  [exp 0,1]& --3&15407250057~E-10 & --3&31837790700~E-10
\\\hline\cline{2-5}
$10^{-4}$
\hfill [exp x,4]& --3&15403282194   ~E-10 & --3&31818461838~E-10
\\
\hfill  [exp 0,4]& --3&15403282194~E-10 & --3&31818461838~E-10
 \\
\hline
$0.001$
\hfill [exp x,6]& --3&15367545429   ~E-10 & --3&31644587150~E-10
\\
\hfill  [exp 0,6]&      --3&15367545429~E-10 & --3&31644587150~E-10
\\
\cline{2-5}
 \hfill      $I_{4,222}^{[d+]^3}$  & --3&15372092999~E-10 & --3&31645245644~E-10
\\\cline{2-5}
 \hfill   $D_{111}$&  --3&15372823537~E-10 &   --3&31635736868~E-10
\\\hline
$0.005$
\hfill [exp x,6]& --3&15208222856   ~E-10 & --3&30874035862~E-10
\\
\hfill [pade x,3]& --3&15208230282   ~E-10 & --3&30874035931~E-10
\\
\cline{2-5}
\hfill  [exp 0,6]&      --3&15208224867~E-10 & --3&30874035862~E-10
\\
 \hfill  [pade 0,3]&      --3&15208230411~E-10 & --3&30874035867~E-10
\\
\cline{2-5}
  \hfill      $I_{4,222}^{[d+]^3}$ & --3&15208269791~E-10 & --3&30874006110~E-10
\\
 \hfill   $D_{111}$ & --3&15208264077~E-10 & --3&30874002667~E-10
\\\hline
$0.01$
\hfill [exp 0,6]& --3&15006665284~E-10 & --3&29915926110~E-10
\\
 \hfill  [pade 0,3]& --3&15007977830~E-10 & --3&29915888075~E-10
\\
\cline{2-5}
\hfill [exp 0,10]& --3&15007991203~E-10 & --3&29915926110~E-10
\\
 \hfill  [pade 0,5]& --3&15007991324~E-10 & --3&29915926110~E-10
\\
\cline{2-5}
 \hfill [exp x,10]& --3&15007991217   ~E-10 & --3&29915926110~E-10
\\
\hfill [pade x,5]& --3&15007991324   ~E-10 & --3&29915936110~E-10
\\
\cline{2-5}
  \hfill      $I_{4,222}^{[d+]^3}$ & --3&15008000292~E-10 & --3&29915916848~E-10
\\
 \hfill   $D_{111}$& --3&15008000292~E-10 & --3&29915915368~E-10
\\
\hline
%
$0.05$
\hfill  [exp 0,6]& --1&34278470211~E-11 & --3&22448580722~E-10
\\
 \hfill  [pade 0,3]& --3&13432516570~E-10 & --3&22580791799~E-10
\\
\cline{2-5}
\hfill  [exp 0,20]& --3&13359445767~E-10 & --3&22448581032~E-10
\\
 \hfill  [pade 0,10]& --3&13365675001~E-10 & --3&22448581024~E-10
\\
\cline{2-5}
\hfill [exp x,20]& --3&13361302214~E-10 & --3&22448581032~E-10
\\
\hfill [pade x,10]& --3&13365674956~E-10 & --3&22448581051~E-10
\\
\cline{2-5}
 \hfill      $I_{4,222}^{[d+]^3}$ & --3&13365675084~E-10 & --3&22448581110~E-10
\\
 \hfill   $D_{111}$& --3&13365675070~E-10 & --3&22448581084~E-10
\\\hline
$0.1$
\hfill [exp 0,26]& --2&49466252165~E-09 & --3&13582331984~E-10
\\
\hfill [pade 0,13]& --3&11144777695~E-10 & --3&13599283949~E-10
\\
\cline{2-5}
\hfill [exp x,26]& --2&34010823441~E-09 & --3&135823319836~E-10
\\
\hfill [pade x,13]& --3&10806582023~E-10 & --3&135870111996~E-10
\\
\cline{2-5}
 \hfill   $I_{4,222}^{[d+]^3}$  & --3&11226750699~E-10 & --3&13582331977~E-10
\\
 \hfill   $D_{111}$& --3&11226750695~E-10 & --3&13582331978~E-10
\\\hline
$1.$  \hfill  $I_{4,222}^{[d+]^3}$& --2&70193791372~E-10 & --2&10251973821~E-10
\\
 \hfill   $D_{111}$ & --2&70193791373~E-10 & --2&10251973821~E-10
\\
\hline
\end{tabular}
\caption[Numerical values for the tensor coefficient $D_{111}$.]{\label{tab-d111}%
\emph{Numerical values for the tensor coefficient $D_{111}$.
Values marked by $D_{111}$ are evaluated with LoopTools, the $I_{4,222}^{[d+]^3}$ is defined in~\eqref{fulld3}.
The labels [exp 0,2n] and  [pade 0,n] denote iteration $2n$ and Pad\'{e} approximant $[n,n]$ when the small Gram determinant expansion starts at $x=0$, and [exp x,2n] and  [pade x,n] are the corresponding numbers for an expansion starting at $x$.}}
}
\end{table}

We have, however, still another option.
With~\eqref{I4L} we provide a sequence of \emph{partial sums} $S^{i}$ of a series expansion for
$F_{4}^{[d+]^L}$.
This is exactly the input needed for the calculation of a Pad\'{e} approximation
for  $F_{4}^{[d+]^L}$.
We apply the $\eps$-algorithm for sequence transformations~\cite{Shanks:1955, Wynn:1956}.
It is described in detail in~\cite{Baker-book:1981}.
The $\eps$-algorithm allows an efficient calculation of elements of the so-called $\eps$-table.
The first column is zero, and the second one consists of the sequence $S^{i}=Z_{4F,i}^L$, the convergence of which shall be improved.
From the first two columns, the others are determined iteratively:
\begin{eqnarray}\label{eq-eps1}
\eps_{-1}^{(i)} &=& 0,
\\\label{eq-eps2}
\eps_{0}^{(i)} &=& Z_{4F,i}^L, ~~~i = 0, \cdots, l_{\max}-L,
\\\label{eq-eps3}
\eps_{k+1}^{(i)} &=& \eps_{k-1}^{(i+1)} + \frac{1}{\eps_{k}^{(i+1)} - \eps_{k}^{(i)}}
.
\end{eqnarray}
The $\eps$-table and the Pad\'{e} table are related:
\begin{eqnarray}\label{eq-eps4}
\eps_{2k}^{(i)} &=& [k+i/k],
\end{eqnarray}
where the symbol $[k+i/k]$ stands for the degrees $k+i$ of numerator and $k$ of denominator polynomials of the corresponding Pad\'{e} approximant $[k+i/k]$.
We took the choice $k=l_{max} -L$ and:
\begin{eqnarray}\label{eq-eps5}
 F_{4}^{[d+]^L} \sim ~~\eps_{2k}^{(0)}~\equiv~ [k/k]_{F_{4}^{[d+]^L}} .
\end{eqnarray}

In the tables we present the Pad\'{e} approximants together with the corresponding sums, for $x=0$
as well as for $x \ne 0$: they are denoted by [pade 0,$i_{max}/{2}$]
and [pade x,${i_{max}}/{2}$], respectively. In general the Pad\'{e}
approximants provide a remarkable improvement of precision compared to the sums such that
we even close up to the values provided by the non-small Gram determinant representation.
It is also remarkable that there is only very little difference between the values obtained by iterations starting
from $x=0$ or from $x \ne 0$.
In fact the Pad\'{e} starting with $x=0$ is only slighly better
than the one calculated with $x \ne 0$.
The difference arises because the integrals $F_4^{[d+]^l}$
change much slower with $x$ than the approximants $Z_{4F,i}^L$ so that for a start at $x=0$
the latter are already much closer to the final values than those starting at $x \ne 0$.
Nevertheless, after a few iterations the difference has already almost disappeared.
At this point we want to remind the reader that we approximate the integrals $I_4^{[d+]^l}$
and they must indeed be very precise since there are considerable cancellations with the
remaining contributions to the tensor coefficients under consideration.

To discuss a few results given in Tabs.~\ref{tab-d1111} and~\ref{tab-d111} we mention
in particular the results for $x=0.05$. In Tab.~\ref{tab-d1111} we can assume the values
of [pade 0,10] to be accurate to $10$ decimals. One reason is that even Pad\'{e} approximants up to
[pade 0,13] (based on 27 terms in~\eqref{finitesum}, not shown here) remain stable in the ${10}^{th}$ and ${11}^{th}$ decimal. With LoopTools we
have an agreement of $8$ decimals - thus it seems that this point is the value where
the representation in terms of non-small Gram determinants starts to loose precision. We also see
that there is even agreement beween [pade 0,10] and [pade x,10] up to $9$ decimals. Just for
curiosity we have also calculated the values for $x=0.1$. Here LoopTools and also our
calculation with the larger Gram determinants work perfectly so that we can say to be
in the domain of larger Gram deteteminants. Nevertheless [pade 0,13] yields a precision
of $3$ decimals - while [exp 0,26] is off by 2 orders of magnitude. It is clear that for
lower $x$ we obtain good results with less terms in the expansion. The results in
Tab.~\ref{tab-d111} are of similar quality - in fact they are slightly
better since the highest dimensional $4$-point function is only $I_4^{[d+]^3}$.

Finally we show in Fig.~\ref{fig-D1111} the smooth behaviour of $D_{1111}(x)$.
The smoothness is in striking contrast to the complexity of its precise calculation.
Indeed, in the critical domain the function is almost constant and one has to
spend quite some effort to calculate it with high enough precision.

\clearpage

\section{Notations and algebraic relations\label{app-nota}}
In the following we quote some relations for the signed minors, which
are of particular relevance for the present work (see~\cite{Melrose:1965kb} and also~\cite{Diakonidis:2008ij}).
First of all we used the sums
\bea
\sum_{i=1}^n {0\choose i}_n =()_n
\label{A1111}
\eea
and
\bea
\sum_{i=1}^n {j\choose i}_n =0,~ (j \ne 0).
\label{A12}
\eea
Next a kind of standard relation, which allows to disentangle signed minors:
\bea
{\left(  \right)_n} {il\choose jk}_n  = {i\choose j}_n {l\choose k}_n - {i\choose k}_n {l\choose j}_n; ~~i,j,k,l = 0, \dots, n
\label{r2}
\eea
and finally, what we call
{\bf "master formula"} in the present context ( (A.13) of~\cite{Melrose:1965kb} )
\begin{eqnarray}
{s\choose i}_n {{\tau}s\choose 0s}_n=
{s\choose 0}_n{{\tau}s\choose is}_n+{s\choose s}_n {{\tau}s\choose 0i}_n,~~~ \tau = 0,1, \dots 5.
\label{Zauber}
\end{eqnarray}

The following correspondences have to be taken into account when comparing our notations to that of e.g.~\cite{Denner:2005nn}:
\bea\label{nota1}
()_5~~~~~~~ &=& -  \tilde{X}_{00}^{(4)} ,
\\\label{nota11}
{0\choose 0}_5~~~&=& \det {X}^{(4)} ,
\\\label{nota12}
{0\choose i}_5~~~ &=& \tilde{X}_{0i}^{(4)} ,
\\\label{nota13}
{i\choose j}_5~~~ &=& -  \tilde{Z}_{ij}^{(4)}  ~=~    \tilde{\tilde{X}}_{(0i)(0j)}^{(4)} ,
\\\label{nota14}
{0i\choose 0j}_5 &=& \tilde{X}_{ij}^{(4)} ,
\\\label{nota15}
{0i\choose kj}_5 &=&  \tilde{\tilde{X}}_{(0i)(jk)}^{(4)} ~=~    \tilde{\tilde{X}}_{(jk)(0i)}^{(4)} .
\eea
The notation in terms of $D_{\cdots}$-functions for $4$-point tensors with even number
of indices reads
\bea
\label{tensorx}
 I_{n}^{\mu\, \nu}& =& \int \frac{d^d k}{{i\pi}^{d/2}} k^{\mu} \, k^{\nu} \, \prod_{j=1}^{n} \, {c_j^{-1}}
 ~~~~~~~~ =  \cdots +
   \, g^{\mu \nu}  \, D_{00} ,
 \\
\label{tensory}
I_{n}^{\mu\, \nu\, \lambda\, \rho} &=&
\int \frac{d^d k}{{i\pi}^{d/2}} k^{\mu} \, k^{\nu} \,  k^{\lambda} \, k^{\rho} \, \prod_{j=1}^{n} \, {c_j^{-1}}
=\cdots  +g^{[\mu \nu} g^{\lambda \rho]}  D_{0000}  ~~~~~ {\rm etc.},
\eea
i.e. the functions $D_{00\cdots}$ correspond in our notation to integrals in higher
dimension $I_4^{[d+]^l}$ ($l=1,2, \cdots$) and our $Z_{4}^{[d+]^l}$-functions correspond
to approximations of the $D_{00\cdots}$-functions. We write
\bea\label{tensory2}
Z_4^{[d+]^l} &=& Z_{4F}^l + \frac{Z_{4D}^l}{\eps} + {\cal O}(\eps^2)
\eea
and the series ~(\ref{final}) thus uses only approximations
for the $D_{00\cdots}$-functions.

\clearpage
\section*{}
\addcontentsline{toc}{section}{References}


\end{document}